\documentclass[twocolumn,aps,pra,amsmath,amssymb,showpacs,superscriptaddress]{revtex4}

\usepackage{ucs}
\usepackage[utf8]{inputenc}  

\usepackage{graphicx}
\usepackage{graphics}
\usepackage{latexsym}
\usepackage{amsmath}
\usepackage{amsfonts}
\usepackage{amssymb}
\usepackage{subfigure}
\usepackage{psfrag}    
\usepackage{relsize}   

\def\openone{\leavevmode\hbox{\small1\kern-3.3pt\normalsize1}}

\begin{document}

\title{
Detection probability ``enhancement'' and ``unfair sampling'' in Bell inequalities\\
(and connections to a theory based on a noisy background)} 

\author{David Rodr{\'\i}guez}
\affiliation{Departamento de F{\'\i}sica Aplicada III, Universidad de Sevilla,
E-41092 Sevilla, Spain}
\email{drodriguez@us.es}

\date{\today}


\begin{abstract}
 \begin{center}  
\emph{
(this version: minor revision)} 
\end{center}

So far great part of the evidence accepted as proof of the
alleged quantum non-locality relied on inhomogeneous Bell inequalities,
involving an additional assumption (no-enhancement) whose role had not
been sufficiently examined
(in homogeneous inequalities, low detection rates play a similarly 
important role which is on the contrary well acknowledged).
Here we provide explicit examples of how a model of hidden local variables
(LHV) defying no-enhancement 
is able to produce a violation of an inhomogeneous inequality, a possibility
so far was suggested only in qualitative terms;
several more general quantitative results accompany these models.
Besides,
recent tests have attempted to overcome this reliance on supplementary
assumptions, but they still show weaknesses, here we focus on two of
them: ``time-biased sampling'' and ``local coincidence counts''.
At least for inequality-based tests, however, such weaknesses could perhaps
be easily bypassed by addressing a more exhaustive set of quantum predictions,
provided of course that the set of quantum states that can be really prepared
in a lab is not bound by local-realism.
\end{abstract}

\maketitle

\tableofcontents


\section{\label{Intro}Introduction} 


The success of Quantum Mechanics (QM) as a physical theory is
beyond any doubt. 
However there is still no incontestable (``loophole-free'') evidence that
quantum states defying local causality are more than just a by-product of
the quantum formalism, with no real physical counterpart.
Such state of affairs has remained unchanged for several decades, and
still does.


Of course there has been plenty of success in reproducing subsets of
quantum predictions corresponding to apparently non-classical states,
but this has been always done in conditions 
that either 
(i) easily allow for a local-realistic model (this is the case of
a certain type of Bell inequalities - homogeneous -, as we will soon
explain), 
or 
(ii) where the visibility (the ``violation'' of an inequality that
we try to observe in an experimental test) 
is still low enough to leave room for one of such local models, 
as we will be arguing here for several situations, in particular
the case of inhomogeneous Bell inequalities.

But moreover, the violations obtained in this second category of
experiments, case (ii), are not only always low in terms of the
visibility of the test, 
but indeed extremely low in comparison to which QM would in theory
be able to attain;
it would look as if the set of quantum mechanical states that
can actually be prepared in a lab was somehow restricted or at least
conditioned by their properties in relation to local-realism.

Usually, that impossibility to obtain high visibilities for inhomogeneous
inequalities has been blamed upon random detection errors.
That is certainly a possible interpretation, but perhaps not the
most plausible one attending at the following reasoning:
if such errors are really random and have nothing to do with the
properties of the quantum state under test, odds are they could
well appear (or become important, imposing a bound on the visibility)
at any other point of the spectrum of violations that QM can 
generate, 
for instance either well before or beyond the region that is critical
in relation to local-realism (LR), and not at is very brink.

Again, provided the detection errors are really random (``fair sampling''),
we could well expect to find this ``resistance against high visibilities'',
in relation to which they are customarily attributed such a main role, 
at any other region.
For instance, we could expect to find it at two thirds of the range
between the maximum values that local-realistic theories and QM can
attain (providing then incontestable evidence of a breach of LR), 
or well below the LR-frontier (which would not imply anything,
neither positive nor negative, in relation to whether local-realism
can be broken), 
or anywhere else within the range of visibilities that the test
theoretically allows. 

The fact that some hypothetical random errors allegedly conspire to keep
things exactly inside the box of local realism, not a smaller
or a bigger one, during so long and in such a variety of situations
(such a variety of tests) should not, and indeed can no longer
be considered a mere coincidence.
While this point does not subtract any merit from any other line of
research (all the opposite perhaps, as it can explain why a conclusive
result is so difficult to obtain),
it certainly provides justification to assert that the present one
is not only reasonable, but necessary too.


But leaving aside now this kind of considerations,
let me resume this introduction by focusing on the so-called Bell 
inequalities \cite{Bell64,CHSH69,CH74}, which have been
the main tools available to investigate the question of QM vs. LR.
In principle, Bell inequalities are derived solely on the hypothesis of
space-like separation and the axioms of probability,
and therefore, their violation implies accepting some sort of instantaneous
action-at-a-distance (which, quite strikingly, does not decay with increasing
distance) or, even more strikingly, a departure of realism itself.
This last point is not superfluous, as a recent experimental test
\cite{Kot_et_al12}, where no locality element is involved, would
point precisely to that.
However, once on experimental grounds things are not so straightforward.

In tests with massive particles, space-like isolation between the parties
is difficult to guarantee, what many refer to as the ``locality loophole''.
In experiments with photons, Pearle and Wigner \cite{Pearle70,Wigner70} already 
noticed early on that caution was needed when not all measurements expected
to produce a detection actually did such, 
a subtlety that left room, in most practical cases, for the existence
of a model making use of local hidden variables (LHV) and able to
reproduce the results of the test.
This is usually known as the ``detection loophole''.

Despite their somehow peculiar status as ``particles'' in nature,
hence perhaps not the ideal candidate to test such fundamental
questions,
Bell tests with photons seem to present clear technical advantages,
as well as very promising applications.
It is then understandable that most of the experimental effort has been
focused on them.
Bell inequalities usually tested in this context (photons) shall be
classified in \textit{homogeneous} and \textit{inhomogeneous}, 
attending to whether they contain terms (either correlations or simple
frequencies) of the same order (double coincidences, for instance)
or not;
see not \cite{hom}, or for instance \cite{Zela09} for recent reference
to these concepts.
Archetypes of these two classes are the Clauser-Horne-Shimony-Holt
\cite{CHSH69} and Clauser-Horne \cite{CH74} inequalities, respectively:
see App.\ref{CH_and_CHSH} for a quick reference.

In a real experiment and as already advanced, each type presents its
own weaknesses.
Homogeneoeus inequalities, while making possible the observation
of substantial violations, rely on a fair sampling of the
physical state being tested, 
something that, as widely recognized, can be challenged at least below
a certain threshold of the (relative) detection rate ($\eta$),
usually known as "critical detection efficiency" ($\eta_{\rm crit}$).

On the other hand,
inhomogeneous inequalities would theoretically seem to by-pass the possibility
of unfair sampling, a property derived from the involvement of probabilities of
different order (marginal and coincidence);
however, low detection rates also pose a problem due to their direct
quantitative effect on the observable frequencies, giving rise to yet another
critical value $\eta_{\rm crit}$ below which no violation is obtainable 
(on the contrary, high violations of homogeneous inequalities can be obtained
even for very low $\eta$'s, see App.\ref{App_2} for more details).
Yet, in spite of the detection loophole (i.e., the way in which low detection
rates burden the inequality) being, as already hinted, not entirely equivalent
in one and other case,
we can still generalize the term ``critical detection efficiency'' by giving
a definition convergent in both cases:
$\eta_{\rm crit}$ is simply the value of $\eta$ such that no LHV model can reproduce
the observed violation.

The problem of determining $\eta_{\rm crit}$, and that of finding scenarios with
the minimum possible $\eta_{\rm crit}$, have, as one could naturally expect,
consistently attracted a lot of attention, 
\cite{Eb93,GM87,Larsson98,LS01,CL07,CRV08,BG08,BGSS07,CLR09,Garbarino2010,PBS2011}.
A recent and particularly exhaustive effort, as well as close to ours here,
can be found in \cite{Joao}.
However, these works usually assume, at least in what regards inhomogeneous
inequalities and with few exceptions \cite{Joao}, 
additional restrictions on the LHV models model they aim to disprove, conditions
that may not be justified as we will later see.
Quite symptomatic is perhaps the fact that the converse question,
i.e., what physics the structure of the LHV models able to account for quantum
predictions may have been giving us hints about, has on the contrary enjoyed
almost no attention, little more than \cite{Risco_PHD}, actually.

In any case, 
critical detection efficiencies pose a severe problem in every single
Bell test making use of photons:
they are usually beyond what it has been achieved so far. 
Whether this is caused just by our technological limitations or it is the
expression of the fact that local-realism may be setting a constraint on the
results that are physically realizable is a matter of opinion.
What it is not is that alternative models have been proposed \cite{WPDC}
that describe the standard technique of Parametric Down Conversion (PDC),
which is of generalized use in (at least all the recent) developments,
and that can potentially explain such low detection rates, as a natural
consequence of their structure and regardless of additional inefficiency
factors one may wish to introduce;
i.e., taking to practice proposals such as \cite{BDDM05} may guarantee
a high detection efficiency, but this does not necessarily mean a
similar increase of the observed rate.
Such models have, too, received little or no attention at all.

At this point, there was no way forward but to modify the inequalities by
including some supplementary assumption that made them more suitable to be
tested experimentally;
this would motivate the distinction between genuine and non-genuine
inequalities, initially proposed by Santos \cite{genuine}:
genuine inequalities would not include supplementary assumptions,
non-genuine would.
The most usual supplementary assumption is probably Clauser and Horne's
\textit{no-enhancement} hypothesis \cite{no_enhancement}, on which we
will concentrate here;
based on \textit{no-enhancement}, several substantial violations of a
non-genuine version of the CH inequality have been accepted as incontestable
evidence of the gap between QM and local realism.
Nevertheless, such validation clearly hinges 
on the validity of the supplementary assumption.

The first 
contribution of this paper is to show clearly and directly how a breach of
\textit{no-enhancement}
can produce a strong violation of the corresponding non-genuine inequality. 
The second contribution is to explore the converse question to that posed by
the Bell inequality literature: 
what kinds of LHV models are able to account for the actually-observed quantum
predictions?  This question has enjoyed
almost no attention, little more than \cite{Risco_PHD} actually.
For example, alternative models have been proposed \cite{WPDC}
that describe the standard technique of Parametric Down Conversion (PDC),
which is used in at least all the recent developments.
Those alternative models can potentially explain the observed low detection
rates, as a natural consequence of their structure and regardless of additional
inefficiency factors one may wish to introduce;
i.e., taking to practice proposals such as \cite{BDDM05} may guarantee
a high detection efficiency, but this does not necessarily mean a
similar increase of the observed rate.
Such models also predict \textit{enhancement} (ENH), a breach of no-enhancement, 
but yet they have, too, received little or no attention at all.

Actually,
\textit{fair-sampling} has been invoked as a necessary supplementary assumption, 
with no other argument than its apparent ``reasonableness'';
however, fair sampling does not stand from the point of view of \cite{WPDC}
either, in particular once the correlation between the intensities originated
in the source of the PDC is taken into account.
The weakness of no-enhancement as a supplementary assumption
is even more compelling than that of fair-sampling, as it hardly requires
the sophistication of models like \cite{WPDC};
as we will argue later, just the presence of a random background, that
recent works in the field now acknowledge as well \cite{Steering12}.

For a quick overview on the main tests related to our work here
see \cite{Aspect02};
of course since then there had been many more experiments, many of
them addressing not a Bell inequality but other alleged properties of
quantum states that are nevertheless related.
It is symptomatic, however, that even those as recent as
\cite{Peruzzo12, Kaiser12} explicitly acknowledge a detection loophole
(which does not necessarily render useless their results);
on the other hand, the series of papers in \cite{WPDC} has explored
in detail, from a local-realistic perspective or at least one somehow
close to us here,
a considerable number of experimental results, though perhaps
failing to address some fundamental questions in a sufficiently direct
way (a local-realistic interpretation of the detection model, for
instance).

Finally, 
there are Bell inequalities which require neither fair sampling nor
no-enhancement as an additional assumption;
one is the so-called Eberhard's inequality which has been used in a
recent test \cite{G13}.
In this inequality, each non-detection is treated as just another proper
result, which negates the effect of the detection loophole;
however,
(i) space-like separation between the observers was not guaranteed,
though this may have possible been corrected in more recent iterations;
(ii) some other potentially relevant issues such as the appearance of
``local coincidences'' (see Sec.\ref{OT}) are also in need of
serious examination.

An exhaustive examination of supplementary assumptions for all
possible inequalities and scenarios being out of the scope
of this paper,
we hope the loss of generality may be somehow compensated by the
gain of credibility from a straightforward, merely algebraic
treatment, 
one that does not require to depart from sophisticated models 
or preconceptions.
With this in mind,
it will be convenient to 
start from a well known model ${\cal M}$
simulating the quantum prediction for the CHSH \cite{CHSH69} inequality
(and hence exhibiting unfair sampling).
From here, our program will include:

\vspace{0.2cm}\noindent
(i) 
showing that a new model ${\cal M}^{\prime}$ can be obtained from ${\cal M}$,
so as to contradict no-enhancement; 

\vspace{0.2cm}\noindent
(ii)
demonstrating with some examples how this model can also lead to a violation
of the Clauser and Horne inequality \cite{CH74} (and presumably of any other
non-genuine inhomogeneous inequality based on no-enhancement);

\vspace{0.2cm}\noindent
(iii)
and finally considering what it would require for these models to go further
than (ii) and adapt, simultaneously, to all quantum predictions for a chosen
state and set of observables:
here we will give some necessary and sufficient conditions for the existence
of such models in some scenarios.

\vspace{0.1cm}\noindent
This program is supplemented with other additional material, which I
consider necessary to place everything into proper context; 
such material includes an analysis of some recent experimental tests 
\cite{G13,Kot_et_al12} which,
though may only have an indirect relation to the main line of this paper,
carry however important implications for the general background issues,
and that from our point of view here should not be left unchallenged.

The paper is structured as follows:
Sec.\ref{Basic} is aimed at providing all the basic concepts, tools and
definitions for our work here.
Departing from the initial LHV model given in Sec.\ref{LHV},
we supplement it with new instructions predetermining detection probabilities
when polarizers are removed, so that the validity no-enhancement assumption 
can be challenged.
Then, 
Sec.\ref{ENH} provides a particular example an LHV of model leading to a
violation of the non-genuine version of the CH inequality;
Sec.\ref{Full_M} addresses the feasibility of the full compliance with quantum
predictions departing from the former model, and
Sec.\ref{Disc} presents a discussion on the origin of ``enhancement''
as a physical phenomenon, and possible tests that may be performed in
this regard.

Sec.\ref{OT} addresses additional questions, and 
Sec.\ref{RT} addresses recent reports of experimental evidence not
directly related to ENH.
Sec.\ref{UF_test} proposes a very basic test of unfair sampling, which 
is not a necessary condition but may help clarify things.
Finally, some further discussion is included in Sec.\ref{FD},
and conclusions and last comments are provided in Sec.\ref{Conc}.
The Appendix provides auxiliary proofs and some other supplemental
material which may be of use.

\begin{figure}[ht!] 
\includegraphics[width=0.95 \columnwidth,clip]{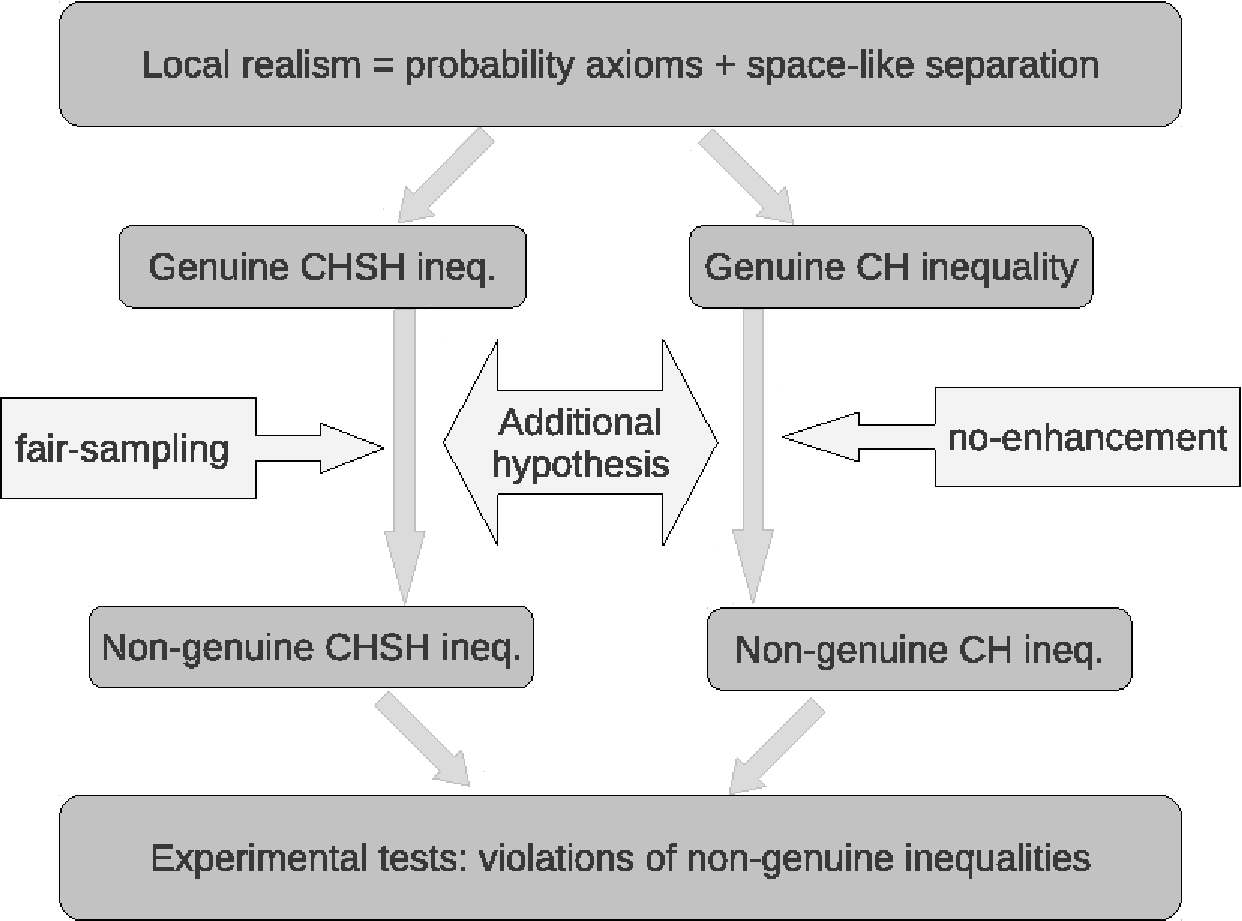} 
\caption{
At least so far,
experimental tests of Bell inequalities always include, implicitly
or explicitly, additional assumptions: (at least for our models here)
those assumptions are the ones that are violated, instead of local-realism.
The diagram does not intend to be exhaustive.
} \label{Scheme_Bell} \end{figure}

\section{\label{Basic}Basic concepts}

\subsection{\label{LHV}
An LHV model for the CHSH (optimal) correlations}

In an LHV description, the results of two pairs $A_1,A_2$ and $B_1,B_2$ of
space-like separated measurements can be expressed as a function of a vector of
hidden variables $\lambda$, and the respective orientations $\phi_i,\phi_j$
of the measuring apparatuses:
\begin{eqnarray}
A_i = A ( \phi_i, \lambda), \quad 
B_j = B ( \phi_j, \lambda), 
\label{obs}
\end{eqnarray}
assuming a deterministic description that will suffice for our purposes here
(any indeterminism can be modeled by adding new random variables to $\lambda$,
whether this are defined at the source or at each detector, see \cite{note_det}). 
We now need to introduce an LHV model that reproduces, for the former two
pairs of observables, the quantum correlations giving rise to a maximal
violation of the CHSH (\ref{CHSH}) inequality.
This model can be obtained as a particular case of the family of models
given in \cite{CLR09}, for the case $N=2$ (two observers, two observables
per observer).

In this model, every pair of particles is in a ``state'' $(A_1,A_2;B_1,B_2)$ that
determines the response of particle 1 when $A_1$ or $A_2$ is measured, and the
response of particle 2 when $B_1$ or $B_2$ is measured.
Each particle has 3 possible responses to the local measurements: being detected by the
detector $-1$, being detected by the detector $+1$, or being undetected.
We denote them as $-1$, $+1$, and $0$, respectively.
For instance, $(+1,-1;+1,0)$ denotes the state in which if $A_1$ ($A_2$) is measured,
then particle $A$ will give the result $+1$ ($-1$), and if $B_1$ ($B_2$) is measured,
particle $B$ will give $+1$ (will not be detected).

Let us also make use of the following conventions:
$P(A_i)$ is the probability that particle $A$ is detected (giving either $1$ or $-1$)
when $A_i$ is measured, $P(A_i|B_j)$ is the probability that particle $A$ is detected
when $A_i$ is measured conditioned to the fact that particle $B$ has been detected
when $B_j$ has been measured, $P(A_i,B_j)$ is the probability that particle $A$ is
detected when $A_i$ is measured and particle $B$ is detected when $B_j$ is measured.
Later we will use $P(A)=P(A=1)$, $P(B)=P(B=1)$ to denote that particle $A$,$B$,
respectively, is detected when the polarizer is removed, with $A=0$, $B=0$, denoting
absence of detection.

Now, assuming that all the detectors have identical detection efficiency $\eta$, and
that this efficiency is independent of the observable measured, any LHV model must
satisfy the following restrictions:
\begin{eqnarray}
P(A_i)=P(B_j) & = & \eta, \label{marginal} \\
P(A_i|B_j)=P(B_i|A_j) & = & \eta, \label{conditional} 
\end{eqnarray}
and, redundantly, $P(A_i, B_j) = \eta^2$ too, for all $i,j \in \{1,2\}$.
Besides, if the LHV model must reproduce the results of the Bell experiments on a
maximally entangled state, the following additional restrictions must be satisfied:
\begin{equation}
\langle A_i \rangle = \langle B_j \rangle = 0, \label{max_ent}
\end{equation}
for all $i,j \in \{1,2\}$, and, if the LHV model must reproduce the
maximum violation,
\begin{equation}
\langle A_1 B_1 \rangle = \langle A_1 B_2 \rangle = \langle A_2 B_1 \rangle = -\langle A_2 B_2 \rangle.
\label{maximalviolation}
\end{equation}
Defining now the following subsets of states:
\begin{eqnarray}
&&{\cal M}_P \equiv \{
(\pm 1,\pm 1;\pm 1,\pm 1),
(\pm 1,\pm 1;\pm 1,\mp 1), \nonumber\\
&&\quad\quad\quad\quad
(\pm 1,\mp 1;\pm 1,\pm 1),
(\pm 1,\mp 1;\mp 1,\pm 1)\}, \\ \nonumber\\
&&{\cal M}_Q \equiv \{
(\pm 1,\pm 1;\pm 1,0),
(\pm 1,\mp 1;0,\pm 1), \nonumber\\
&&\quad\quad\quad\quad\quad
(\pm 1,0;\pm 1,\pm 1),
(0,\pm 1;\pm 1,\mp 1)\}, \\ \nonumber\\
&&{\cal M}_R \equiv \{(\ 0,\ 0;\ 0,\ 0)\},
\end{eqnarray}
where for instance
$(\pm 1,\pm 1;\pm 1,0)$ actually means two states, $(+1,+1;+1,0)$ and
$(-1,-1;-1,0)$,
and letting 
$\beta = \langle A_1 B_1 + A_1 B_2 + A_2 B_1 - A_2 B_2 \rangle$
correspond to the value obtained in an experimental test of the CHSH
inequality (\ref{CHSH}), then, for
\begin{eqnarray}
\eta_{\rm crit}(\beta) = 2 / \left(1 + \tfrac{\beta}{2} \right), \label{eta_beta}
\end{eqnarray}
and 
\begin{subequations}
\begin{align}
p = \eta_{\rm crit}(\beta)  \ [\ 3 \eta_{\rm crit}(\beta)-2 \ ], \label{p} \\
q = 4 \eta_{\rm crit}(\beta)\ [\ 1-\eta_{\rm crit}(\beta) \ ],   \label{q}
\end{align}
\end{subequations}
the LHV model in which each of the states in ${\cal M}_P$ appears with frequency $p/8$,
each of the states in ${\cal M}_Q$ appears with frequency $q/8$, and the state in
${\cal M}_R$ appears with frequency $1-p-q$,
satisfies (\ref{marginal})--(\ref{maximalviolation}), and gives a $\beta$ consistent
with (\ref{eta_beta}) (all for $\eta = \eta_{\rm crit}$, see \cite{note_LHV_CHSH}).
Specifically, the maximal violation allowed by QM ($\beta = 2\sqrt{2}$) is obtained
when $p \approx 0.40$ and $q \approx 0.57$.
On the other hand, for the same $\beta$, other models (they are not unique) can be
obtained for $\eta < \eta_{\rm crit}(\beta)$ \cite{models_lower_heta}.

The sets ${\cal M}_P,{\cal M}_Q,{\cal M}_R$
cannot be experimentally discriminated, as this would require performing
the four measurements on a single pair or particles (photons);
nonetheless, they are a valid hypothetical construction once assumed the
existence of some vector of hidden variables $\lambda$, to which the occurrence
of one or other result is conditioned.
This point is important as from here on we may play with quantities such as
$P(A_1|A_2)$ which are clearly inaccessible from the physical point of view,
but yet perfectly defined from the purely mathematical.

Finally, we note that ${\cal M}$ violates ``fair-sampling'', as defined for
instance by Clauser \textit{et al} themselves \cite{fs}: it is enough to
see that even with the model satisfying (\ref{marginal})--(\ref{conditional}), 
we come across with that, in general, 
$P(A_i|B_j=b) \neq \eta$, $P(B_j|A_i=a) \neq \eta$. 
Indeed, from the model, for instance
\begin{eqnarray}
P(A_1|B_2=+1) = \tfrac{1}{3} \eta^2 + \tfrac{2}{3} \eta \neq \eta.
\label{UF_LHV}
\end{eqnarray}
In other words: restricted to the subset of pairs for which one of the particles has
a particular polarization ($B_j=b$ for particle $B$ or $A_i=a$ for particle $A$), the
probability of detection of the other particle is variable on the choice of observable,
clearly contradicting \cite{fs}.

\subsection{\label{Ext_LHV}
Detection probabilities without polarizers and the no-enhancement hypothesis}

We will now simply add two last instructions to each ``state'' of ${\cal M}$ (see
Sec.\ref{LHV}), obtaining a new model ${\cal M^{\prime}}$;
each state $s\in{\cal M^{\prime}}$ is now defined by a list of six (and not only four)
values: 
\begin{eqnarray}
s \equiv (A_1,A_2;B_1,B_2;A,B).
\end{eqnarray}
The last two instructions $A,B\in \{0,1\}$ simply tell if the corresponding particle
would be detected ('$1$') or not ('$0$') if no polarizer was placed on its way.
The LHV so defined should now also abide to the following set of (experimentally
testable) restrictions:
\begin{eqnarray}
P(A)=P(B) & = & \eta, \label{marginal_2} \\
P(A|B)=P(B|A) & = & \eta, \label{cond_2} \\
P(A_i|B)=P(B|A_i) & = & \eta, \label{cond_2_mix_1} \\
P(B_j|A)=P(A|B_j) & = & \eta, \label{cond_2_mix} 
\end{eqnarray}
and of course we would also have, this time redundantly, 
$P(A, B) = \eta^2$ and $P(A_i,B) = P(A,B_j) = \eta^2$, all $\forall i,j$;
all (\ref{marginal_2})--(\ref{cond_2_mix}) are conditions on the whole ensemble
of states $s \in {\cal M}^{\prime}$.
Indeed, let us consider, amongst them, $P(A_i|B)= \eta$, which, let $P_s(\cdot)$ be
a probability conditioned to a state $s$, really means
\begin{eqnarray}
\int_{\Lambda} P_s(A_i|B) \cdot P(s|\lambda) \cdot \rho(\lambda) \ d\lambda; \label{av_cond}
\end{eqnarray}
i.e., we do not need to satisfy $P_s(A_i|B) = \eta$, $\forall s \in {\cal M}^{\prime}$,
but simply do it ``on average''.
On the other hand,
the room for variability in $P_s(A_i|B),P_s(B_j|A)$ is obviously also there for $P_{s}(A_i|A),
P_{s}(B_j|B)$ (these last are of course not experimentally accessible, but are indeed
perfectly defined from the mathematical point of view), which in addition are not
even constrained by an ``average condition'' such as (\ref{av_cond}).
Therefore, in general we will have
$P_{s_1}(A_i|A) \neq P_{s_2}(A_i|A)$ 
and 
$P_{s_1}(B_j|B) \neq P_{s_2}(B_j|B)$, 
for $s_1 \neq s_2$ and $s_1,s_2 \in {\cal M}^{\prime}$. 
Following \cite{Risco_PHD}, we will call this a ``variable detection probability'' (VDP).


Now, the \textit{no-enhancement} assumption in \cite{CH74} stands for a
restriction on that VDP; in particular it stands simply for
\begin{eqnarray}
P_{s}(A_i) \leq P_{s}(A), \ P_{s}(B_j) \leq P_{s}(B), \ \forall s\in {\cal M}^{\prime}.
\label{n_enh}
\end{eqnarray}
The requirement on every state is already present in Clauser and Horne's 
original formulation \cite{no_enhancement}: ``for \emph{every emission $\lambda$}...'';
here each state $s$ corresponds to a particular $\lambda$.
Whenever a breach of (\ref{n_enh}) takes place, we will refer to it as
``enhancement'' (ENH).

\section{\label{ENH}ENH in experimental tests}

From here on, $P_{{\cal M}}(\cdot)$ will stand for a probability defined on
(any) LHV model (or subset of states) ${\cal M}$, 
and we will also assume that ${\cal M}$ defines all required probabilities, with
and without polarizers;
in absence of subscript, we will assume that by defect probabilities are defined
over the full model.
The following two quantities correspond to what one would be able to observe
in the respective tests of the CH inequality and its operational (non-genuine)
expression, on any model ${\cal M}$,
what we will respectively call, following Santos' classification \cite{genuine},
the ``genuine'' (GEN) and ``non-genuine'' (NG) tests:
\begin{eqnarray}
&&\beta_{gen}^{CH}({\cal M}) = \nonumber\\
&&\quad\quad
P_{{\cal M}}(A_1=B_1=1) + P_{{\cal M}}(A_1=B_2=1) \nonumber\\
&&\quad\quad\quad
+  P_{{\cal M}}(A_2=B_1=1) - P_{{\cal M}}(A_2=B_2=1) \nonumber\\
&&\quad\quad\quad\quad
- P_{{\cal M}}(A_1=1)
- P_{{\cal M}}(B_1=1), 
\label{gen_beta}
\end{eqnarray}
\begin{eqnarray}
&&\beta_{ng}^{CH}({\cal M}) = 
\tfrac{1}{\eta^2} \times [\ \nonumber\\
&&\quad\quad\quad
P_{{\cal M}}(A_1=B_1=1) + P_{{\cal M}}(A_1=B_2=1) \nonumber\\
&&\quad\quad\quad\quad
+  P_{{\cal M}}(A_2=B_1=1) - P_{{\cal M}}(A_2=B_2=1) \nonumber\\
&&\quad\quad\quad\quad\quad
- P_{{\cal M}}(A_1=1,B) - P_{{\cal M}}(A, B_1=1) \ ].
\nonumber\\ \label{ng_beta}
\end{eqnarray}
Further details on the derivation of these two expressions, can be found in
App.\ref{App_2}; 
in particular we refer to expressions (\ref{CH}) and (\ref{CH_op}),
respectively.
Right now we will see how ENH, while unable to alter the behavior
of the first of them, $\beta_{gen}^{CH}({\cal M})$, it definitely conditions
that of $\beta_{ng}^{CH}({\cal M})$.

\subsection{\label{Ex}An example of ``enhancement''}

There are many different models ${\cal M^{\prime}}$ consistent with restrictions
(\ref{marginal_2})--(\ref{cond_2_mix}); a particularly simple one is obtained from the
following assignations:\\
(i) $A=B=1$ for states in ${\cal M}_P$, obtaining an extended subset ${\cal M^{\prime}}_P$,\\
(ii) $A=0$ where some $A_i=0$ (for instance, when $A_2=0$, but not when $A_1=0$), and
$B=0$ where some $B_j=0$ (for instance, when $B_2=0$, but not when $B_1=0$), for states
in ${\cal M}_Q$, obtaining ${\cal M^{\prime}}_Q$,\\
(iii) $A=B=0$ for states in ${\cal M}_R$, obtaining ${\cal M^{\prime}}_R$.

\begin{figure}[ht!] 
\includegraphics[width=0.85 \columnwidth,clip]{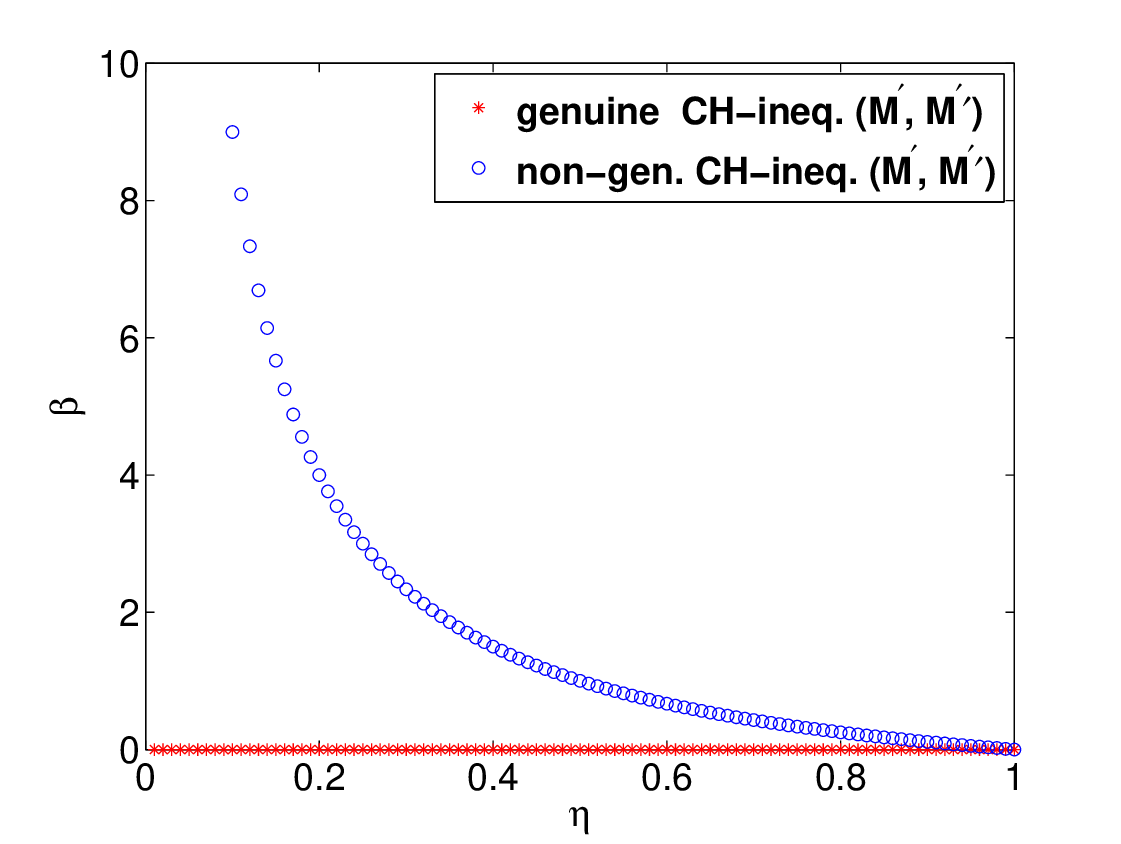} 
\caption{
Results in the genuine ($\beta_{gen}^{CH}$) and non-genuine ($\beta_{ng}^{CH}$)
experiments for the Clauser-Horne inequality, upon the LHV model 
${\cal M}^{\prime},{\cal M}^{\prime\prime}$, for a ``detection efficiency'' $\eta$.
As advanced, $\beta_{gen}^{CH}({\cal M}^{\prime},{\cal M}^{\prime\prime}) \leq 0$ (i.e,
below local bound) for all values of $\eta$,
while $\beta_{ng}^{CH}({\cal M}^{\prime},{\cal M}^{\prime\prime}) > 0$ for $\eta < 1$;
for $\eta = 1$ no VDP can take place (and therefore no ENH either) so we have,
necessarily, $\beta_{ng}^{CH}({\cal M}^{\prime},{\cal M}^{\prime\prime}) = 0$.
} \label{CH_comp} \end{figure}

\vspace{0.2cm}\noindent
With (i)--(iii), 
${\cal M}^{\prime}$ already satisfies (\ref{marginal_2})--(\ref{cond_2_mix}),
from the fact that ${\cal M}$ already did the same with 
(\ref{marginal})--(\ref{conditional}),
and all provided that
$P({\cal M^{\prime}}_P) = p$ and $P({\cal M^{\prime}}_Q) = q$, with $p$ and $q$
retaining, for a given $\eta$, their former values in (\ref{p})--(\ref{q}).
Actually we have:
\begin{eqnarray}
&&{\cal M^{\prime}}_Q \equiv \{
(\pm 1,\pm 1;\pm 1,0;  1, 0),
(\pm 1,\mp 1;0,\pm 1;  1, 1), \nonumber\\
&&\quad\quad\quad\quad\quad
(\pm 1,0;\pm 1,\pm 1;  0, 1),
(0,\pm 1;\pm 1,\mp 1;  1, 1)\}, \nonumber\\
\label{M_Q_prime}
\end{eqnarray}
where we note that some of the states now clearly defy (\ref{n_enh});
that is indeed the case of, for instance, the third pair (fifth and sixth
states) in ${\cal M^{\prime}}_Q$, where $A = 0$ but $A_1 = \pm 1$.
%
%
These quantities will be of interest in a moment:
\begin{eqnarray}
A \equiv P_{{\cal M^{\prime}}}(A_1=+1,B_1=+1) &=& \tfrac{3}{8}p + \tfrac{1}{4}q, \\
B \equiv P_{{\cal M^{\prime}}}(A_1=+1,B_2=+1) &=& \tfrac{3}{8}p + \tfrac{1}{4}q, \\
C \equiv P_{{\cal M^{\prime}}}(A_2=+1,B_1=+1) &=& \tfrac{3}{8}p + \tfrac{1}{4}q, \\
D \equiv P_{{\cal M^{\prime}}}(A_2=+1,B_2=+1) &=& \tfrac{1}{8}p, 
\end{eqnarray}
\begin{eqnarray}
E \equiv P_{{\cal M^{\prime}}}(A_1=+1) &=& \tfrac{1}{2}p + \tfrac{3}{8}q, \\
F \equiv P_{{\cal M^{\prime}}}(B_1=+1) &=& \tfrac{1}{2}p + \tfrac{3}{8}q, \\
E^{\prime} \equiv P_{{\cal M^{\prime}}}(A_1=+1,B) &=& \tfrac{1}{2}p + \tfrac{1}{4}q, \\
F^{\prime} \equiv P_{{\cal M^{\prime}}}(A,B_1=+1) &=& \tfrac{1}{2}p + \tfrac{1}{4}q.
\end{eqnarray}
Once here we can rewrite (\ref{gen_beta})--(\ref{ng_beta}) as
\begin{eqnarray}
\beta_{gen}^{CH} ({\cal M^{\prime}}) = A + B + C - D - E - F, 
\end{eqnarray}
\begin{eqnarray}
\beta_{ng}^{CH}  ({\cal M^{\prime}}) = 
\frac{1}{\eta^2}\left[ A + B + C - D - E^{\prime} - F^{\prime} \right].
\end{eqnarray}
As seen in Fig.\ref{CH_comp},
the introduction of ENH in ${\cal M}$ (obtaining
${\cal M}^{\prime}$) has been enough to qualify it to violate
the non-genuine version of the CH inequality.

\subsection{\label{Ex_2}A second example}

It has been very convenient, for simplicity, to ``lock'', in each particular
state $s \in {\cal M}_Q$, 
the fate of a detection with the polarizer removed, $P_s(A)$ and $P_s(B)$,
to that of a detection when one of the observables is measured, in this case 
$A_2$  and $B_2$, respectively.
Aside from looking rather unnatural, this feature is easy to disprove
experimentally \cite{Joao_2};
however, we will show now that such a choice is not at all necessary.
We do not need much sophistication, just a slight redefinition of our
states, now given by
\begin{eqnarray}
s \equiv (A_1,A_2;B_1,B_2; p_A, p_B), \label{redef_s}
\end{eqnarray}
with $p_A = P_s(A=1)$ and $p_B = P_s(B=1)$ (i.e., $s$ is in general
indeterministic in $A$ and $B$; a fully deterministic description can always
be recovered by re-expressing $s$ as a union of sub-states $s_i$ with
$p_A,p_B \in \{0,1\}$ and such that $\sum_i P(s_i) = P(s)$).
We also redefine our former subset ${\cal M^{\prime}}_Q$, in a way
also consistent with (\ref{marginal_2})--(\ref{cond_2_mix}), as
\begin{eqnarray}
&&{\cal M^{\prime\prime}}_Q \equiv \{
(\pm 1,\pm 1;\pm 1,0;  1, \tfrac{1}{2}),
(\pm 1,\mp 1;0,\pm 1;  1, \tfrac{1}{2}), \nonumber\\
&&\quad\quad\quad\quad\quad
(\pm 1,0;\pm 1,\pm 1;  \tfrac{1}{2}, 1),
(0,\pm 1;\pm 1,\mp 1;  \tfrac{1}{2}, 1)\}, \nonumber\\
\label{M_Q_primeprime}
\end{eqnarray}
and, finally,
consider a new model ${\cal M^{\prime\prime}}$ containing subsets
${\cal M^{\prime\prime}}_P \equiv {\cal M^{\prime}}_P$, ${\cal M^{\prime\prime}}_Q$
and ${\cal M^{\prime\prime}}_R \equiv {\cal M^{\prime}}_R$
with exactly the same frequencies $p,q,r$ as in ${\cal M^{\prime}}$.
It is easy to see that 
$\beta({\cal M^{\prime\prime}}) = \beta({\cal M^{\prime}})$ for all $\eta$'s
in Fig.\ref{CH_comp}, both for the genuine and non-genuine version, 
but now we also have what looks as a more reasonable behavior,
\begin{eqnarray}
P_{\cal M^{\prime\prime}}(A_1|A) = P_{\cal M^{\prime\prime}}(A_2|A), \label{J_r_1}\\
P_{\cal M^{\prime\prime}}(B_1|B) = P_{\cal M^{\prime\prime}}(B_2|B). \label{J_r_2}
\label{reasonable}
\end{eqnarray}
Anyway, from our point of view and as a difference with \cite{Joao}
(which we will discuss in Sec.\ref{Disc}),
(\ref{J_r_1})--(\ref{J_r_2}), or similar,
are demands that shall be introduced exclusively as \textit{a property of the state
under probe}, an element that we have not even considered yet at this stage of our
treatment: all we wanted to show is that they are compatible with ENH.
Besides, we note (\ref{J_r_1})--(\ref{J_r_2}) are, again, not experimentally
accessible but as an average estimation; they are ``average'' conditions, of
the type we have already seen in (\ref{av_cond}): 
particular states (in this case particular sub-states $s_i$ within each 
$s \in {\cal M^{\prime\prime}}_Q$) will in general defy it.

\subsection{\label{S_int}A mathematical interpretation}

Any LHV ${\cal M}$ defines a probabilistic space
$(\Lambda,\rho)$ where $\rho(\lambda):\lambda \in \Lambda \rightarrow [0,1]$;
let us now consider the subsets of events (or in LHV terminology, states)
$\Lambda_A, \Lambda_B \subset \Lambda$ as the ones where always $A=1$ and $B=1$,
respectively.
Then, (\ref{ng_beta}) can be rewritten as
\begin{eqnarray}
&&\beta_{ng}^{CH}({\cal M}) = 
\tfrac{1}{\eta^2} \times [\ \nonumber\\
&&\quad
P_{{\cal M}}(A_1=B_1=1|\Lambda) +  P_{{\cal M}}(A_1=B_2=1|\Lambda) \nonumber\\
&&\quad\quad
+  P_{{\cal M}}(A_2=B_1=1|\Lambda) - P_{{\cal M}}(A_2=B_2=1|\Lambda) \nonumber\\
&&\quad\quad\quad
- \eta P_{{\cal M}}(A_1=1|\Lambda_B) - \eta P_{{\cal M}}(B_1=1|\Lambda_A) \ ],
\nonumber\\ 
\label{ng_beta_2}
\end{eqnarray}
which clearly shows that
$\beta_{ng}^{CH}({\cal M}) \leq 0$, 
is not in general a legitimate Bell inequality (which means the bound can be
violated), 
because in general $\Lambda_A \neq \Lambda_B \neq \Lambda$, i.e., in general
the corresponding estimates are done on different subsamples, which means that
in general they do not keep statistical significance with respect to $\Lambda$.

It is important to advance that ENH (basically, as we will later see in
Sec.\ref{Disc}, a process that is statistically independent between the
two arms) 
cannot produce, by itself, the sort of correlations leading to unfair sampling
(at least as needed for a violation of a Bell inequality);
however, expression (\ref{ng_beta_2}) clearly shows how ENH can act as an 
``enabler'' of an unfair sampling that should be ultimately occasioned, as
we will also argue later, by the correlation between intensities arriving
from the source.
In other words, unfair sampling is a direct manifestation of the properties
of the state under probe, while ENH arises as a consequence of new vacuum
noise inserted at the polarizers.
This said,
the no-enhancement assumption \cite{no_enhancement} does not assure the
statistical significance of the ``marginal'' terms either, but only that
$\beta_{ng}^{CH} \leq 0$ remains a legitimate inequality:
see either \cite{CH74} or our App.\ref{NEN_conseq}.

Finally, the divergence of the (non-genuine) curves in the region of low
$\eta$ (Fig.\ref{CH_comp}) shall not cause concern:
indeed, the divergence would disappear if the models were forced
to satisfy restrictions on all probabilities involved,
such as (\ref{cond_joint}) in App.\ref{Full_M}: see \cite{div}.
Anyway, divergent or not, ${\cal M}^{\prime}, {\cal M}^{\prime\prime}$ 
are well defined LHV models and violate an inhomogeneous inequality;
this was so far thought impossible, as genuine and non-genuine expressions
were wrongly regarded, resting on the alleged validity of \textit{no-enhancement},
as equivalent at all effects.
Also symptomatic is the fact that a violation $\beta > 0$ can still be attained
for values of $\eta$ very close to unity.

\section{\label{Full_M}LHV models in full compliance with QM}

Once seen the role of ENH in inequalities, the natural question to ask is
if it has so much effect when the full set of quantum predictions are considered.
The models
${\cal M}^{\prime},{\cal M}^{\prime\prime}$ simulated just the CHSH correlations
as well as marginal probabilities of detection (with and without polarizers), 
in consistency with all experimentally testable restrictions;
however, they do not necessarily correspond to any legitimate quantum state
and set of observables.
For any model ${\cal M}$ to do so, we would have to demand:
\begin{eqnarray}
&&P_{{\cal M}}(A_i=a,B_j=b) = \nonumber\\
&&\quad\quad\quad\quad\quad\quad\quad
\eta^2 P_{QM}(A_i=a,B_j=b), \label{cond_joint}
\end{eqnarray}
as well as, depending on the test,
either
\begin{eqnarray}
P_{{\cal M}}(A_i=a) &=& \eta P_{QM}(A_i=a), \label{cond_g_A}\\
P_{{\cal M}}(B_j=b) &=& \eta P_{QM}(B_j=b), \label{cond_g_B}
\end{eqnarray}
or, instead,
\begin{eqnarray}
P_{{\cal M}}(A_i=a\ |\ B) &=& \eta P_{QM}(A_i=a), \label{cond_ng_A}\\
P_{{\cal M}}(B_j=b\ |\ A) &=& \eta P_{QM}(B_j=b), \label{cond_ng_B}
\end{eqnarray}
all of them $\forall i,j$ and for all $\forall a,b \in \{\pm 1\}$,
and where $P_{QM}$'s refer to a particular quantum mechanical state.
We will call (\ref{cond_g_A})--(\ref{cond_g_B}) the ``genuine'' (GEN) 
conditions, and (\ref{cond_ng_A})--(\ref{cond_ng_B}) the ``non-genuine''
(NG) ones;
we will also define, in consistency, two different critical parameters
\begin{eqnarray}
\eta_{\rm crit}(all;gen), \quad \eta_{\rm crit}(all;ng),
\end{eqnarray}
as higher bounds on the values of $\eta$ for which a proper model
can be obtained in the GEN and NG problems, respectively.
Actually, our former models ${\cal M^{\prime}},{\cal M^{\prime\prime}}$
already satisfy
\begin{eqnarray}
P_{{\cal M}}(A_i=a) 
= 
P_{{\cal M}}(B_j=b) 
= \tfrac{1}{2} \eta, 
\label{g_max_ent}
\end{eqnarray}
something not surprising because ${\cal M}$ in Sec.\ref{LHV} already
did, and
\begin{eqnarray}
P_{{\cal M}}(A_i=a\ |\ B) 
= 
P_{{\cal M}}(B_j=b\ |\ A) 
= \tfrac{1}{2} \eta, 
\label{ng_max_ent}
\end{eqnarray}
which would coincide with (\ref{cond_g_A})--(\ref{cond_ng_B}) for the case
of maximal entanglement (in that case, QM predicts an equally probable mix of
$\pm 1$'s): see (\ref{max_ent}).

\begin{figure}[ht!] 
\includegraphics[width=1.0 \columnwidth,clip]{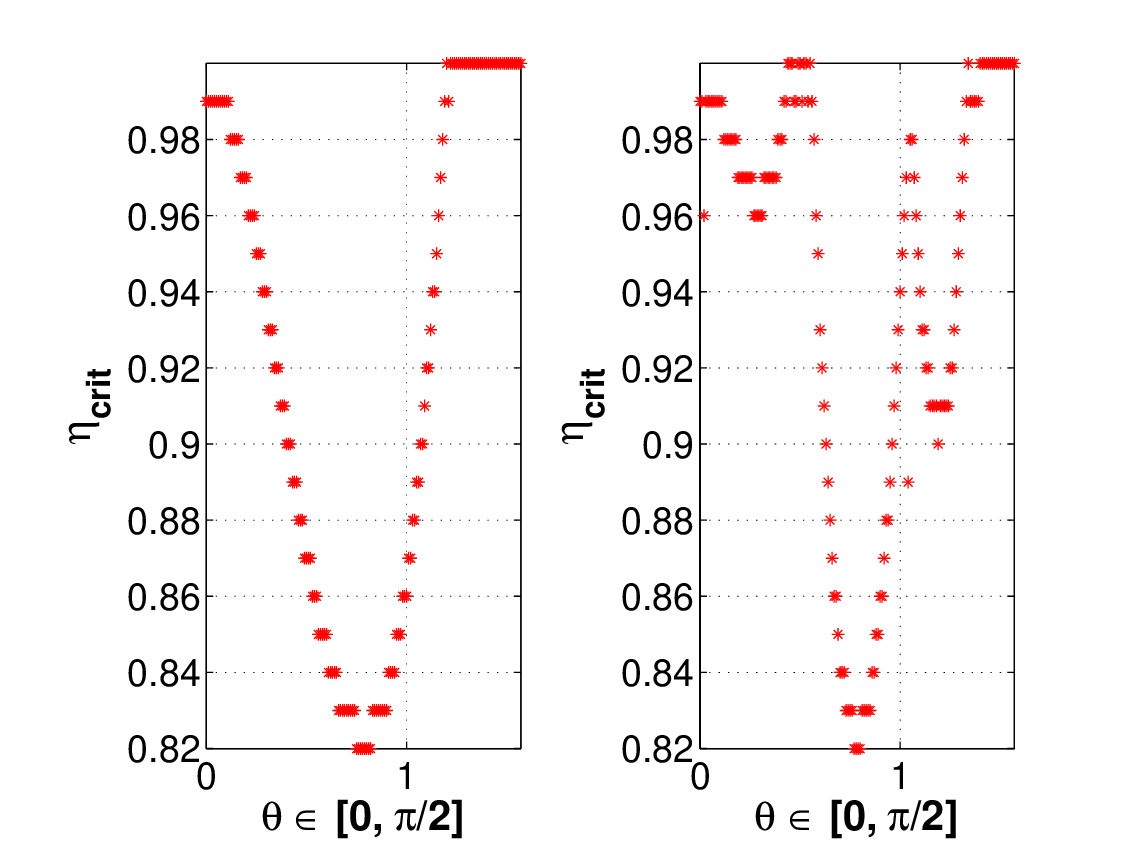} 
\caption{
Estimation of the critical values of $\eta$ guaranteeing the existence of ENH-based
LHV models \textit{for all quantum predictions} for the states $|\psi_1\rangle$ (left)
and $|\psi_2\rangle$ (right), defined in (\ref{q_state}),
and the four observables given in (\ref{A1})--(\ref{B2}).
Numerical calculation according to our layout in App.\ref{App_4}.
Genuine and non-genuine conditions (or combination of both) produce the same
curves: however, this should not be presumed for a single inequality alone
(see Fig\ref{CH_comp}), neither for other choices of state and observables.
Clearly, critical values for any test on $|\psi_{1,2}\rangle$ involving just
a subset of the quantum predictions considered here (for instance the test
of a particular CH or CHSH inequality) would necessarily be higher than
these ones.
} \label{s_y_t} \end{figure}

Beyond particular cases,
in App.\ref{Model_1by2} a proof is given that such an LHV exists for
any state and set of observables, under the hypothesis of balanced
(symmetrical) detection rates, 
for $0 \leq \eta \leq \tfrac{1}{2}$.
The condition there is sufficient, as it does not exclude other
possible models for $\eta > \tfrac{1}{2}$.
Moreover, App.\ref{App_4} provides a recipe for calculating
models for higher (in many cases optimal) $\eta$'s,
therefore at least a lower bound on the critical value.
I have applied it to the (maximally entangled) states
\begin{eqnarray}
|\psi_{1,2}\rangle = \tfrac{1}{\sqrt{2}}
\left(\ 
|\uparrow_{A,z} \downarrow_{B,z} \ \rangle \mp  |\downarrow_{A,z} \uparrow_{B,z}\ \rangle 
\ \right),
\label{q_state}
\end{eqnarray}
and the set of four observables
\begin{eqnarray}
A_1 &=&  \sigma_z, \label{A1}\\
A_2 &=&  \sin(2\theta)\cdot\sigma_x + \cos(2\theta)\cdot\sigma_z, \label{A2}\\
B_1 &=&  \sin(\theta)\cdot\sigma_x  + \cos(\theta)\cdot\sigma_z,  \label{B1}\\
B_2 &=&  \sin(3\theta)\cdot\sigma_x + \cos(3\theta)\cdot\sigma_z, \label{B2}
\end{eqnarray}
each one defined of course so as to act on the degrees of freedom of the
appropriate particle $A$ or $B$
(as rotations of the corresponding polarizer or PBS in ordinary space,
the former angles should be halved).
This choice of observables produces, for some values of $\theta$, maximal
violations of both the CHSH and CH inequalities, 
though we would need to write them with a different convention: see 
Figs.\ref{CH_and_CHSH_vs_theta_A} and \ref{CH_and_CHSH_vs_theta_B} below.

Up to our numerical resolution (which is not much, see the outliers),
the curves in Fig.\ref{s_y_t} seem to be exactly the same whether we choose to
impose NG or GEN conditions, i.e., 
(\ref{cond_ng_A})--(\ref{cond_ng_B}) or (\ref{cond_g_A})--(\ref{cond_g_B}),
or both sets of restrictions at the same time.
Anyway, we cannot presume such an equivalence for models on just the subset
of the quantum predictions involved in a particular inequality
(see Fig.\ref{CH_comp}).
Besides,
both $|\psi_1\rangle$ and $|\psi_2\rangle$ are maximally entangled, with
a symmetric distribution of $\pm 1$ that may play a role in relation
to this behavior.
Of particular interest is the case of the quasi-product states used in
\cite{LS01}; this seems to present some numerical difficulties
and will be examined elsewhere.

\section{\label{Disc}Discussion: plausibility of ENH}

In this section we provide physical support for what we have so far
just sketched mathematically;
we explore as well the possibility of an experimental test, and also
analyze in some detail the relation with other alternative theoretical
frameworks.

\subsection{ENH as a background effect}

A phenomenon such as ENH arises as a rather natural consequence of a wave-like
description of light, for instance of the kind of the one developed in \cite{WPDC}.
The same is also true for any other alternative one acknowledging the presence
of a background of random fluctuations of the electromagnetic field, either with
or without the properties of that predicted by Quantum ElectroDynamics (QED).

Let us consider a Bell experiment based on the technique of (Spontaneous)
Parametric Down Conversion: 
as modeled in \cite{WPDC}, the intensities arriving at detector $x$ ($x = o,e$:
``ordinary'' and ``extraordinary'', or alternatively simply $+1$ and $-1$) 
placed after the corresponding polarizer or polarizing beam splitter (PBS)
can be expressed as
\begin{eqnarray}
I^{(det)}_{A,x} = I^{(sc)}_A - I^{(ref)}_{A,x} + I^{(pol)}_{A,x}, \label{Idet_A}\\
I^{(det)}_{B,x} = I^{(sc)}_B - I^{(ref)}_{B,x} + I^{(pol)}_{B,x}, \label{Idet_B}
\end{eqnarray}
where $I^{(sc)}$, $I^{(ref)}$ and $I^{(pol)}$ are the intensities emitted
by the source, reflected by the polarizer (or diverted by the PBS) and, finally,
inserted (as a result of the existence of a background) at the exit channel
of the polarizer/PBS, respectively.

Now, the crucial point to grasp is that $I^{(pol)}_{A,x}$, $I^{(pol)}_{B,x}$
depend on a particular realization of the background amplitudes, and such
a realization will in general have components at both polarizations (whatever
the choice of basis):
as a result and as opposed to what it is generally assumed, \textit{each of
the detectors will receive intensity at both polarizations}.
Then, the overall intensities arriving at the detectors determine \cite{Glauber}
the corresponding probabilities of detection,
\begin{eqnarray}
P_{A,x}  \propto   I^{(det)}_{A,x} + C, \quad
P_{B,x}  \propto   I^{(det)}_{B,x} + C, \label{Pdet}
\end{eqnarray}
where $C$ is a constant arising from the normal order of field operators
in Glauber's expression \cite{WPDC};
indeed, $C = - I_0$ with $I_0$ the expectation value of the background
intensity for each particular frequency mode $\omega$
(realistic detection expressions involve an integral over the bandwidth of
the incoming wavepacket, as well as spatial and temporal integration).

Actually, that subtraction introduces an apparent problem due to a possible
negativity of the former expressions, which we will not address here and
on which we have already argued anyway \cite{DR_detec};
in any case, all we really need to support the phenomenon of
ENH as described in former sections is an expression for the probability
of detection that shows some dependence on the incoming intensity.

Leaving details aside, expressions (\ref{Idet_A})--(\ref{Idet_B})
clearly show two sources of variability:\\
(i)
the terms $I^{(ref)}_{A,B}$, clearly dependent on the polarization state
of the incoming wave (the analogue of the quantum mechanical state);\\
(ii)
the terms $I^{(pol)}_{A,B}$, introducing additional random components, 
which can now compensate the loss in (i), and even increase the overall
intensity producing ENH.\\
The usual models, where photons are regarded at all effects as particles whose
polarization state is governed by the algebra of half-integer spin (though
carrying integer units of angular momentum), 
can accommodate (in a certain way) the variability in (i), but ignore that
of (ii).

Though not indispensable for our purposes here, it may be interesting to
comment just some more details in (\ref{Pdet}).
On one side, $C \equiv - I_0$ does not compensate with $I^{(pol)}$ in
(\ref{Idet_A})--(\ref{Idet_B}), as these last refers to a particular realization
of the random background;
on the other, $I_0$ allows for the vacuum contribution to the energy to be
``subtracted'' on average.
To see this last we must consider that, according to \cite{WPDC} at
least,\\
(a)
oversimplifying again for the case of a perfect monochromatic signal of frequency
$\omega$,
\begin{eqnarray}
\langle I^{(pol)}_{A,x}(\omega) \rangle = \langle I^{(pol)}_{B,x}(\omega) \rangle = I_0(\omega);
\end{eqnarray}
(b) 
the intensities $I^{(sc)}_{A,B}$ (which also contain other background
components inserted at the crystal \cite{WPDC})
are already satisfying energy conservation with respect to the intensity of
the ``pump'' (laser).

According to (a)--(b), 
from the point of view of the experimenter the net effect of the
background should be to add a zero-mean variability to the observable rates.
On the other hand, the components in (ii) depend only on the properties of
the background, not on the angle of the polarizer or the PBS; 
therefore and as we had already advanced, (ii) cannot introduce
unfair sampling, in the sense of a variability of the following rates
\begin{eqnarray}
P(A_i|B_j = b), \quad P(B_j|A_i = a), \label{cond_fs}
\end{eqnarray}
where $a,b \in \{\pm 1\}$, and in respect to a range of angles in
setups $A$ and $B$.
However, the variability of (\ref{cond_fs}) is a necessary condition for
LHV models able to reproduce the observed violations of  homogeneous inequalities
(such as the CHSH inequality), and their associate set of quantum mechanical
correlations: see (\ref{UF_LHV}).

Such variability must therefore be due to (i) alone: rather than being
a source of unfair sampling, the role of ENH is that of ``creating room''
for it to manifest (in the non-genuine version of the inequality, under
the no-enhancement assumption);
in a model such as that of \cite{WPDC}, that unfairness of the sampling
is generated as result of source-induced correlations of the field
intensities arriving at the detectors 
(i.e., correlation between the fluctuations of the intensities that each
detector ``sees'').
An experimental test of the variability of (\ref{cond_fs}) does not
seem so difficult and could settle a lot of questions.






Additionally,
in what regards our treatment of LHV models in previous sections,
a formal analogy can be established between my approach here and that of
\cite{Joao}, and in consistency part of our results and interpretations
are also convergent.
The analogy is not complete, though:
in Sec.III, \cite{Joao}, all restrictions involving powers of $\eta$
higher than two go beyond the mere ``average'' independence of errors 
(see App.\ref{App_2}) that we are demanding here.

However, according to \cite{WPDC} and as just mentioned,
detection probabilities are not just determined by some efficiency parameter,
but also depend (through Glauber's expression) on the intensities arriving
at the detectors, intensities which are correlated between ``signal'' and
``idler'' arms.
More specifically, in \cite{WPDC}
part of the vector of hidden variables $\lambda$ can be identified with
the set of vacuum amplitudes $\{\alpha\}$ 
entering in the crystal, 
as well as an additional set of (uncorrelated) amplitudes per polarizer
or PBS, in this case $\{\alpha_{A,B}\}$, respectively.
This allows to rewrite (\ref{Idet_A})--(\ref{Idet_B}) as
\begin{eqnarray}
I^{(det)}_{A,x} = I^{(sc)}_{A}(\{\alpha\}) - I^{(ref)}_{A,x}(\{\alpha\}) + I^{(pol)}_{A,x}(\{\alpha_{A}\}), 
\label{Idet_Ax_alpha}\\
I^{(det)}_{B,x} = I^{(sc)}_{B}(\{\alpha\}) - I^{(ref)}_{B,x}(\{\alpha\}) + I^{(pol)}_{B,x}(\{\alpha_{B}\}), 
\label{Idet_Bx_alpha}
\end{eqnarray}
for $x = o,e$ 
(again, subindexes refer exclusively to how the label of the associated
detector, not to a determined polarization of the signal received at the
detector $x$),
and taking now into account also (\ref{Pdet}),
\begin{eqnarray}
P_{det}(A)   =    P_A (\{\alpha\}), \quad
P_{det}(B)   =    P_B (\{\alpha\}), 
\end{eqnarray}
but, however,
\begin{eqnarray}
P_{det}(A_i = a) &=&  P_{A_i}(a, \{\alpha\}, \{\alpha_A\}),	\\
P_{det}(B_j = b) &=&  P_{B_j}(b, \{\alpha\}, \{\alpha_B\}),	
\end{eqnarray}
where now $a,b$ ($\pm 1$ or $o,e$) play the role of the former label $x$.
Those expressions clearly show that, in the most general case,
none of the probabilities 
(of course some of them unobservable, but well defined mathematically)
\begin{eqnarray}
P(A_i \neq 0,A), \quad P(B_j \neq 0,B), \quad P(A,B),
\end{eqnarray}
needs be factorisable,
neither for a particular event nor as an average over the whole model, or any
sub-ensemble of it.
Therefore, from \cite{WPDC} full error independence is not only unjustified
(because of its unobservability) but an implausible condition.
Anyway, 
the insertion of $\{\alpha_A\}$ and $\{\alpha_B\}$ at the polarizers is, due to
its statistical independence between polarizers (between $\{\alpha_A\}$ and $\{\alpha_B\}$),
still captured by the attack in \cite{Joao}.
In lack of an exhaustive study, my view is that results in \cite{Joao}
should be considered as a restricted case of ours here; 
further examination will be desirable, too.

I must add that so far to my knowledge all experimental refutations
of the work in \cite{WPDC} are confined to former attempts to
interpret the detection probability expressions in consistency with
local-realism (LR).
For instance I am aware of \cite{Brida_et_al02} disproving the model of
detection proposed in \cite{W_LHV_02}.
Neither such attempts, nor other extensions of the formalism, such as
the one proposed in \cite{pdc10b} (apparently disproved by \cite{BG03}, 
too, only to regain credit again following some recent reports from two
different experimental groups \cite{SPUC})
have any implication on the core of the model (the interaction of
the vacuum amplitudes with a quadratic Hamiltonian, hence one that
preserves the positivity of the Wigner function);
neither on the fact that QED itself predicts ENH.

Finally and as an advance of Sec.\ref{Kot}, neither 
Kot {\it et al}'s result \cite{Kot_et_al12} can be considered conclusive
evidence of the existence of physical states with a negative Wigner
function, in a direct conflict with the principle of realism.
Yet, of course, the framework developed in \cite{WPDC} is no less
in need of experimental validation than are those quantum mechanical
predictions out of the LR frontier, conclusive evidence of
which has proven so elusive.
Neither is \cite{WPDC} something that cannot be refined or
modified at all, for instance by including new ``inputs'' of vacuum
modes.
Actually, it is my understanding that the properties of the QED-background,
some of them as puzzling as an isotropic, homogeneous and also divergent
density of energy, are not an indispensable element either: 
such background can be perhaps substituted by a more realistic source
of noise.


\subsection{\label{Other_E_sources}Other possible sources of ENH}


A recent model \cite{Graft} for the famous experiment by Weihs \textit{et al}
\cite{Weihs98} also makes use of a field-like formulation but
does not require the intervention of any background.
Interestingly, it can also easily be extended so as to exhibit
``enhancement'', 
choosing an intermediate value for the corresponding detector threshold
(see \cite{Graft}) when the polarizer/PBS is removed.
Assuming we could somehow extrapolate results (Weihs \textit{et al}
address a CHSH-type inequality, which does not require measurements
without polarizers), 
such threshold-induced ENH could potentially explain violations of the
non-genuine expression of the Clauser-Horne inequality, 
this time without the need to resort to background effects beyond a mere
second order correction.
It is obviously a point well deserving of attention; besides,
threshold calibration for the detectors has already been considered
in \cite{K11} as well.

\section{\label{OT}On recent tests}



\subsection{\label{RT}General comments}

Recently, efforts \cite{G13,Chr13} seem to have been focused on inequalities
that would in principle not suffer the detection loophole:
this is the Eberhard inequality \cite{Eb93}, or equivalently, a genuine
Clauser-Horne inequality which can be obtained directly from the former
and allows to evaluate it on a single channel setup.
None of these two inequalities can be violated by the occurrence of unfair
sampling alone
(which does not mean that their violation excludes unfair sampling, this
is another issue that should also deserve some comments).
However, they are still conditioned to other loopholes, in particular the 
locality one.
Besides, both \cite{G13,Chr13} both show similar hints of inconsistency, 
as detailed in Appendix \ref{RT_details}.
In the case of \cite{G13}, such inconsistency was first pointed out
by E.Santos in \cite{Santos_arxiv}, 
whose calculations I presume convergent with mine.

A recent post in \cite{Zeil_last} offers a possible explanation for
the anomalies of \cite{G13}, based on an alleged corruption of the prepared
state which must be hence described by a mixed state (which means
more degrees of freedom to adjust the results).
Another recent report in \cite{Chr13} also showed, at least in its first
version, similar hints of inconsistency (see App.\ref{RT_details});
apparently a similar strategy as in \cite{G13} can be applied to solve
these difficulties (private communication from \textit{nightlight},
and probably new post in arXiv or publication by Kwiat's group themselves).

Models as \cite{Zeil_last} prove the capacity of QM to fit (approximately)
the observations,
but in terms of the non-locality issue they are not, at least in my view,
necessarily convincing,
mainly because the violations obtained are of a very small magnitude
in regard to what QM could in principle achieve, even in the absence
of loopholes
(significant violations of Bell inequalities have indeed
been achieved upon homogeneous inequalities, but these admit a
local-realistic interpretation through the detection loophole).
The usual \emph{number of standard deviations} is irrelevant a criteria,
as it refers to mere random errors, not to possible
systematic effects such as the aforementioned ``local coincidences'' in
Sec.\ref{LC}, a possible loophole that has not been yet properly
explored.

The fact that both the observed violations and the so-called detection
efficiencies (which I would call detection rates) remain in the critical
region cannot be considered,
in particular after several decades of attempts, just mere coincidence.

There have been other recent tests aimed at proving non-classicality not in
terms of a Bell inequality but through perhaps (this is also
my opinion) more direct criteria:
for instance negativity of the Wigner function has been probed upon a photonic
state \cite{Kot_et_al12}, 
and some related ``quadrature quasi-probability upon the state of an ensemble
of massive particles \cite{Kiesel_et_al12}, 
in this last case by means of allegedly ``non-destructive'' measurements.
In absence of more detailed examination, such criteria use, as usual, a
non-commuting set of observables,
and in both cases a measurement procedure that I believe could open room
for a ``detection-loophole'', in the particular form that I will just
sketch here for the first;
attending at the indirect nature of the measurements in the second (which 
includes some alleged ``inefficiency'' of detection), I would also presume
it might be possible to approach a local-realistic interpretation in
a similar fashion, but this will be left for elsewhere.

\subsection{\label{LC}Local coincidence counts}

So far, all our previous models assume that a measurement $A_i$ or $B_j$
produces either a count on one of the detectors after the polarizing
beam splitter (PBS) or no detection at all;
however, it is a well known experimental fact that some events produce 
counts at both the $\pm 1$ detectors (alternatively, both the $o$ - ordinary -
and $e$ - extraordinary - exit channels of the PBS).

The usual approach is to regard such events as mere ``accidental
coincidences'' due to the presence of a noisy background \cite{Eb93}
(a ``conventional'' one, mere noise that does not need any of the
properties of the QED vacuum state);
this again ignores that models such as \cite{WPDC} do predict
those counts as the result of the PDC-generated states not being the alleged
``2-photon'' states, but a mixture of multi-photon states produced by
overlap of the emission times
(this overlap becomes particularly relevant if we consider wave-packets
with a certain time duration).
Such prediction has been found to be consistent with the observed statistics 
of PDC-pairs, characterizing the corresponding states as 
a Poissonian mixture of $2$-photon states: see \cite{Teich95}.

In any case, any quantum electrodynamical model making use of the Glauber
expression \cite{Glauber} predicts a non-negligible local coincidence
rate as the result of a non-negligible intensity arriving at the detectors
from both exit channels of the PBS.

\subsubsection{Inclusion in our models}

Now,
for the purpose of analysis let us suppose such ``local coincidences''
are included in our former models by means of a new hidden instruction
that we will denote $\gamma$.
Clearly, there are two basic types of events we have to be aware of:\\
(i) a local coincidence in one of the arms:
$\{ A = \gamma; B = \pm 1,0 \}$ or $\{ A = \pm 1,0; B = \gamma \}$;\\
(ii) double local coincidences: $\{ A = \gamma; B = \gamma \}$.

A double channel setup (monitoring both channels at the exit of the PBS, 
with their respective detector) would easily allow to discard events 
of the type (i) and (ii);
this is however very difficult in a single channel test such as the
recent ones in \cite{G13,Chr13}.
Let us first abbreviate 
\begin{eqnarray}
P_{i,j}(a,b) &\equiv& P(A_i=a,B_j=b), \\ 
P_{i,B}(a) &\equiv& P(A_i=a,B), \\ 
P_{A,j}(b) &\equiv& P(A,B_j=b), 
\end{eqnarray}
\begin{eqnarray}
P^{A}_{i}(a) &\equiv& P(A_i=a), \\ 
P^{B}_{j}(b) &\equiv& P(B_j=b), \\ 
P_{A,B} &\equiv& P(A,B),
\end{eqnarray}
as well as
\begin{eqnarray}
Q_{i,j}(a,b) &\equiv& P_{QM}(A_i=a,B_j=b), \\ 
Q^{A}_i(a) &\equiv& P_{QM}(A_i=a), 	   \\ 
Q^{B}_j(b) &\equiv& P_{QM}(B_j=b),
\end{eqnarray}
these last denoting quantum predictions.

\subsubsection{Local coincidences in the CH inequality}

Now, in order to keep the legitimacy of (\ref{gen_beta}) the following
substitutions are in order, 
\begin{eqnarray}
P_{i,j}(a,b) &\rightarrow & \eta^2 \cdot Q_{i,j}(a,b) - \Delta_{i,j}(a,b), 
\end{eqnarray}
\begin{eqnarray}
P_{i}(a)     &\rightarrow & \eta \cdot Q^{A}_{i}(a) - \Delta^{A}_{i}(a), \\
P_{j}(b)     &\rightarrow & \eta \cdot Q^{B}_{j}(b) - \Delta^{B}_{j}(b), 
\end{eqnarray}
all defined for $a,b \in \{\pm 1\}$ (i.e., $a,b \neq 0,\gamma$), with
\begin{eqnarray}
\Delta_{i,j}(a,b) 	&\equiv& P_{i,j}(\gamma,b) + P_{i,j}(a,\gamma) + P_{i,j}(\gamma,\gamma), \\
\Delta^{A}_{i}(a)     	&\equiv& \Delta_{i,j}(a,b) + P_{i,j}(\gamma,\bar{b}) + P_{i,j}(\gamma,0), \\
\Delta^{B}_{j}(b)     	&\equiv& \Delta_{i,j}(a,b) + P_{i,j}(\bar{a},\gamma) + P_{i,j}(0, \gamma),
\end{eqnarray}
defining $\bar{x} = \pm 1$ for $x = \mp 1$.
In the former expressions, none of the $\gamma$ events are accomodable by
the usual ($\tfrac{1}{2}$ spin algebra)-based quantum description of the
state, i.e., the $Q$'s 
(as already said, this would not at all be the case if we adopted a quantum
electrodynamical one).
Defining now the quantity
\begin{eqnarray}
&&M \equiv 
+ \Delta_{1,1}(1,1) + \Delta_{1,2}(1,1) + \Delta_{2,1}(1,1) \nonumber\\
&&\quad\quad\quad\quad\quad\quad\quad\quad\quad
- \Delta_{2,2}(1,1) - \Delta^{A}_{1}(1) - \Delta^{B}_{1}(1),
\nonumber\\\label{M_CH_def}
\end{eqnarray}
the ``selection'' ($\equiv SEL$, $NS \equiv$ ``no-selection'') of events
leads, for $\beta^{CH}_{gen}$ as in (\ref{gen_beta}), 
to an expression such as
\begin{eqnarray}
\beta^{CH}_{gen}(NS) = \beta^{CH}_{gen}(SEL) + M \leq M,
\label{M_CH_def_2}
\end{eqnarray}
by using, at the last step, $\beta^{CH}_{gen}(SEL) \leq 0$ (which was
the true legitimate inequality).

Though each term in (\ref{M_CH_def}) depends on the state under probe
and the choice of observables, 
from some general considerations (for instance imagine the angular
dependence is equal for all terms, so similar ones can annihilate each
other) 
the likelihood of $M < 0$ becomes more or less clear;
in particular in view that
$\Delta^{A}_{i}(a), \Delta^{B}_{j}(b) \geq \Delta_{i,j}(a,b)$.
Therefore, the positive bound of the inequality ($\beta \leq 0$) is very
unlikely to be compromised as a consequence of the $\gamma$-events alone;
a different thing may occur for the negative ($\beta \geq -1$).

On the other hand, 
for the non-genuine expression (\ref{ng_beta}) similar substitutions
are due:
\begin{eqnarray}
P_{i,B}(a) &\rightarrow & \eta^2 \cdot Q^{A}_{i}(a) - \Delta_{i,B}(a), \\
P_{A,j}(b) &\rightarrow & \eta^2 \cdot Q^{B}_{j}(b) - \Delta_{A,j}(b). 
\end{eqnarray}
Assuming again
$\Delta_{i,B}(a), \Delta_{A,j}(b) \geq \Delta_{i,j}(a,b)$,
we arrive to results analogous to those of the genuine case.
In any case,
\textit{local coincidences should not be ignored in a test of either
the genuine or non-genuine Clauser-Horne inequality}, 
something particularly sensitive when such test is focused on the
lower bound (i.e., when we evaluate if $\beta \geq -1$).
Usually this issue is (allegedly) taken care of by correcting $\eta_{\rm crit}$ 
(critical detection rate) based on some ``background'' estimation;
again, not enough as argued at the beginning of the section.


\subsubsection{Local coincidences in the Eberhard inequality}


As a difference with other proposed inequalities,
Eberhard's inequality \cite{Eb93} does not ``select'' events with regard to
detection, neither it assumes an additional hypothesis such as 
\textit{no-enhancement}.
Nevertheless, Eberhard's inequality requires space-like separation, which is
not guaranteed in the recent test by Giustina \textit{et al} \cite{G13};
but beyond that ``locality loophole'', the same considerations can also be
applied here.
In \cite{Eb93} and \cite{G13} the Eberhard inequality is written as:
\begin{eqnarray}
&& J = 
- n_{oo}(\alpha_1,\beta_1)
+ n_{oe}(\alpha_1,\beta_2)
+ n_{ou}(\alpha_1,\beta_2)
\nonumber\\
&&\quad\quad\quad
+ n_{eo}(\alpha_2,\beta_1)
+ n_{uo}(\alpha_2,\beta_1)
+ n_{oo}(\alpha_2,\beta_2)
 \geq 0,
\nonumber\\
\label{Eb_NS_prev}
\end{eqnarray}
with $n_{ab}(\alpha_i,\beta_j)$ the number of counts registered as
corresponds to results $a,b \in \{o,e,\emptyset\}$ in the associated
side and detector 
($o,e$ are labels at all effects equivalent to $\pm 1$).
To subtract the double coincidences, we would have to perform, on
the former expression, the substitutions
\begin{eqnarray}
n_{ab}(\alpha_i,\beta_j) \rightarrow n_{ab}(\alpha_i,\beta_j) - \bar{n}_{ab}(\alpha_i,\beta_j),
\end{eqnarray}
which lead in this case to
\begin{eqnarray}
\beta_{Eb.}(NS) = \beta_{Eb.}(SEL) + M \geq M,
\label{M_def}
\end{eqnarray}
and where
\begin{eqnarray}
&&M = 
- \bar{n}_{oo}(\alpha_1,\beta_1)
+ \bar{n}_{oe}(\alpha_1,\beta_2)
+ \bar{n}_{ou}(\alpha_1,\beta_2)
\nonumber\\
&&\quad\quad\quad\quad
+ \bar{n}_{eo}(\alpha_2,\beta_1)
+ \bar{n}_{uo}(\alpha_2,\beta_1)
+ \bar{n}_{oo}(\alpha_2,\beta_2),
\nonumber\\
\end{eqnarray}
is with great probability positive this time.
Therefore, in principle \textit{local coincidences} are unlikely to produce
a violation in (\ref{Eb_NS_prev}), though once again such a possibility cannot 
really be discarded ``a priori'' 
(once more, please notice that all terms in the former expression depend
on the choice of state and observables) either.

On the other hand, 
the operational expression used in \cite{G13} (eq.4) is actually equivalent
to a (sign-reversed) genuine Clauser-Horne inequality (\ref{gen_beta}),
which reduces to the case already examined in Sec.\ref{LC}.
The likelihood of local coincidences compromising the $0$-bound is 
dependent on the choice of state and observables, however low it may be
we cannot discard this possibility either.


\subsection{\label{Kot}
Kot \textit{et al}: time-biased sampling?}

The recent test in \cite{Kot_et_al12}, apparently disproving the existence
of a well defined probability density function for a set of local observables,
could be perhaps interpreted under a formal equivalence with an homogeneous
Bell inequality: only one observable is measured at each ``run'' of the
test (a measurement upon one pair of emitted photons).
However, two differences are clear:\\
(i) it does not require remote observers;\\
(ii) it makes use of analogical measurements in one of the arms of the PDC
scheme, in a way that would seem to exclude the detection loophole, 
at least as we think of it regarding the usual tests.

Basically,
Kot {\it et al}'s proposal in \cite{Kot_et_al12} rests on the probing
of some ``test function'' 
\begin{eqnarray}
F \equiv F(Q_{1},Q_{2},\ldots Q_{N}),
\end{eqnarray}
where $Q_{m}$ is an outcome obtained when an observable $\hat{Q}_{m}$ is
measured, and where $\{\hat{Q}_{m}\}$ is a set of mutually exclusive
observables
(more precisely, according to eq.8 in \cite{Kot_et_al12}, $F$ involves
a set of powers $\{ (Q_{m})^{2n}, \quad m,n = 1,\ldots N\}$),
as well as the test of an inequality 
\begin{eqnarray}
\beta = \langle F \rangle \geq 0,
\end{eqnarray}
that local realism (LR) could not in principle violate.
However, each $\hat{Q}_{m}$ may be then sampled on a different subset
$\Lambda_m \subset \Lambda$... which would be no problem as long as
all $\{\Lambda_m\}$ are statistically faithful to $\Lambda$.

To show that such a thing may happen even in view of (ii), we must go
to Glauber's expression \cite{Glauber}.
The key is that, for a given time-stamp, in general $P(t) < 1$:
the hypothetical $0$-instructions in a hypothetical LHV model 
would no longer have anything to do with some ``detection
inefficiency'', but simply express the fact that for some given time-stamp
and observable, $P(t) \neq 1$.

Once here, the physical connection with the test in \cite{Kot_et_al12} can be
established by assigning to each time stamp $t$ a different set ${\cal M}(t)$
of hidden instructions,
\begin{eqnarray}
t_1 \rightarrow {\cal M}_1 = {\cal M}(t_1), \quad
t_2 \rightarrow {\cal M}_2 = {\cal M}(t_2), \quad
\ldots
\end{eqnarray}
i.e., a different LHV model for each $t$.
In particular, following \cite{prev07}, a detection on the ``signal'' arm, at a
time-stamp $t$, prompts the analysis of the signal (coming from the homodyne setup
and entering a high frequency oscilloscope) at the ``idler'' one, over a fixed
time window.
Yet, a correlation between the detection time-stamp at the
signal arm and the choice of observable at the idler would seem to require
communication or ``signaling'' between the two measurement setups:
the conditions of the test do not exclude such cross-talk.

Anyway, there are reasons to presume that such signaling is not 
necessary either: 
again in consistency with \cite{WPDC},
the set of relevant vacuum electromagnetic modes inserted at the source are
still contained in the fields arriving at each detector, and may make such
correlation possible.
As the simplest possible mathematical example, 
let us again consider an observable $\hat{Q}$ that (always) produces
two possible results $Q = \{q_1,q_2\}$, and let us denote $P_{Q}(q|\lambda)$
the probability of an outcome $q$ when $Q$ is measured on the state $\lambda$
(according to \cite{WPDC}, $\lambda$ includes the vacuum amplitudes
inserted at the source).
Then,
\begin{eqnarray}
\langle Q \rangle_{\Lambda} = \frac{ 
q_1 \cdot P_{Q}(q_1|\Lambda) + q_2 \cdot P_{Q}(q_2|\Lambda)}
{P_{Q}(det|\Lambda)};
\end{eqnarray}
however, if each result is associated to a detection at a different set of
time-stamps $\{t_{i,1}\},\{t_{i,2}\}$
(realistically, this correspondence would be only in terms of unbalanced
probability), 
then the experimentally accessible quantity is
\begin{eqnarray}
\langle Q \rangle_{ob} = \frac{ 
q_1 \cdot P_{Q}(q_1|\Lambda^{(Q)}_1) + q_2 \cdot P_{Q}(q_2|\Lambda^{(Q)}_2)}
{P_{Q}(det|\Lambda)}.
\end{eqnarray}
In this last expression,
$\Lambda^{(Q)}_1,\Lambda^{(Q)}_2 \subset \Lambda$ are again two
different sets selected by the correlation between the time-stamp at
one arm and the result of the measurement at the other when $\hat{Q}$
is measured.

The normalization factor $P_{Q}(det|\Lambda)$ simply generalizes to
the case where some attempts to measure an observable may fail, 
which may or may not be the case: my argument works anyway 
(of course as a mere formal device, further examination is needed to see
whether it can really be applied to the results of \cite{Kot_et_al12}).

%

\section{\label{UF_test}
A basic test of unfair sampling}

As already said,
the problem of finding scenarios with the minimum possible $\eta_{\rm crit}$, 
has attracted a lot of attention,
but the converse question,
i.e., what physics the structure of the LHV models able to account for quantum
predictions may have been giving us hints about, has on the contrary enjoyed
almost none.
Let us once more look at the family of LHV models given in Sec.\ref{LHV}; 
for the case $N = 2$ (two observers) and the set of observables that
maximizes the value of the Clauser-Horne-Shimony-Holt inequality,
we come across with the fact that in general
\begin{eqnarray}
P(A_i|B_j = b) &=& f(b), \\
P(B_j|A_i = a) &=& f(a), \label{uf_condition}
\end{eqnarray}
for some particular choice of $a,b \in \{-1,+1\}$, 
an instance of which we can find in
\begin{eqnarray}
P(A_1|B_2 = +1) = \tfrac{1}{3} \eta^2 + \tfrac{2}{3} \eta \neq \eta,
\end{eqnarray}
and where $\eta$ is a function of the overall value of the inequality
$\beta$ (i.e., $\eta = f(\beta)$) as indicated in \cite{CLR09}.

In other words: restricted to the subset of pairs for which one of the particles has
a particular polarization ($B_j=b$ for particle $B$ or $A_i=a$ for particle $A$), the
probability of detection of the other particle is variable on the choice of
observable, something that contradicts Clauser and Horne's own definition of
unfair sampling \cite{CHSH69}:
``given a pair of photons emerges from the polarizers, the probability of their
joint detection is independent of the polarizer orientations''.


In view of the former, it might be a good idea to dispose a double channel
setup (instead of mere polarizers, we use polarizing beam splitters or PBSs
and we monitor both their output $o \equiv +1$ and $e \equiv -1$ channels) and
quantify (\ref{uf_condition}), 
for a battery of different angles (additionally, a full set of
angles could be probed for each instance of the prepared quantum state).
Adopting the notation in \cite{G13},
with $n_{ab}(\alpha_i,\beta_j)$ is the number of counts registered for the 
angles $\alpha_i,\beta_j$ and where the sub-indexes admit the values $o,e$
but also $u \equiv$ undetected, we would write
\begin{eqnarray}
P(A_i|B_j = b) =  \frac{ 
  \mathlarger{\mathlarger{‎‎\sum}}_{a = o,e} n_{ab}(\alpha_i,\beta_j) }
{ \mathlarger{\mathlarger{‎‎\sum}}_{a = o,e,u} n_{ab}(\alpha_i,\beta_j) }, 
\label{A_a_B}
\end{eqnarray}
as well as the analogous expression for $P(B_j|A_i = a)$, for $a,b \in \{o,e\}$
and $i,j \in \{1,2\}$.
Unfair sampling would then manifest as a non-negligible, statistically significant
variability of (\ref{A_a_B}) and other similar expressions, as we rotate 
one of the PBSs.
Notice that this variability cannot be explained with an ``external'' efficiency
factor $\eta$ due to the dependence on the polarization state manifested at
the other side:
this dependence shall not be attributed to some signaling between the parties
(or even some ``non-locality'') but merely to the fact that detection rates
are at least partially determined by the set of hidden variables $\lambda$ that
characterizes the state generated at the PDC-source.

Of course, additional factors may cause variability of (\ref{A_a_B}), for
instance non-stationarity of the prepared state.
Besides, our condition is not even necessary: the absence of variability 
in (\ref{A_a_B}) does not exclude unfair sampling either.
But nevertheless, it seems a necessary step: \textit{how are we suppose to believe
a sophisticated criteria such as a Bell inequality has been fairly evaluated 
without knowing for clear what is happening at more basic levels?}

The question retains interest even if a conclusive test of non-locality
is achieved:
contrary to what it seems to be taken for granted in \cite{G13},
even in the event of a ``loophole-free'' violation of local-realism, 
unfair sampling may still remain an important feature of the behavior
of PDC-generated states.

\section{\label{FD}
Further discussion: considerations on the nature of the background}

So far we have been invoking the 
(statistically) homogeneous, isotropic, Lorentz invariant background predicted
by QED as the ``input'' to a hypothetical local-realistic theory that could
provide a more detailed, credible account of the optical tests based on PDC,
a theory that would share its basic features with that of \cite{WPDC};
anyway,
we must admit the idea of a background with such ``supernatural'' properties
(and the problems it carries, one of them an infinite energy density, which
is not to be ignored) looks at least as unrealistic as quantum entanglement
between space-like isolated parties
(nothing to object on entanglement as an expression of local correlation
or ``cross-talk'', as I would assume is taking place in Bell experiments
with massive particles).

A possible solution is to regard the QED-background as a mere ``intermediate
step'', nothing but an abstraction resultant from the mathematical structure of
the quantum formalism, structure that may be a consequence of some very simple
hypothesis.

A discussion on this is nothing more than mere speculation,
but anyway I will say that it is my view that QM is simply the simplest
mathematical formalism imposing angular momentum
quantization (AMQ) over all its states.
Of course here I am depriving quantum states of any ontological
implications but for its suitability to represent, at least approximately,
information about the observable properties of the physical systems.

Since the beginning of what is known as Stochastic ElectroDynamics (SED),
it has been recognized that any stable dynamics of a system of charger
implies an equilibrium between radiated and absorbed power which,
as a consequence, quantizes the value of the average angular momentum
(QM must imply, then, some sort of spatio-temporal average).
Out of bound states this AMQ is of course not justified, but is my
conjecture that then, there, such assumption is quantitatively irrelevant
at least at the observational level.

So far, enthusiasts of Stochastic Electrodynamics (SED) seem to accept
the properties of the QED-background as the necessary element making 
possible (the stability of) that equilibrium;
for the reasons above, my view is that it might not be a false step at all
to try to look for such element somewhere else: the most obvious case,
the dynamical properties of elementary particles as complex, sub-structured
entities, capable of storing and releasing, therefore exchanging energy
with a ``realistic'' background, and doing this in a way that it induces
such stability.
Such micro-dynamics would have been completely ``traced out'' by QM
as a mere effective theory expressing the probability distribution of
observables that are nothing but time-averages (therefore abiding to
the former equilibrium);
only the average features caused by the dynamics of such substructure 
remain in the theory, for instance the appearance of ``spin''.

This last route does not need more than a much more ``modest'' background,
without a divergent or even considerable energy density, and perhaps
also highly inhomogeneous and dependent on the distribution of matter, 
this last a particularly plausible conjecture if the ultimate origin
of this vacuum noise is to be tracked down (as suggested by some authors,
Puthoff for instance) to the very same micro-dynamics of the substructure
of charges that QM may be averaging. 
Yet again, a pure field ontology for the photon also carries many
other difficulties, 
as is how and in which conditions a wave-packet is able to  travel long
distances with almost no spread, and why observable energy exchanges
between matter and the electromagnetic field are constrained to a ``quanta''
$h\omega$.  

None of those problems seem, however, unsolvable from a the framework of
classical electromagnetism:
for instance we know that systems with a rich spatial structure can give
rise both to highly directional radiation patterns and, again, the appearance of
``dynamical attractors'' in their phase space, with an associated discrete
spectrum of observable energies as well as of apparent state transitions.
The geometrical symmetry of quantum energy eigenstate wave-functions
would appear to render impossible that directionality; however, neither a
particular realization of those densities, nor some hypothetical 
underlying substructure micro-dynamics (averaged both in time and space by QM)
would be bound to that symmetry.

There is plenty of work on the local-realist, stochastic electrodynamical side
which is extraordinary (Casado and co-workers, previously that
of Marshall and Santos on their own, related works by Boyer, Puthoff or D.C.Cole,
just to name a few), ignored over decades now and to which we may eventually
have to go back to.
More recently, \cite{FH} proposes a unified view departing from a
background, too, though this time its nature is not necessarily known: 
in particular and amongst many other results, an explanation of those
striking ``double slit'' phenomena which looks rather satisfying.

\section{\label{Conc}Conclusions, and last comments}

The results of this paper suggest that the role of additional assumptions
in Bell tests, however implicit or explicit these were, has been grossly
overlooked.
For instance, so far tests of inhomogeneous inequalities making use of the
no-enhancement hypothesis were perceived as solid evidence of quantum
non-locality; here we have shown that this can be challenged.

At the purely mathematical level, we have provided models allowing for what we
have called ``enhancement'' (ENH), a breach of the no-enhancement assumption;
these show a compelling consistency with existing models of local hidden
variables (LHV) for the case of homogeneous inequalities, where
the need for fair-sampling as an additional assumption is, in contrast with
what happens for no-enhancement, widely acknowledged.
But beyond that, there are also physical arguments to support such models:
for instance the existence of a random background,
whether we choose to take the QED-predicted Zero Point Field of vacuum
fluctuations \cite{WPDC} or some other background of even an uncertain
origin \cite{FH}.

Remarkably, in a model such as \cite{WPDC} (and other possible ones based on
Glauber's expression) the expectation value of the vacuum field intensity is
subtracted at the detector, 
which gives rise to a variability of the detection probability;
interestingly, it does not require a net average energy contribution from the
background (clearly a desirable property).
Besides, ENH can be generated from other sources, too: following recent developments, 
from a (quantum electrodynamical) model based on choices of the detectors' thresholds,
either involving also the background \cite{K11}, or not \cite{Graft}.
Threshold variability is a refinement that could also be applied to
\cite{WPDC}, relegating (or not) background effects to a secondary level.

Conveniently, there is a unifying property of all LHV-based models for Bell tests,
and also related ones such as that of \cite{Kot_et_al12}: 
whatever the situation, their existence is only possible for values of the
detection rates below a threshold known as ``critical detection efficiency''. 
This tern is in my opinion misleading: it implies a loss of generality by
implicitly assuming that the values of the rates have nothing to do with the
physical state under probe,
something that can be challenged again from \cite{WPDC}, but possibly also
from a careful examination of exhaustive sets of data.
I therefore propose a substitution in favor or ``critical detection rate''.

Actually, 
in a bipartite test (two observers $A,B$) only the following detection
rates can be defined without normalization factors or additional hypothesis (such
as fair sampling),
in plain words, only these detection rates make sense as a quantitative element
or comparison:
\begin{eqnarray}
&\eta_{(A_i|B_j)} = P(A_i \neq 0 |B_j \neq 0), \label{cond_rate_A}\\
&\eta_{(B_j|A_i)} = P(B_j \neq 0 |A_i \neq 0), \label{cond_rate_B}
\end{eqnarray}
as well as, when non-genuine expressions are considered, $P(A_i|B)$, $P(B_j|A)$
and $P(A|B)$ too.
Whenever $\eta_{(A_i|B_j)},\eta_{(B_j|A_i)} < 1$, 
fair-sampling is not guaranteed and any expression other than the genuine
inequality requires a cautious interpretation.
%

According to our work here such critical values should be revised to
account for the possibility of ENH, whenever the corresponding test has made use
of the no-enhancement hypothesis.
The phenomenon of ENH provides additional room for the LHV models to adapt to
the quantum predictions, so we would expect those critical ``detection efficiencies''
to be increased.

In particular, in App.\ref{Model_1by2} we provide proof that such 
$\eta_{\rm crit} \geq \tfrac{1}{2}$ always, for any bipartite test based on
observer-symmetric efficiencies.
Interestingly, our (non-exhaustive) 
simulations also seem to suggest that experimental tests could perhaps
be validated, without the need to assume any other hypothesis (i.e., with
or without ENH, with or without fair sampling, etc), by demanding not only the
violation of a particular inequality, 
but also compliance with the full set of quantum predictions for the
choice of state and observables involved in it,
plus the condition $\eta > \eta_{\rm crit}(all;ng)$ (see Sec.\ref{Full_M});
actually, our simulations suggest the equivalence
\begin{eqnarray}
\eta_{\rm crit}(all;ng) \equiv \eta_{\rm crit}(all;gen), \label{equiv}
\end{eqnarray}
i.e., the equivalence of genuine and non-genuine critical detection rates when
the full set of quantum predictions involved in the inequality is tested as well.
This suggests that, the consequences of ENH can be therefore bypassed by
the demand of all quantum predictions at once.

To complete the picture, 
in Sec.\ref{OT} we proposed a re-examination of some recent tests,
which does not intend to be exhaustive but just be able to convey a realistic
view of the subtlety of things,a view that the current hype regarding non-locality
has completely obscured.


Of course, this work can also be understood just as yet another
``loophole'' (as is the absence of strict space-like separation in tests
with massive particles, for instance), or collection of loopholes.
Yet,
let me then insist once more on the hints that a framework such as \cite{WPDC}
provides about the possible relation between the observed low detection rates
and the real properties of the state under probe.
Finally, 
let me also insist on the fact that evidence of ENH would clearly
point toward the need to depart from the usual particle-like models of
the photon,
merely based on the correspondence between the $\tfrac{1}{2}$-spin
algebra for massive particles and the polarization states of a plane wave, 
in favor perhaps of a fully quantum-electrodynamical description.


\section*{Acknowledgments}

First let me acknowledge the ground-breaking work done in the past by 
R. Risco-Delgado \cite{Risco_PHD},
whom I also thank, along with A. Casado and R. O'Reilly, for valuable
comments and encouragement.
I thank, too, PRA referees for suggestions on the content and structure
of the paper.
I thank Jo\~{a}o N.C. Especial \cite{Joao} for very useful discussions;
though both our works were independently derived and motivate from
different directions, they share important points of convergence and
it is my opinion they should be regarded as complementary, see
Sec.\ref{Disc}).
Finally, I thank ``nightlight'' for drawing my attention to the issue
commented in Sec.\ref{LC}.


\appendix

\section*{\label{App}Appendix}

\subsection{\label{CH_and_CHSH}The CH and CHSH inequalities}

Let again $A_1,A_2$ and $B_1,B_2$ be, respectively, pairs of dichotomic
observables ($A_i,B_j \in \{\pm 1\}$) at two distant sides, a possible
formulation of the Clauser-Horne (CH) inequality \cite{CH74} 
(there are equivalent ones, see below, and also be aware of \cite{CH_conv})
would read as
\begin{eqnarray}
& P(A_1 = B_1 = 1) + P(A_1 = B_2 = 1) + P(A_2 = B_1 = 1) \nonumber\\
& - P(A_2 = B_2 = 1) - P(A_1 = 1) - P(B_1 = 1) \leq 0.   \nonumber\\
\label{CH}
\end{eqnarray}
On the other hand, the Clauser-Horne-Shimony-Holt (CHSH) \cite{CHSH69}
inequality can be written as
\begin{equation}
|\langle A_1 B_1 \rangle + \langle A_1 B_2 \rangle +
\langle A_2 B_1\rangle - \langle A_2 B_2 \rangle| \le 2,
\label{CHSH}
\end{equation}
or simply, for any four values $A_1,A_2,B_1,B_2 \in \{\pm 1\}$, as
\begin{equation}
A_1 B_1 + A_1 B_2 + A_2 B_1 - A_2 B_2 \le 2.
\end{equation}
However,
with some slight changes in relation to (\ref{CH}) and (\ref{CHSH}),
the CHSH and CH inequalities can be written in the following form,
consistent for instance with that of \cite{Aspect02},
\begin{eqnarray}
&& -1 \leq P(A_1 = B_1 = 1) - P(A_1 = B_2 = 1) \nonumber\\
&& \quad\quad\quad\quad\quad + P(A_2 = B_1 = 1) + P(A_2 = B_2 = 1) \nonumber\\
&& \quad\quad\quad\quad\quad\quad\quad\quad\quad\quad
- P(A_2 = 1) - P(B_1 = 1) \leq 0, \nonumber\\
\label{CH_alt}
\end{eqnarray}
and
\begin{eqnarray}
|\langle A_1 B_1 \rangle - \langle A_1 B_2 \rangle 
+ \langle A_2 B_1 \rangle  + \langle A_2 B_2 \rangle| \leq 2.
\label{CHSH_alt}
\end{eqnarray}
Such choices do not compromise the consistency of any of the
results of the paper.
Both the  CH and CHSH inequalities are applicable either for \emph{deterministic}
or \emph{indeterministic} theories.

Different conventions for the inequalities yield different violations for the
choice of state and observables we are working with;
take into account all angles are defined as rotations in spin space and therefore
must be halved to make the correspondence with polarizer orientations.

\begin{figure}[ht!] 
\includegraphics[width=1.0 \columnwidth,clip]{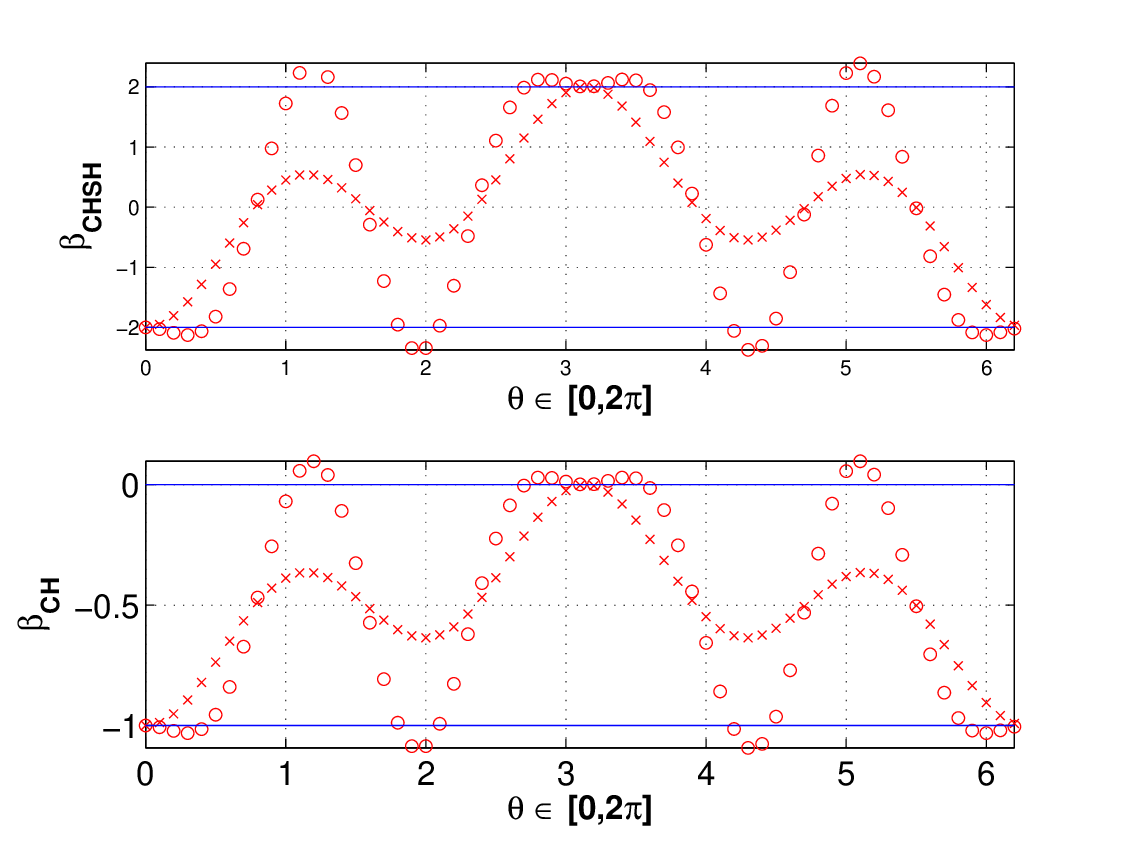} 
\caption{
Ideal ($\eta = 1$) quantum mechanical predictions for the CHSH and CH
inequalities, as written in (\ref{CHSH}) and (\ref{CH}), respectively
for the states $\psi_{1}$ (x) $\psi_{2}$ (o) and the set of observables
defined in (\ref{A1})--(\ref{B2});
solid lines are the upper and lower local bounds for each inequality.
Our convention in this paper is not the optimal in terms of maximizing violations
for this set of observables (see next figure);
angles must be halved to obtain rotations in ordinary space.
} \label{CH_and_CHSH_vs_theta_A} \end{figure}

\begin{figure}[ht!] 
\includegraphics[width=1.0 \columnwidth,clip]{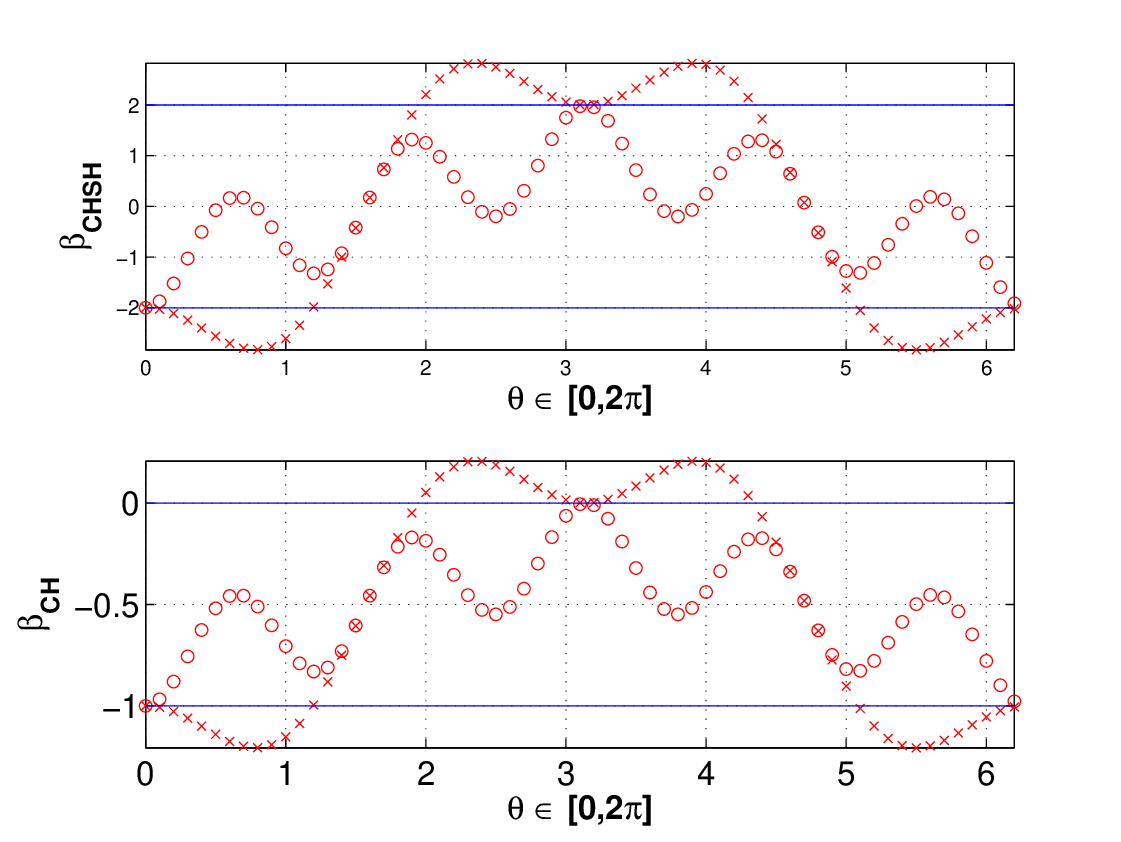} 
\caption{
Same as in Fig.\ref{CH_and_CHSH_vs_theta_A}, 
but now we adopt the alternative conventions in (\ref{CHSH_alt}) and (\ref{CH_alt}), 
consistent with those of \cite{Aspect02}.
While for this case the ``singlet'' state $|\psi_1\rangle$ produces the maximum
violations for the inequalities,
$\beta_{CHSH} = 2\sqrt{2}$ and $\beta_{CH} = (\sqrt{2}-1)/2$ respectively, 
the state $|\psi_2\rangle$), which is also maximally entangled, 
does not even violate the inequalities (the same happened for $|\psi_1\rangle$
in the former figure);
angles must be halved to obtain rotations in ordinary space.
} \label{CH_and_CHSH_vs_theta_B} \end{figure}

Figs.\ref{CH_and_CHSH_vs_theta_A} and \ref{CH_and_CHSH_vs_theta_B} 
suggest that Bell inequalities are useful but just partial tools to investigate
the incompatibility of QM and LR;
the full set of quantum predictions clearly provides a more accurate
characterization of such incompatibility.

\subsection{\label{App_2}
Clauser-Horne inequality and the need for additional assumptions}

Suppose now we want to test (\ref{CH}):
in contrast to what happened with the CHSH inequality (\ref{CHSH}), we do
not need reject the events with non-coincident detections, but simply
treat non-detections as a proper result (assigning a numerical value,
for instance $-1$), and compute directly the probabilities involved as
number of counts in each corresponding device.
That would clearly prevent unfair sampling, but is no solution
in practice:
again in contrast with homogeneous inequalities, which operate on coincidence
events alone and whose $\beta$ is therefore unaffected by the value $\eta$
of the detection rate 
($\beta$ depends on the state and observables, but not in $\eta$),
for inhomogeneous inequalities $\beta = \beta(\eta)$, 
and the equation
\begin{eqnarray}
&& \eta^2 P_{QM}(A_1 = B_1 = 1) + \eta^2 P_{QM}(A_1 = B_2 = 1) \nonumber\\
&&\quad\quad  + \eta^2 P_{QM}(A_2 = B_1 = 1) - \eta^2 P_{QM}(A_2 = B_2 = 1) \nonumber\\
&&\quad\quad\quad - \eta P_{QM}(A_1 = 1) - \eta P_{QM}(B_1 = 1) > 0,  
\nonumber\\
\label{CH_eta}
\end{eqnarray}
defines, for each pair (state, set of observables), yet another parameter
$\eta_{\rm crit}$, one that is, again, prohibitively high:
from \cite{LS01}, $\eta_{\rm crit} = 0.66$ for some very particular quasi-product
states.

We need, therefore, some strategy to extract a violation (if this is indeed
the case) even for $\eta < \eta_{\rm crit}$.
Looking at (\ref{CH_eta}), one might be tempted to try to alienate
the effect of the alleged ``detection efficiency'' by simply normalizing
bipartite and marginal probabilities by the corresponding factor 
(we take $\eta$ as a data, estimated for instance from a quotient of rates), 
writing
\begin{eqnarray}
&&\beta_{norm}^{CH} = 
\tfrac{1}{\eta^2} \cdot [\ 
P(A_1=B_1=1) + P(A_1=B_2=1) \nonumber\\
&&\quad\quad\quad\quad\quad\quad
+  P(A_2=B_1=1) - P(A_2=B_2=1) \nonumber\\
&&\quad\quad\quad\quad\quad\quad\quad
- \eta P(A_1=1) - \eta P(B_1=1) \ ];
\nonumber\\
\label{beta_norm}
\end{eqnarray}
however, the inequality 
$\beta_{norm}^{CH} < 0$ 
is not legitimate unless we explicitly acknowledge an implicit assumption,
that of \textit{independent errors} (or, equivalently, \textit{fair-sampling}): 
had we worked with perfect detectors, the probabilities would have
not changed but for the former ``apparent'' factors (either in $\eta$
for marginal or $\eta^2$ for bipartite).

\textit{Independent errors} may be true on average detection probabilities
such as $P(A_i,B_j)$, $P(A_i)$ and $P(B_j)$, or also $P(A,B)$, $P(A)$
and $P(B)$; indeed this ``average'' sense is the one in which the
term ``independent errors'' is usually invoked.
However, it definitely cannot be assumed, for some particular 
$a,b \in \{\pm 1\}$, about $P(A_i = a,B_j = b)$, $P(A_i=a)$ or $P(B_j=b)$,
as has been argued throughout this paper: see for instance (\ref{UF_LHV}).

The rejection of full \textit{independent errors} (full $\equiv$ on
every event) as a reliable (or at least testable) supplementary hypothesis
could in principle be bypassed by the following expression \cite{CH74}, 
dealing directly with the number of ``counts'' registered at each
detector, and which is now assuming \textit{no-enhancement} instead:
see (\ref{n_enh}).
We adopt for clarity the same notation as in \cite{CH74}, but with
the necessary changes on the sub-indexes to make it fully consistent
with ours:
\begin{eqnarray}
&&\beta_{op}^{CH} = \tfrac{1}{N(\infty, \infty)} \times [ \nonumber\\
&&\quad\quad\quad\quad\quad
N(a_1, b_1) + N(a_1, b_2) + N(a_2, b_1)  \nonumber\\
&&\quad\quad\quad\quad\quad\quad
- N(a_2, b_2) - N(a_1, \infty) - N(\infty, b_1) ], \nonumber\\
\label{CH_op}
\end{eqnarray}
where $a_i$ ($b_j$) is the orientation of the measuring device when the
observable $A_i$ ($B_j$) is measured, and
$N(a_i,b_j)$ the number of coincidence detections (outcome $+1$) for
orientations $a_i,b_j$;
$\infty$ stands for the case where the polarizer is removed.
App.\ref{NEN_conseq} can actually be considered a (reversed) re-derivation
of (\ref{CH_op}).

Finally, writing probability estimates in terms of the number of
registered counts, with
\begin{eqnarray}
P(A_i = 1, B_j = 1) &\approx & N(a_i, b_j) / N(\infty, \infty),    \\
P(A_i = 1)          &\approx & N(a_i, \infty) / N(\infty, \infty)  \\
P(B_j = 1)          &\approx & N(\infty, b_j) / N(\infty, \infty),
\end{eqnarray}
we can go from (\ref{CH_op}) to our former (\ref{ng_beta}).
Complementarily, the former discussions should have made evident why we
cannot write, for a double channel experiment, simply something like:
\begin{eqnarray}
&&\beta_{2-ch.}^{CH} = 
\tfrac{N(A_1=B_1=1)}{N(A,B)} + \tfrac{N(A_1=B_2=1)}{N(A,B)} \nonumber\\
&&\quad\quad\quad\quad\quad\quad\quad\quad
+ \tfrac{N(A_2=B_1=1)}{N(A,B)} - \tfrac{N(A_2=B_2=1)}{N(A,B)} \nonumber\\
&&\quad\quad\quad\quad\quad\quad\quad\quad\quad\quad\quad\quad\quad\quad
- \tfrac{N(A_1=1)}{N(A)} - \tfrac{N(B_1=1)}{N(B)};
\nonumber\\ \label{2D_beta}
\end{eqnarray}
with $N(\cdot)$ meaning again number of counts for the corresponding
observable/s and result/s,
and where now, taking advantage of the double channel setup, we do not
observe counts in absence of polarizers, but rather use
$N(A) \equiv N(A_1=+1) + N(A_1=-1)$, $N(B) \equiv N(B_1=+1) + N(B_1=-1)$.
The quantities involved in (\ref{2D_beta}) are not real probabilities, but
mere estimations under the assumption of independent errors, and therefore
$\beta_{2-ch.}^{CH} \leq 0$ is, yet again, not a genuine inequality.

\subsection{\label{NEN_conseq}
Test under the no-enhancement assumption}

Under the no-enhancement (NEN) assumption (\ref{n_enh})
the non-genuine version of the CH inequality, i.e.,
\begin{equation}
\beta_{ng}^{CH}({\cal M}) \leq 0, \label{ng_ineq}
\end{equation}
is still a legitimate inequality.
Our derivation goes along different lines to those of \cite{CH74};
Clauser and Horne's original derivation relies on the so-called ``factorability
condition'' \cite{CH_fac}, which in our notation translates (we omit model
subscripts for simplicity) to
\begin{eqnarray}
P(A_i,B|\lambda) = P(A_i|\lambda) \cdot P(B|\lambda),  \\
P(A,B_j|\lambda) = P(A|\lambda)   \cdot P(B_j|\lambda), 
\end{eqnarray}
which we can use to arrive to
\begin{eqnarray}
P(A_i|B) 
&=& P(A_i,B)/P(B) \nonumber\\ 
&=& \frac{1}{P(A)} \int_{\Lambda} P(A_i|\lambda) \cdot P(B|\lambda) \ \rho(\lambda)\ d\lambda,\nonumber\\
\end{eqnarray}
with $P(A) = P(B)$.
From here, assuming a reasonable condition such as $P(A|\lambda) = P(B|\lambda)$
(for instance justified by the \textit{rotational symmetry of each of the detector
setups in the absence of polarizers}),
we can get to
(we recall our definition, in Sec. IV.C of the paper, of $\Lambda_A,\Lambda_B$ as the
subsets of states where $A = 1$ and $B = 1$, respectively):
\begin{eqnarray}
P(A_i|\Lambda_B) &=& P(A_i|\Lambda_A), \label{imp_A} \\
P(B_j|\Lambda_A) &=& P(B_j|\Lambda_B), \label{imp_B}
\end{eqnarray}
or, equivalently, simply $P(A_i|B) = P(A_i|A)$ and $P(B_j|A) = P(B_j|B)$, the
``subset'' notation being convenient for convergence with Sec.\ref{ENH}.

Finally, (\ref{imp_A})--(\ref{imp_B}) lead (see intermediate calculations
in \cite{note_leads}), 
together with \textit{no-enhancement} (\ref{n_enh}), to:
\begin{eqnarray}
\eta P_{{\cal M}}(A_1=1|\Lambda_A) \geq P_{{\cal M}}(A_1=1|\Lambda), \label{ne_A}\\
\eta P_{{\cal M}}(B_1=1|\Lambda_B) \geq P_{{\cal M}}(B_1=1|\Lambda), \label{ne_B}
\end{eqnarray}
which, taking into account that marginal probabilities enter in the
inequality with negative sign, turn $\beta_{ng}^{CH}({\cal M}) \leq 0$
into legitimate again;
indeed, looking at (\ref{gen_beta}) and (\ref{ne_A})--(\ref{ne_B}) we
see that $\beta_{gen}^{CH} \geq \eta^2 \beta_{ng}^{CH}$,
and therefore the legitimacy of $\beta_{gen}^{CH} \leq 0$ assures the
one of $\eta^2 \beta_{ng}^{CH} \leq 0$, and hence that of $\beta_{ng}^{CH} \leq 0$.


\subsection{\label{Model_1by2}
A model for all (genuine and non-genuine) quantum predictions when $\eta \leq 2$}

For $i,j = 1,2$, $a,b \in \{;,-\}$ and $\Theta = A,B$,
let us now consider the following classes of states:

\vspace{0.2cm}\noindent
(i) $s \in S^{AB}_{i,j;a,b}$ iff

\vspace{0.1cm}\noindent
$A_i = a, B_j = b$, $A_i^{\prime} = B_j^{\prime} = 0$ for
$i^{\prime} \neq i, j^{\prime} \neq j$, 
and $P_s(A) = P_s(B) = \tfrac{1}{2}$ always, which we will denote, following
our convention in (\ref{redef_s}), as
\begin{eqnarray}
S^{AB}_{1,1;+,-} &=& (+ \ 0\ ;  \ - \  0\ ; \tfrac{1}{2} \ \tfrac{1}{2}), \\
S^{AB}_{1,2;+,-} &=& (+ \ 0\ ;  \ 0 \  -\ ; \tfrac{1}{2} \ \tfrac{1}{2}), \\
S^{AB}_{2,1;+,-} &=& (0 \ +\ ;  \ - \  0\ ; \tfrac{1}{2} \ \tfrac{1}{2}), \\
S^{AB}_{2,2;-,-} &=& (0 \ -\ ;  \ 0 \  -\ ; \tfrac{1}{2} \ \tfrac{1}{2}). 
\end{eqnarray}

\vspace{0.2cm}\noindent
(ii) $s \in S^{A}_{i;a}$ ($S^{B}_{j;b}$) iff

\vspace{0.1cm}\noindent
$A_i = a$ ($B_j = b$), all rest of instructions equal to zero, and
again $P_s(A) = \tfrac{1}{2}$ ($P_s(B) = \tfrac{1}{2}$).
For instance:
\begin{eqnarray}
S^{A}_{1;+} &=& ( +   \ 0\ ; \ 0  \ \ 0\ ; \tfrac{1}{2} \ 0), \\
S^{A}_{2;-} &=& ( 0   \ -\ ; \ 0  \ \ 0\ ; \tfrac{1}{2} \ 0), \\
S^{B}_{1;-} &=& ( 0 \ \ 0\ ; \ -    \ 0\ ; 0 \ \tfrac{1}{2}), \\
S^{B}_{2;+} &=& ( 0 \ \ 0\ ; \ 0    \ +\ ; 0 \ \tfrac{1}{2}).
\end{eqnarray}

\vspace{0.2cm}\noindent
(iii) $s \in S^{(0)}$ iff all $A_i = B_j = A = B = 0$, i.e., 
\begin{eqnarray}
S^{(0)} = (\ 0 \ 0\ ; \ 0 \ 0\ ; \ 0 \ 0).
\end{eqnarray}
%
Now, defining the shorthands
\begin{eqnarray}
Q^{AB}_{i,j;a,b} &=& P_{QM}(A_i=a,B_j=b), \\
Q^{A}_{i;a} 	 &=& P_{QM}(A_i=a), \\
Q^{B}_{j;b} 	 &=& P_{QM}(B_j=b), 
\end{eqnarray}
and assigning probabilistic weights $\rho$ to each of the former
possible states,
we can write the (complete) set of equations for the model as:
\begin{eqnarray}
\rho^{(0)} + \sum_{i,a} \rho^{A}_{i,a} + \sum_{j,b} \rho^{B}_{j,b} + \sum_{i,j,a,b} \rho^{AB}_{i,j;a,b} = 1; 
\label{r_1}
\end{eqnarray}
\begin{eqnarray}
\rho^{A}_{i,a} + \sum_{j,b} \rho^{AB}_{i,j;a,b} &=& \eta Q^{A}_{i;a},  \quad\quad\quad \forall i,a 
\label{r_2}\\
\rho^{B}_{j,b} + \sum_{i,a} \rho^{AB}_{i,j;a,b} &=& \eta Q^{B}_{j;b},  \quad\quad\quad \forall j,b
\label{r_3}\\
\rho^{AB}_{i,j;a,b} &=& \eta^2 Q^{AB}_{i,j;a,b}, \quad \forall  i,j,a,b
\label{r_4}
\end{eqnarray}
that we constrain with the additional axiomatic restrictions (the
upper bound for all $rho$'s is already implemented in eq.\ref{r_1}
above):
\begin{eqnarray}
& \rho^{(0)} \geq 0, 
\ \rho^{A}_{i,a} \geq 0, 
\ \rho^{B}_{j,j} \geq 0, 
\ \rho^{AB}_{i,j;a,b} \geq 0, 
\nonumber\\
& \forall i,j,b,a.
\label{r_5}
\end{eqnarray}
It can be seen (though it requires patience) that given (\ref{r_1})--(\ref{r_4}), 
any other restriction on the model, for instance
(\ref{marginal})--(\ref{conditional}) and (\ref{marginal_2})--(\ref{cond_2_mix}),
is also satisfied provided that
\begin{eqnarray}
\sum_{a} Q^{A}_{i;a} &=& \eta,  \quad \forall i \label{q_cond_1}\\
\sum_{b} Q^{B}_{j;b} &=& \eta,  \quad \forall j \label{q_cond_2}\\
\sum_{a,b} Q^{AB}_{i,j;a,b} &=& \eta^2,  \ \ \forall i,j. \label{q_cond_3}
\end{eqnarray}
So far we have $1 + 4 + 4 + 16 = 25$ restrictions for exactly
$4 \times 4 + 4 \times 2 + 1 = 25$ states, which means the corresponding linear
system is not over or under-determined: 
the solution, if existent, is unique for each $\eta$.

Besides, (\ref{r_1})--(\ref{r_4}) already express such system in
diagonal form, that can be then straightforwardly solved by direct
Gauss substitution:
the set of equations (\ref{r_4}) can be substituted into the (sets
of) equations (\ref{r_2})--(\ref{r_3}),
\begin{eqnarray}
\rho^{(0)} + \sum_{i,a} \rho^{A}_{i,a} + \sum_{j,b} \rho^{B}_{j,b} + \sum_{i,j,a,b} \rho^{AB}_{i,j;a,b} = 1; 
\label{r_1b}\\
\rho^{A}_{i,a} = \eta Q^{A}_{i;a} - \sum_{j,b} \eta^2 Q^{AB}_{i,j;a,b}  \quad \forall i,a 
\label{r_2b}\\
\rho^{B}_{j,b} = \eta Q^{B}_{j;b} - \sum_{i,a} \eta^2 Q^{AB}_{i,j;a,b}  \quad \forall j,b
\label{r_3b}
\end{eqnarray}
and this in turn into (\ref{r_1}), yielding, finally,  
\begin{eqnarray}
&& 
\rho^{(0)} = 1 
- \sum_{i,a} \left[ \eta Q^{A}_{i;a} - \sum_{j,b} \eta^2 Q^{AB}_{i,j;a,b} \right] \nonumber\\
&& \quad\quad\ \ 
- \sum_{j,b} \left[ \eta Q^{B}_{j;b} - \sum_{i,a} \eta^2 Q^{AB}_{i,j;a,b} \right] 
- \sum_{i,j,a,b} \eta^2 Q^{AB}_{i,j;a,b};
\nonumber\\
\end{eqnarray}
which develops to
\begin{eqnarray}
\rho^{(0)} = 1 
- \sum_{i,a} \eta Q^{A}_{i;a} 
- \sum_{j,b} \eta Q^{B}_{j;b} 
+ \sum_{i,j,a,b} \eta^2 Q^{AB}_{i,j;a,b}. 
\nonumber\\
\end{eqnarray}
Finally, by use of (\ref{q_cond_1})--(\ref{q_cond_3}) we obtain
\begin{eqnarray}
\rho^{(0)} = 1 - 4\eta + 4 \eta^2. 
\end{eqnarray}
Yet, (\ref{r_5}) has not been enforced so far, and can be used to obtain
the condition:
\begin{eqnarray}
\rho^{(0)} \geq 0 \Rightarrow \eta \leq \tfrac{1}{2}, \label{fin_eq}
\end{eqnarray}
far from surprising given that no subset composed of the proposed states
can produce $P(A_i),P(B_j)$ over $\tfrac{1}{2}$.

We can now go back and check that (\ref{fin_eq}) also assures 
$\rho^{A}_{i,a} \geq 0$ $\forall i,a$ in (\ref{r_2}), and 
$\rho^{B}_{j,b} \geq 0$ $\forall j,b$ in (\ref{r_3}):
a family of models  $\{{\cal M}(\eta), \eta\}$ can be then obtained,
adapting to all predictions on any given state and set of observables, 
under the \textit{sufficient condition} $\eta \in [0,\tfrac{1}{2}]$.

Yet, for the non-genuine model, which is the experimentally accessible
one, restrictions (\ref{r_2})--(\ref{r_3}) should be substituted by our
equivalent here for (\ref{cond_ng_A}) and (\ref{cond_ng_B}).
The operation is unnecessary though, thanks to the fact that
all former states are designed so as to guarantee that 
$P_s(A_i=a|B) = P_s(A_i=a)$ and $P_s(B_j=b|B) = P_s(B_j=b)$ for all
$a,b,i,j$;
hence, so is the case for the overall model, too.

\subsection{\label{App_4}
An approach to models maximizing $\eta$}

Values of $\eta$ beyond $\tfrac{1}{2}$ do not forbid the existence of
${\cal M}(\eta)$, though.
Determining the true ``critical value'' $\eta_{\rm crit}(all)$ beyond
which no LHV can exist (for either the genuine or non-genuine conditions)
requires to solve a high-dimensional problem;
however, by using some smart choice ($\equiv$ maximal in $\eta$) of a
subset of the overall space
of states we can, at least, determine a lower bound for such critical
value.
For instance, for obvious reasons a good point of departure may be
a model built from a combination of three subsets of states
${\cal M}_P, {\cal M}_Q, {\cal M}_R$, containing states with no
$0$'s, one $0$ and all $0$'s, respectively, therefore obeying the set
of equations:
\begin{eqnarray}
P({\cal M}_R) + 
\sum_{s \in {\cal M}_Q} \rho_{s} + 
\sum_{s \in {\cal M}_P} \rho_{s} 
= 1; 
\label{r_op_1}
\end{eqnarray}
plus
\begin{eqnarray}
\sum_{s \in {\cal M}_Q} P_s(\Theta) \cdot \rho_{s} +
\sum_{s \in {\cal M}_P} P_s(\Theta) \cdot \rho_{s} 
= f_{\Theta}(\eta) \cdot P_{QM}(\Theta);
\nonumber\\
\label{r_op_2}
\end{eqnarray}
in the genuine case with
$\Theta \in \{A_i = a, B_j = b ,A_i = a \cap B_j=b,\}$,
and in the non-genuine with
$\Theta \in \{A_i = a \cap B=1, A = 1 \cap B_j = b, A_i = a \cap B_j=b\}$,
all for $a,b \in \{\pm 1\}$, $i,j \in \{1,2\}$,
and where $f_{\Theta}(\eta) \in \eta,\eta^2$ (depending on the case).

As said, all
(\ref{marginal})--(\ref{conditional}) and (\ref{marginal_2})--(\ref{cond_2_mix})
are already taken care of as long as $P_{QM}(\Theta)$ are consistent
with (\ref{q_cond_1})--(\ref{q_cond_3}).
Whether we choose any of the last two options (genuine or non-genuine 
conditions), we have $2^4 + 2 \times 4 \times 2^3 + 1 = 81$
states (hence independent variables) for only $25$ (non-redundant)
restrictions.
This procedure provides a good lower bound for either $\eta_{\rm crit}(all;gen)$
or $\eta_{\rm crit}(all;ng)$ (see Sec.\ref{Full_M}),
with which we can later calibrate the problem for the full space of
states (see Appendix for further details).

\subsection{\label{App_num}On numerical calculations}

Data in Fig.\ref{CH_comp} is easy to generate;
for Fig.\ref{s_y_t}, we use a search algorithm based on quadratic programming
(MATLAB function quadprog) which minimizes the error 
$||A \cdot \mathbf{x} - \mathbf{b}||^2$:
$A$ is a coefficient matrix, $\mathbf{x}$ is the vector of absolute
frequencies for all states (the independent variables), and $\mathbf{b}$
is a vector of terms which depend solely on the corresponding quantum
predictions and either $\eta$ or $\eta^2$.
We note that the presence of $\eta^2$ in restrictions involving coincidences
$P(A_i,B_j),P(A_i,B),P(A,B_j)$ disrupts the possibility of using linear
programming to maximize in $\eta$.
The algorithm then descends from $\eta = 1$ down to some $\eta_{\rm crit}$
where the error is under a chosen tolerance parameter.
The appropriateness of such choice is conditioned to the numerical
sensibilities of the problem, which do not seem to be important for states
with rotational ($|\psi_1\rangle$) or some smooth (along $\theta$) inversion
symmetries ($|\psi_2\rangle$).

In order to control such numerical issues, we used a reduced
subset of states such as the one described in App.\ref{App_4}; 
as mentioned there, the optimal $\eta$ in that set is at least a lower bound
for the critical $\eta$ in the full space of states.
Anyway, results for $|\psi_{1,2}\rangle$ in Fig.\ref{s_y_t} are further
confirmed by the decrease, at the critical point, of the error in relation
to the preceding values, in one or two orders of magnitude.
Specifically, we define a relative error criteria as
\begin{eqnarray}
\mu = \frac{ |A \cdot \mathbf{x} - \mathbf{b}| } {|A \cdot \mathbf{x}|},
\label{mu}
\end{eqnarray}
which is always well defined as the denominator is at least
$|A \cdot \mathbf{x}| \geq 1$ for $\eta \geq 0$.
Once these preliminary results were obtained, we could tune appropriately
the tolerance parameter for the full problem, confirming that the former curves
are actually optimal;
my simulations also confirm the existence, for $|\psi_{1,2}\rangle$, of LHV
models for $\eta = \tfrac{1}{2}$ and all values of $\theta$ built upon the
reduced set of states proposed in App.3.


\subsection{
GEN and NG equivalence, when all quantum
predictions are considered}
As already detailed before, we consider the states $|\psi_{1,2}\rangle$ defined
in (\ref{q_state}),
and the set of four observables already given in (\ref{A1})--(\ref{B2}),
and we find LHV models demanding (\ref{cond_joint}),
together with
either (\ref{cond_g_A})--(\ref{cond_g_B}),
or (\ref{cond_ng_A})--(\ref{cond_ng_B}),
We call the first option the ``genuine'' (GEN) case, and the second
the ``non-genuine'' (NG) case.
Our simulations seem to suggest that once all quantum predictions are 
considered, both the GEN and NG cases are equivalent, in the sense that they
do not require a different degree of freedom on the model to be implemented:
critical detection rates seem to be equivalent in both cases
(this does not happen, as said in the paper, with LHV models for just the
subset of quantum predictions involved in a particular inequality).

\begin{figure}[ht!] 
\includegraphics[width=0.9 \columnwidth,clip]{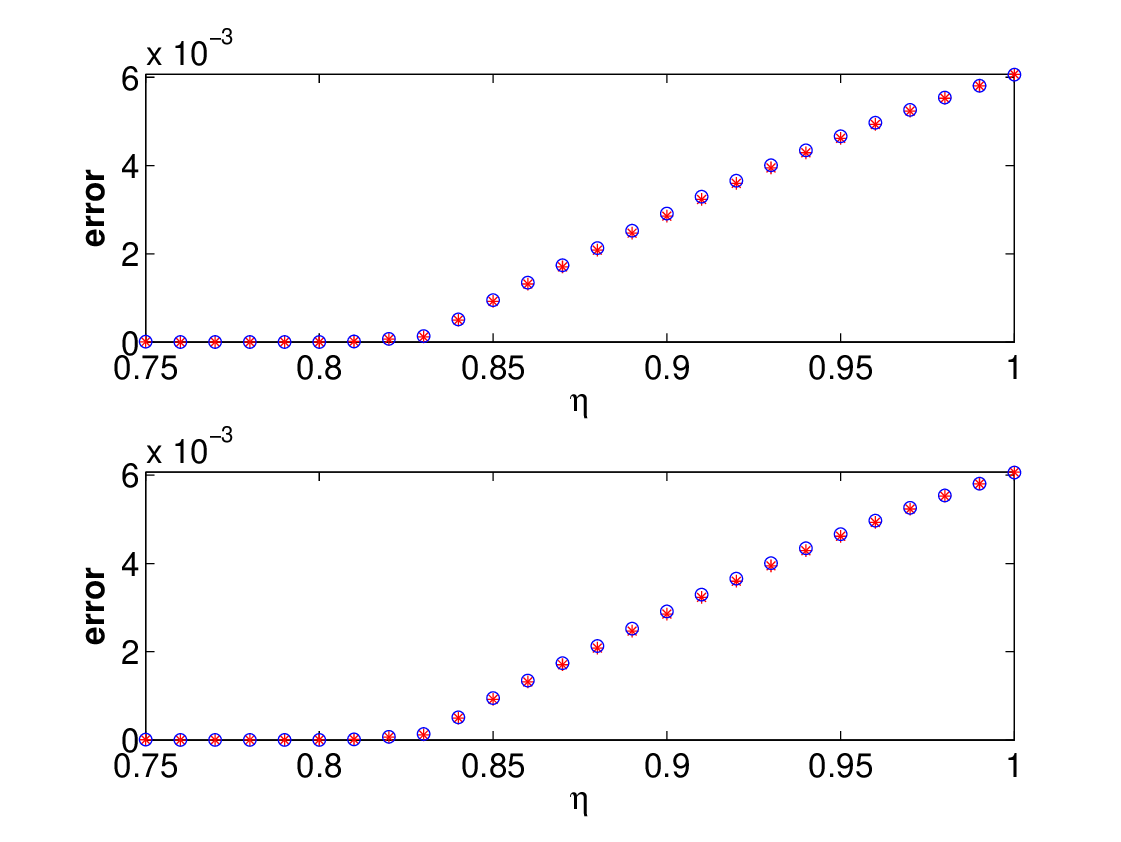} 
\caption{
Numerical simulation: evolution of the error (see App.\ref{App_num}) vs. $\eta$,
in a model simulating all quantum predictions for the GEN (stars, red) and NG 
(circles, blue) cases, 
for the states $|\psi_1\rangle$ (up) and $|\psi_2\rangle$ (down) and $\theta = \pi/4$.
As we would expect, once surpassed a critical threshold the error begins to grow,
monotonically, as $\eta$ is increased.
Again, in absence of more exhaustive results the equivalence between GEN and NG
conditions when the full set of quantum predictions involved in the inequality
is tested as well ($\eta_{\rm crit}(all;gen) \equiv \eta_{\rm crit}(all;ng)$)
should not be presumed as a general property.
} \label{error_vs_eta_SYT} \end{figure}

The treatment of non-maximally entangled states is not so numerically friendly,
because of the lack of symmetry; in particular, that of asymptotically untangled
states such as the ones proposed in \cite{LS01}, eq.10, which we would write,
following our conventions (see note \cite{CH_conv}),
\begin{eqnarray}
&&|\delta\rangle = 
C \{ 
(1 - 2\cos(\xi)) \cdot |0_{A^\prime_2} 0_{B^\prime_2} \rangle + \nonumber\\
&&\quad\quad\quad\quad
\sin(\xi) \cdot (|1_{A^\prime_2} 0_{B^\prime_2} \rangle + |0_{A^\prime_2} 1_{B^\prime_2} \rangle) 
\},
\end{eqnarray}
where $C$ is just a normalization constant and where the observables
$\{A^\prime_i.B^\prime_j\}$ (we use primes to distinguish them from our previous
set of observables, unprimed)
are defined such that $A^\prime_1,B^\prime_1$ can be obtained from 
$A^\prime_2,B^\prime_2$ by a rotation by exactly the angle $\xi$
(see \cite{LS01}, eq.11, again be aware of note \cite{CH_conv}).

Now, the case of interest is $\xi$ close to zero: while this seems to minimize
$\eta_{\rm crit}$, it also introduces some numerical bad conditioning because, whether
there is an LHV model or not, the error for such a state will always be low.
To see this clearly we simulate with $\xi = 0.1$, with the state $|\delta\rangle$
built such that $B^\prime_2$ is chosen as rotated $\pi/2$ with respect to
$A^\prime_2$ (this is consistent with \cite{LS01}, eq.11, once more be aware of
note \cite{CH_conv}), 
and we observe the evolution of the error versus $\eta$ for a model attempting
to reproduce the set of quantum prediction on two choices of the set of
observables:\\
(i) the first (left below), our ``symmetrical'' choice: $\{A_i,B_j\}$ as defined in
(\ref{A1})--(\ref{B2}), with $\theta = \pi/4$;\\
(ii) the second (right below), $\{A^\prime_i,B^\prime_j\}$ as defined in \cite{LS01}
(this is the choice that according to \cite{LS01} is supposed to yield 
$\eta_{\rm crit} \approx 0.67$).

\begin{figure}[ht!] 
\includegraphics[width=1.0 \columnwidth,clip]{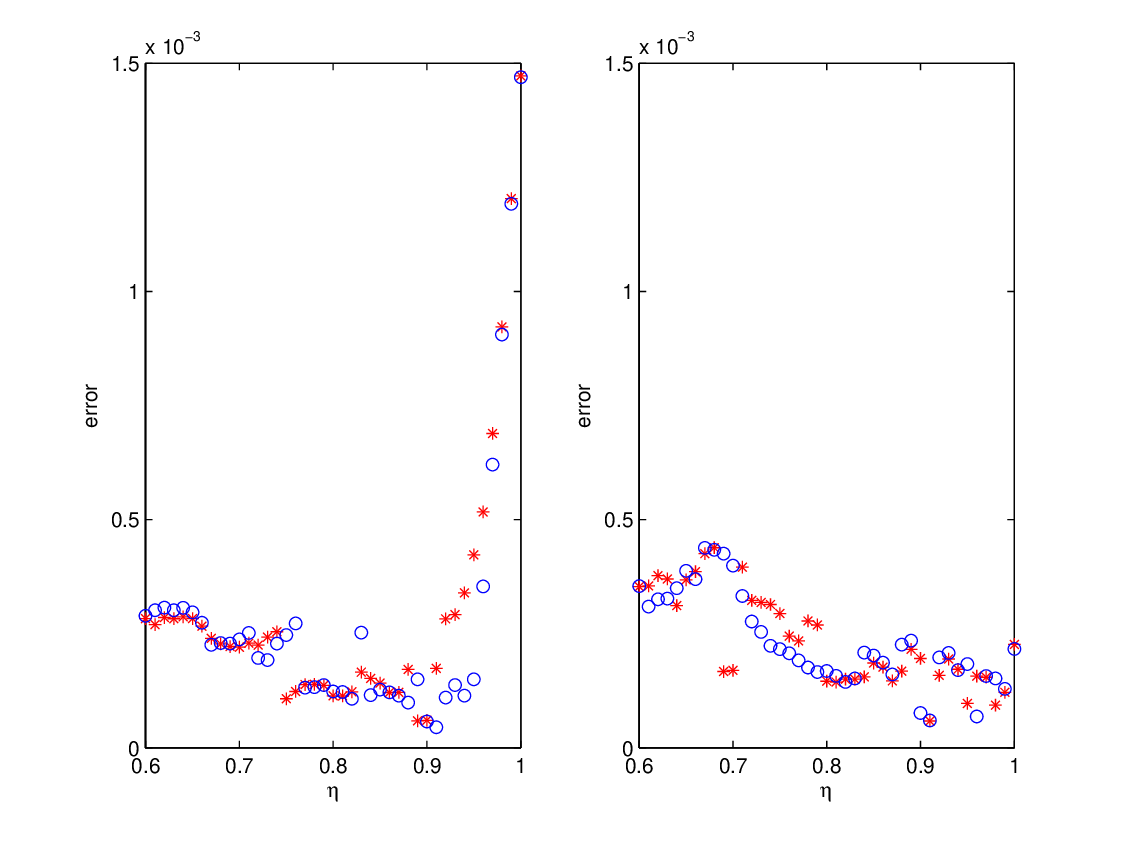} 
\caption{
Error vs. $\eta$ for a model reproducing the full set of quantum predictions for 
$|\delta\rangle$ ($\xi = 0.1$) in cases (i), left, and (ii), right, respectively.
Stars (red) are GEN conditions; circles (blue) are NG conditions.
Simulation on the left suggests a clear critical parameter;
the absence of a clear inflection point on the error curve at the right suggests that
the problem is numerically ill-defined: even when an LHV model ceases to exist, the
numerical error between QM and the model still remains very low, as corresponds
to an asymptotically untangled state.
The simulation does not consider however the full space of states, but just a subset
biased towards the higher $\eta$'s: 
consideration of all the $324$ possible different states does not  alter much the
results and adds more outliers.
Anyway,
playing with reduced subset of states, it can be seen that, necessarily, when
$\xi \rightarrow 0$ then $\eta_{\rm crit} \rightarrow 0.67$, just as proven
in \cite{LS01}:
we shall not include detailed arguments on this, for the sake of brevity, bu
merely insist on the fact that the error between QM and the (hypothetical)
LHV model is very low in these cases, which is impractical for experimental purposes.
} \label{error_vs_eta_Larsson_theta_pi4_xi_01_noAll} \end{figure}


\subsection{\label{RT_details}Detailed analysis on recent related tests}

We include out initial analysis of two recent tests; recently new models
based on a mix state (instead of the initial pure ones) have been proposed
that can ``absorb'' (I would not say ``solve'') these initial inconsistencies,
but at the expense of having now equal number of parameters and restrictions,
a situation that is clearly unsatisfying for such an important issue.

First, in the case of Giustina's test the inconsistencies where already
pointed out in \cite{Santos_arxiv} and probably other subsequent
communications (perhaps just private),
but it is the point of view of the author that it may be very convenient
to include the following calculations,
not only to facilitate and encourage further research (there is
more information here and from a more general approach),
but also and more importantly because they help put into context the objective
magnitude of the violations (of local-realism) that have been allegedly obtained:
they are rather weak, or very weak, in terms of what QM could allegedly
produce.
Of course, we are referring to the loophole-free case: once a loophole is
allowed the frontier that local-realism imposes is displaced opening room
for farther excursions into the quantum realm.

Again,
we consider two sets $\{A_i\}$, $\{B_j\}$ of remotely measured observables,
each of them accepting three possible outcomes $\pm 1$ and $0$ (this
last standing for a ``non-detection'').
We can denote then the probabilities of one or other result as
$P(A_i=a)$, $P(B_j=b)$ and $P(a_i=a,B_j=b)$, for $a,b \in \{\pm 1,0\}$; 
for overall detections (results either $\pm 1$) we will also use the shorthands 
$P(A_i) \equiv P(A_i \neq 0)$, 
$P(B_j) \equiv P(B_j \neq 0$ and
$P(A_i|B_j) \equiv P(A_i \neq 0 | B_j \neq 0)$.


\vspace{0.3cm}\noindent
(A) \textit{Test by Giustina et al \cite{G13}:}
Abbreviating with $P(A_i=a,B_j=b) \equiv P_{i,j}(a,b)$,
the Eberhard's inequality can be written as 
\begin{eqnarray}
&& \beta_{Eb.} \propto 
- 
P_{1,1}( o, o) + 
P_{1,2}( o, e) +  
P_{1,2}( o, \emptyset) +  \nonumber\\
&&\quad\quad\quad
P_{2,1}(e, o) +  
P_{2,1}(\emptyset, o) + 
P_{2,2}(o, o)
\geq 0,  
\label{Eb_equiv}
\end{eqnarray}
where $o \equiv +1$, $e \equiv -1$ are labels for the two detectors at
the exit of the PBSs, and $\emptyset$'s denote, as usual, the absence
of detection.

In order to estimate a quantum mechanical prediction for the results of
a test upon the inequality,
we now also abbreviate $P_{QM}(\cdot) \equiv Q(\cdot)$
and accommodate imperfect detection (or what I would merely call ``realistic
detection rates'') as it is usual through the hypothesis of 
\textit{independent errors}, with
\begin{eqnarray}
&&\beta_{Eb.} \equiv 
\frac{N}{4} \cdot \bigg(  
- 
\eta^2 \cdot Q_{1,1}( o, o) + 
\eta^2 \cdot Q_{1,2}( o, e)   \nonumber\\
&&\quad\quad\quad\quad\quad\quad
+ \ 
\eta (1-\eta) \cdot Q_{1,2}( o,  \emptyset) +  
\eta^2 \cdot Q_{2,1}(e, o)     \nonumber\\
&&\quad\quad\quad\quad\quad\quad
+ \ 
\eta (1-\eta) \cdot Q_{2,1}( \emptyset, o) + 
\eta^2 \cdot Q_{2,2}( o, o) \bigg),
\nonumber\\
\label{Eb_eta}
\end{eqnarray}
where $N$ is the overall number of emitted pairs, a quantity we cannot in principle
have access to. 

\begin{figure}[!t]
\centering
 \subfigure[$ $QM pred. convention 1]{
  \includegraphics[width=0.95 \columnwidth,clip]{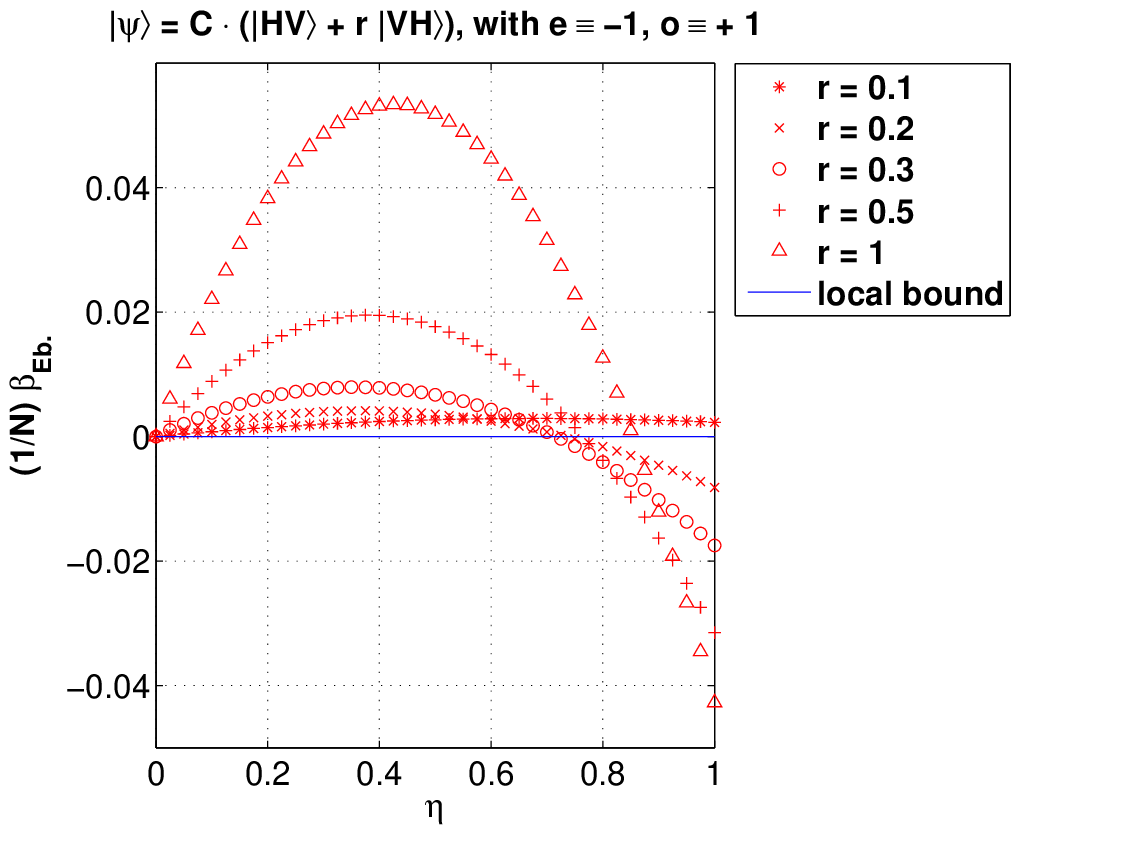}
   \label{Fig_1}
   }
\centering
 \subfigure[$ $Detail of (a)]{
 \includegraphics[width=0.9 \columnwidth,clip]{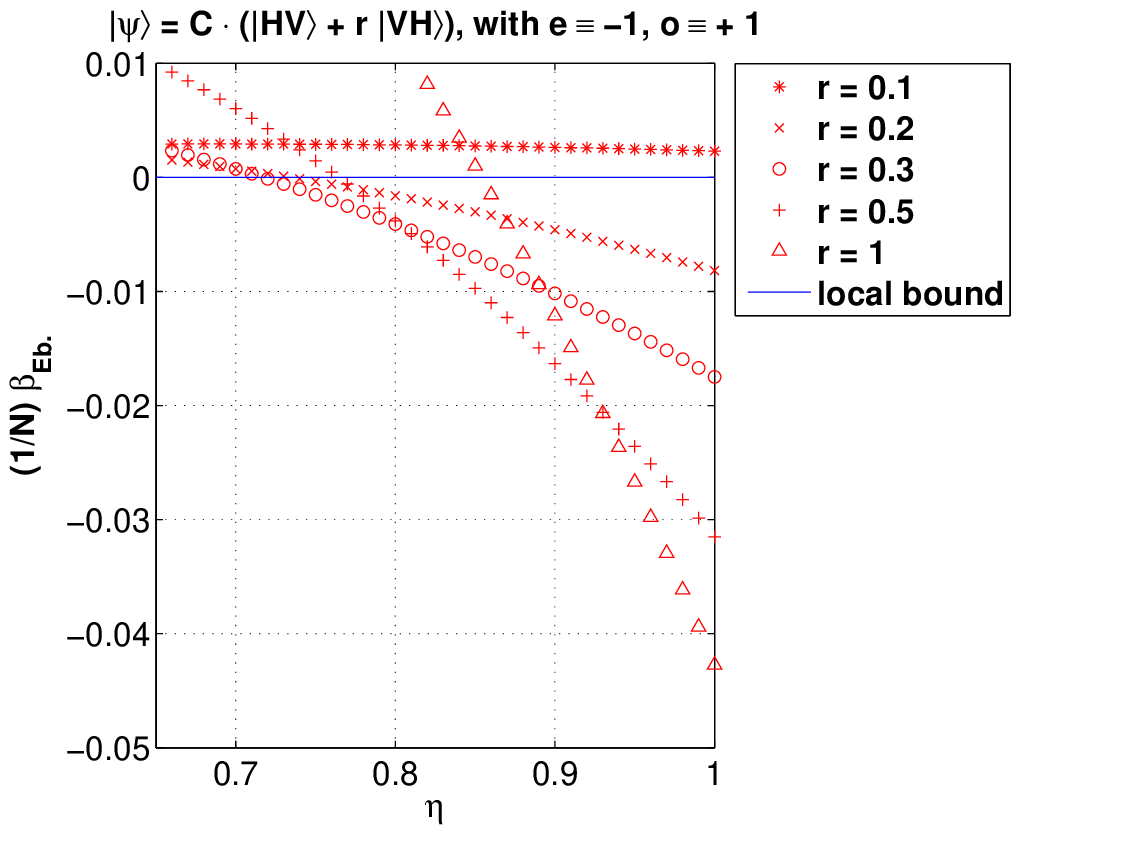}
   \label{Fig_1_det}
   }
\centering
 \subfigure[$ $QM pred. convention 2]{
 \includegraphics[width=0.95 \columnwidth,clip]{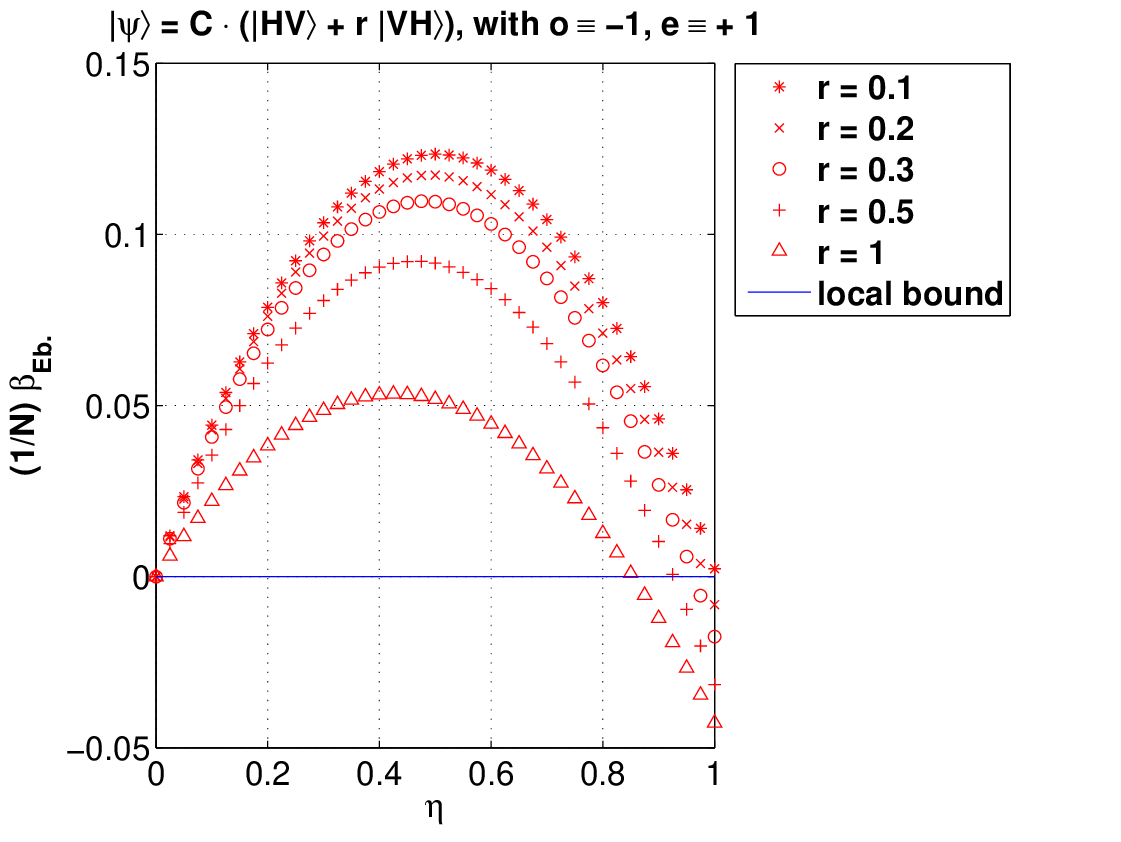}
   \label{Fig_1_alt}
   }
\caption{
Quantum prediction, under the hypothesis of independent errors, of the value of
the inequality (\ref{Eb_eta}) for $|\psi_r\rangle$ in (\ref{psi_r}),
with the two complementary choices of conventions.
Subfigure \ref{Fig_1_alt} can be obtained from (a) by:
(i) eigenvalue (``label'') permutation: $o \leftrightarrow e$;
(ii) direction permutation: $H \leftrightarrow V$;
both (i)--(ii) at the same time leave (a) invariant.
} \label{Fig_all} \end{figure}

Now we associate, as usual, horizontal and vertical polarizations to the
eingenvectors of $\sigma_z$ with eigenvalue $+1$ and $-1$, respectively,
and define
\begin{eqnarray}
A_i &\equiv&  \sin(2\alpha_i)\cdot\sigma^{A}_x +  \cos(2\alpha_i)\cdot\sigma^{A}_z, \label{Ai}\\
B_j &\equiv&  \sin(2\beta_j)\cdot\sigma^{B}_x  +  \cos(2\beta_j)\cdot\sigma^{B}_z.  \label{Bj}
\end{eqnarray}
The $2$-factor is included to consider the angles directly as
rotations of the polarizer in real space;
the report provides the values
$\alpha_1 = + 85.6^{\circ}$,  
$\alpha_2 = + 118.0^{\circ}$, 
$\beta_1  = - 5.4^{\circ},$
and
$\beta_2  = + 25.9^{\circ}$,
so as to produce measures upon the (family of) states \cite{G13}:
\begin{eqnarray}
|\psi_r\rangle = C \cdot \bigg( |HV\rangle +  r |VH\rangle \bigg). \label{psi_r}
\end{eqnarray}
Numerical simulations are provided in Figs. \ref{Fig_1} and \ref{Fig_1_det},
the ``critical value'' $\eta_{\rm crit}$ can be defined as that where $\beta$
crosses the local bound;
with the present choice of conventions, lower values of $r$ yield lower violations 
(less negative $\beta$'s), but, very conveniently, such violations are reachable at
lower $\eta_{\rm crit}$'s.
This makes the family of states $\{|\psi_r\rangle\}$ 
particularly appropriate for a Bell test; 
in Fig.\ref{Fig_1_alt}, however, a low $r$ does not decreases $\eta_{\rm crit}$.

\begin{figure}[t!] 
 \centering
  \subfigure[ Confidence interval $\Gamma_{2,2}$ vs. bounds for $\Gamma_{A}$, $\Gamma_{B}$]{
 \includegraphics[width=1.0 \columnwidth,clip]{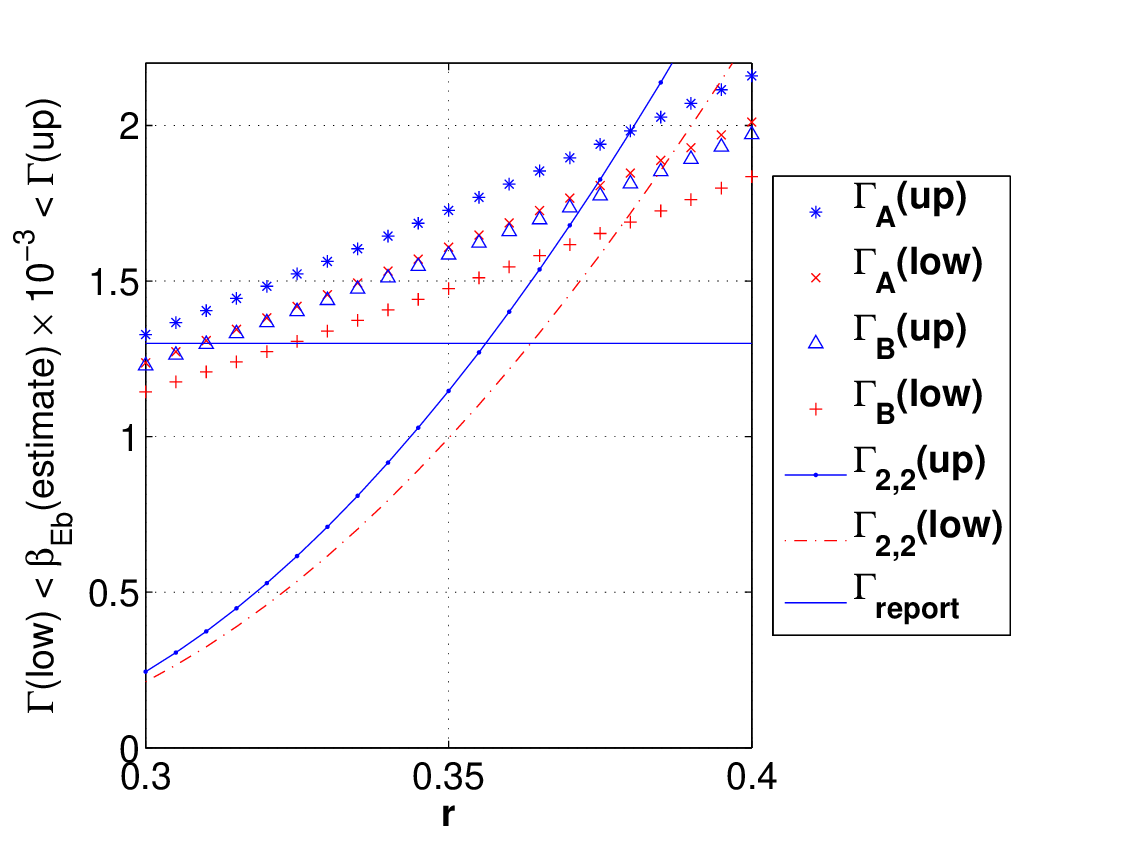}
   \label{Fig_bound_S}
    }
 \centering
  \subfigure[ Bounds $\Gamma_{2,2}$ vs. bounds for $\Gamma_{1,1}$, $\Gamma_{1,2}$, $\Gamma_{2,1}$]{
  \includegraphics[width=1.0 \columnwidth,clip]{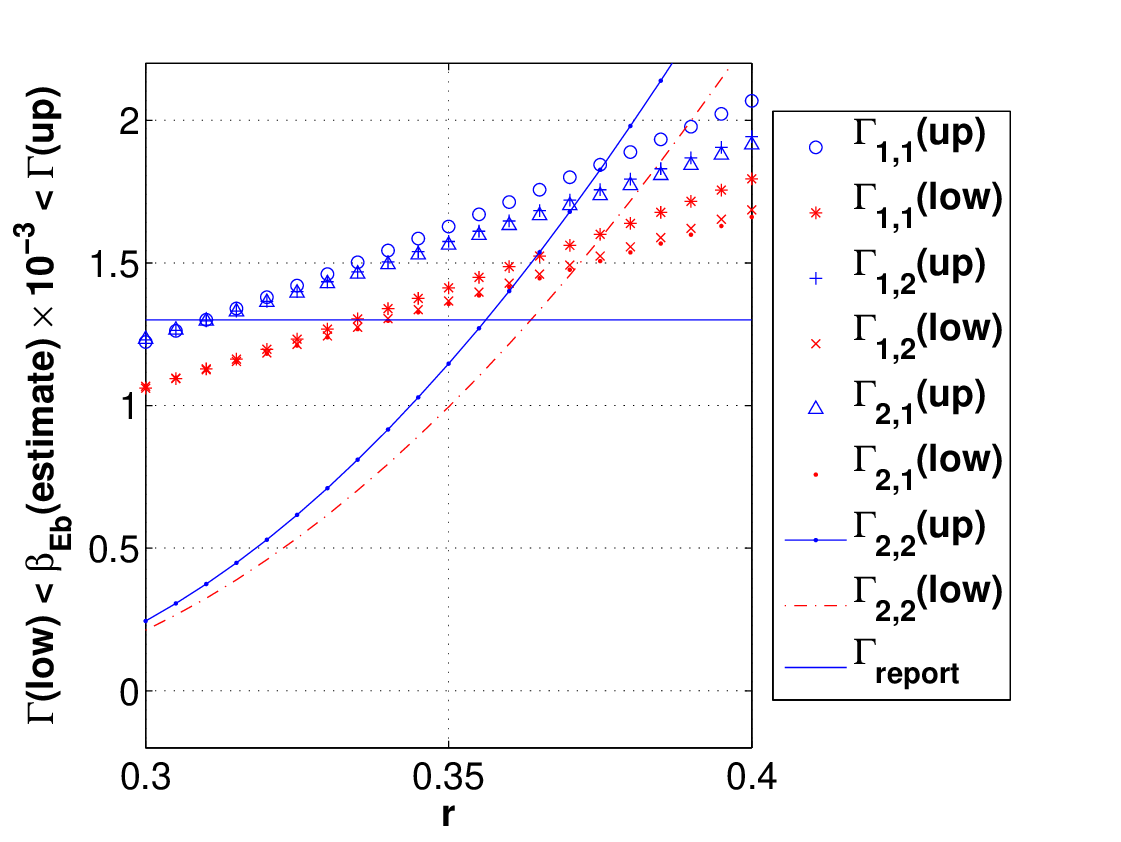}
    \label{Fig_bound_D}
    }
\caption{
Analysis of Giustina \textit{et al} \cite{G13}:
lower (red) and upper (blue) bounds on $\beta_{Eb}$ as defined below.
All but one $\Gamma$-confidence intervals seem to agree (though there is
some minor discrepancy) with the value $\Gamma_{report}$, as defined on
the reported observations (also pointing at $r \approx 0.3$ as reported
in \cite{G13};
however, $\Gamma_{2,2}$ (playing a crucial role in the inequality as
it enters with negative sign) does not fit well:
QM and context-independent (symmetric or asymmetric) efficiency factors
are not enough to explain the inconsistency.
} \label{Fig_beta_vs_eta} \end{figure}

Given the data available the only direct estimations of the detection
rates at both arms, $\eta_A,\eta_B$, would come from the expressions
\begin{eqnarray}
\eta_1 \equiv \frac{C_{oo}(\alpha_1,\beta_1)}{S^A_o(\alpha_1)} \cdot (1/P_{QM}(B_1 = o \ | A_1 = o)), \label{eta_1}\\
\eta_2 \equiv \frac{C_{oo}(\alpha_2,\beta_1)}{S^B_o(\beta_1)}  \cdot (1/P_{QM}(A_2 = o \ | B_1 = o)), \label{eta_2}
\end{eqnarray}
where $C_{ab}(\alpha_i,\beta_j)$ and $S^A_o(\alpha)$, $S^B_o(\beta)$ is
notation inherited directly from \cite{G13}.
Those expressions that let us establish a ``confidence interval'',
for instance $0.68 \leq \eta_A,\eta_B \leq 0.73$, at least for $r \leq 4$
($r = 3$ according to the original report);
the lower bound bound is actually artificially decreased to consider
possible accidental counts produced by a noisy background.

For now, for higher values of $r$ we could still reconcile the
estimations of $\eta_1,\eta_2$ by considering unbalance of losses
between the two arms.
These two preliminary values can be now used to perform a further analysis.
Now, their $N$ corresponding to $N/4$ in our framework: we use the total number
of events in (\ref{Eb_eta}), which already contains the factor $4$ as well, i.e., 
$N_{ours} < 4 \times 25 \times 10^6 = 10^8$.
Adding that $J \approx - 127 \times 10^3$ \cite{G13} we can then write
\begin{eqnarray}
\Gamma_{report} \equiv |(\beta_{Eb.}/N)|_{(report)} = J/N_{ours} < 0.0013.
\end{eqnarray}
Now we can also write expressions like, for instance,
\begin{eqnarray}
N_{ours} \approx 4 S^{A}_{o}(\alpha_1) / (\eta_A P_{QM}(A_1 = o)), 
\end{eqnarray}
which together with $\beta_{Eb.} = J/N_{ours}$ allow us to define
\begin{eqnarray}
\Gamma_{A} \equiv
\frac{\eta_A \cdot Q^{A}_{1}(o) }{4 S^{A}_{o}(\alpha_1)} \cdot J, \quad
\Gamma_{B} \equiv
\frac{\eta_B \cdot Q^{B}_{1}(o) }{4 S^{B}_{o}(\beta_1)} \cdot J, 
\end{eqnarray}
where $Q^{\Lambda}_{i}(o) \equiv P_{QM}(\Lambda_i = o)$,
and similarly,
\begin{eqnarray}
\Gamma_{i,j} &\equiv&
\frac{\eta_A \eta_B \cdot Q_{i,j}(o,o) }{4 C_{o,o}(\alpha_i,\beta_j)} \cdot J, 
\end{eqnarray}
with $Q_{i,j}(o,o) \equiv P_{QM}(A_i=B_j=o)$.
Using the former $0.68 < \eta_A, \eta_B < 0.73$, the $\Gamma$'s provide
bounds on the violation that we would theoretically expect, by means of
$\Gamma(\eta_A = \eta_B = 0.68) \leq \beta_{Eb} \leq \Gamma(\eta_A = \eta_B = 0.73)$.

The inconsistency in Fig.\ref{Fig_beta_vs_eta} cannot be bypassed by introducing
any additional ``efficiency'' factor, unless such factors are allowed to show a
dependence on the choice of settings. There are two possibilities,
(i): assuming
$\eta_A \equiv \eta_A(\alpha_i)$ and $\eta_B \equiv \eta_B(\beta_j)$, or 
(ii): a ``full contextuality''
$\eta_A \equiv \eta_A(\alpha_i,\beta_j)$, $\eta_B \equiv \eta_B(\alpha_i,\beta_j)$.

While (i) does not invalidate the basic assumption in
Eberhard's derivation (the fate of one photon is independent of the
choice of angle in the remote setup), expressions (ii) would do it.
Attending to the fact that a major discrepancy between QM and the observations is
only present for one of the four coincidence rates, specifically for
$n_{oo}(\alpha_2,\beta_2)$, we have to give credit to (ii).
A convergent analysis can be found in \cite{Santos_arxiv}.


\vspace{0.3cm}\noindent
(B) \textit{Test by Christensen et al \cite{Chr13}:}
This very recent test also gives hints of a similar context-dependence
upon the detection rates, as seen in the following simple calculation
(again we use the same notation as in the paper we comment upon, in this case
that of \cite{Chr13}):
\begin{eqnarray}
p_{2}(b)|_1 &\equiv& \frac{S_2(b)}{C(a,b)} 	  = 0.001696, \\
p_{2}(b)|_2 &\equiv& \frac{S_2(b)}{C(a^\prime,b)} = 0.005157,
\end{eqnarray}
two alternative possible estimations of the same quantity that give rise,
respectively, to a violation ($p_{2}(b)|_1$, by very little margin) or
no violation ($p_{2}(b)|_2$, very far from the bound) of the tested inequality.
Once more we seem to encounter a context-dependence of the detection rates,
one that cannot be reduced to a mere dependence on the local observable;
again, full details can be obtained from the author.


\vspace{0.3cm}
Again, both (A) and (B) can be accounted by models of
the style of \cite{Zeil_last} (the same strategy can also be applied to the
test by Christensen \textit{et al}), 
something not surprising as these models increase the number of degrees of 
freedom to be fitted by experimental data, to the point that the inconsistency
is not really solved but just diluted.
I do not think it is unreasonable to find such state of affairs not good
enough: 
all quantum predictions that can be reliably tested at once should be tested, 
in order to make possible systematic errors (such as the local coincidences
we will describe right now) manifest, or on the contrary to discard them
definitely.


\subsection{Explicit examples of LHV models}

The following models correspond to the GEN $+$ NG case, i.e., they satisfy 
both (\ref{cond_g_A})--(\ref{cond_g_B}) and (\ref{cond_ng_A})--(\ref{cond_ng_B}).
For the models below, each choice of sign corresponds to one state,
i.e., for instance $(\pm 1, \pm 1, \pm 1, \pm 1; 1, 1)$ 
actually means the two states
$(+1, +1, +1, +1; 1, 1)$ and $(-1, -1, -1, -1; 1, 1)$,
both with equal probability given in the respective
column (a given $\theta$ and $\eta$). 
For a given event $\Theta$ we can calculate
(recall the last row is all $0$'s, therefore there is not choice of sign)
\begin{eqnarray}
P(\Theta) = \sum_{i=1}^{N} \Gamma_i(\Theta) \cdot \rho_i;
\end{eqnarray}
where the coefficient $\Gamma_i(\Theta) = 0,1,2$,  
depending on whether $\Theta$ occurs for $0$, $1$ or $2$ of the possible
choices of sign for that $i$-th row (always two possible choices, except,
again, for the last row).
Models are of course not exact, and they are calculated on a reduced set
of states under symmetry restrictions (the error is, as to be expected,
considerably higher in the case of $|\psi_2\rangle$);
additional simulations on the full space of states (no symmetry restrictions)
confirm these results.
%


\clearpage

\begin{table}[h]
\centering 
\begin{tabular}{ c c c c c c c c c c c c c c c c c}
$A_1$ & $A_2$ & $B_1$ & $B_2$ & $A$ & $B$ & $\theta = 0.1$  & $\theta = 0.2$  & $\theta = 0.3$  & $\theta = 0.4$  & $\theta = 0.5$  & $\theta = 0.6$  & $\theta = 0.7$  & $\theta = 0.8$  & $\theta = 0.9$  & $\theta = 1$  & $\theta = 1.1$  \\
& & & & &  & $\eta = 0.99$  & $\eta = 0.97$  & $\eta = 0.94$  & $\eta = 0.9$  & $\eta = 0.87$  & $\eta = 0.84$  & $\eta = 0.83$  & $\eta = 0.82$  & $\eta = 0.83$  & $\eta = 0.86$  & $\eta = 0.91$  \\
\hline
 $\pm 1$ & $\pm 1$ & $\pm 1$ & $\pm 1$ & 1 & 1 & 5.68e-05 & 5.2e-05 & 5.53e-05 & 0.000162 & 0.000149 & 0.000186 & 6.8e-05 & 0.000145 & 0.000141 & 0.000136 & 0.000127 \\
 $\pm 1$ & $\pm 1$ & $\pm 1$ & $\mp 1$ & 1 & 1 & 5.16e-05 & 4.48e-05 & 4.76e-05 & 0.000137 & 9.3e-05 & 0.000178 & 6.48e-05 & 0.000145 & 0.000148 & 0.000108 & 0.000133 \\
 $\pm 1$ & $\pm 1$ & $\mp 1$ & $\pm 1$ & 1 & 1 & 0.000511 & 0.00407 & 0.00916 & 0.0136 & 0.0213 & 0.028 & 0.0395 & 0.0487 & 0.0629 & 0.0829 & 0.11 \\
 $\pm 1$ & $\pm 1$ & $\mp 1$ & $\mp 1$ & 1 & 1 & 0.478 & 0.429 & 0.358 & 0.274 & 0.201 & 0.134 & 0.0843 & 0.0419 & 0.0142 & 0.000314 & 0.000427 \\
 $\pm 1$ & $\mp 1$ & $\pm 1$ & $\pm 1$ & 1 & 1 & 0.000511 & 0.00407 & 0.00916 & 0.0136 & 0.0213 & 0.028 & 0.0395 & 0.0487 & 0.0629 & 0.0829 & 0.11 \\
 $\pm 1$ & $\mp 1$ & $\pm 1$ & $\mp 1$ & 1 & 1 & 5.15e-05 & 4.48e-05 & 4.76e-05 & 0.000137 & 9.65e-05 & 0.000179 & 6.49e-05 & 0.000145 & 0.000148 & 0.000108 & 0.000133 \\
 $\pm 1$ & $\mp 1$ & $\mp 1$ & $\pm 1$ & 1 & 1 & 0.000513 & 0.00406 & 0.00917 & 0.0136 & 0.0212 & 0.028 & 0.0395 & 0.0487 & 0.0629 & 0.0828 & 0.11 \\
 $\pm 1$ & $\mp 1$ & $\mp 1$ & $\mp 1$ & 1 & 1 & 5.16e-05 & 4.48e-05 & 4.76e-05 & 0.000137 & 9.3e-05 & 0.000178 & 6.48e-05 & 0.000145 & 0.000148 & 0.000108 & 0.000133 \\
 +0 & $\pm 1$ & $\pm 1$ & $\pm 1$ & 0 & 1 & 6.48e-05 & 5.81e-05 & 6.35e-05 & 0.000202 & 0.000163 & 0.000212 & 8.32e-05 & 0.000188 & 0.000191 & 0.000181 & 0.000282 \\
 +0 & $\pm 1$ & $\pm 1$ & $\pm 1$ & 1 & 1 & 5.91e-05 & 5.27e-05 & 5.54e-05 & 0.000162 & 0.000124 & 0.000186 & 6.88e-05 & 0.000151 & 0.000152 & 0.000148 & 0.00019 \\
 +0 & $\pm 1$ & $\pm 1$ & $\mp 1$ & 0 & 1 & 0.000117 & 0.000105 & 0.000112 & 0.000338 & 0.000316 & 0.000384 & 0.000144 & 0.000322 & 0.000328 & 0.000315 & 0.000447 \\
 +0 & $\pm 1$ & $\pm 1$ & $\mp 1$ & 1 & 1 & 0.000103 & 9.01e-05 & 9.51e-05 & 0.000277 & 0.000212 & 0.000342 & 0.000124 & 0.000256 & 0.000267 & 0.000272 & 0.000315 \\
 +0 & $\pm 1$ & $\mp 1$ & $\pm 1$ & 0 & 1 & 0.000117 & 0.000104 & 0.000111 & 0.000339 & 0.00028 & 0.000384 & 0.000144 & 0.000322 & 0.000323 & 0.000312 & 0.000447 \\
 +0 & $\pm 1$ & $\mp 1$ & $\pm 1$ & 1 & 1 & 0.000103 & 9.02e-05 & 9.51e-05 & 0.000277 & 0.00017 & 0.000342 & 0.000124 & 0.000255 & 0.000265 & 0.00027 & 0.000316 \\
 +0 & $\pm 1$ & $\mp 1$ & $\mp 1$ & 0 & 1 & 0.00217 & 0.0072 & 0.0141 & 0.0217 & 0.0223 & 0.0306 & 0.0352 & 0.0358 & 0.0358 & 0.0311 & 0.0197 \\
 +0 & $\pm 1$ & $\mp 1$ & $\mp 1$ & 1 & 1 & 0.00222 & 0.00692 & 0.0136 & 0.0217 & 0.033 & 0.0347 & 0.0348 & 0.0365 & 0.0332 & 0.0276 & 0.0193 \\
 $\pm 1$ & +0 & $\pm 1$ & $\pm 1$ & 0 & 1 & 0.000114 & 0.000101 & 0.000113 & 0.000335 & 0.000324 & 0.000376 & 0.000147 & 0.00032 & 0.000325 & 0.000316 & 0.000448 \\
 $\pm 1$ & +0 & $\pm 1$ & $\pm 1$ & 1 & 1 & 0.000107 & 9.23e-05 & 0.000102 & 0.000287 & 0.000235 & 0.000339 & 0.000128 & 0.000255 & 0.000262 & 0.000271 & 0.000275 \\
 $\pm 1$ & +0 & $\pm 1$ & $\mp 1$ & 0 & 1 & 6.75e-05 & 5.77e-05 & 6.21e-05 & 0.000196 & 0.000161 & 0.000211 & 8.24e-05 & 0.000188 & 0.000192 & 0.000136 & 0.00023 \\
 $\pm 1$ & +0 & $\pm 1$ & $\mp 1$ & 1 & 1 & 5.77e-05 & 4.84e-05 & 5.15e-05 & 0.000149 & 0.000125 & 0.000182 & 6.7e-05 & 0.000152 & 0.000155 & 0.000112 & 0.000159 \\
 $\pm 1$ & +0 & $\mp 1$ & $\pm 1$ & 0 & 1 & 0.00218 & 0.00687 & 0.0136 & 0.0215 & 0.0327 & 0.0346 & 0.0347 & 0.0364 & 0.0331 & 0.0277 & 0.0191 \\
 $\pm 1$ & +0 & $\mp 1$ & $\pm 1$ & 1 & 1 & 0.00221 & 0.00727 & 0.0142 & 0.0219 & 0.0226 & 0.0307 & 0.0352 & 0.0359 & 0.036 & 0.0313 & 0.0202 \\
 $\pm 1$ & +0 & $\mp 1$ & $\mp 1$ & 0 & 1 & 0.000124 & 0.000103 & 0.000108 & 0.000324 & 0.000234 & 0.000383 & 0.000143 & 0.000325 & 0.000333 & 0.000198 & 0.000319 \\
 $\pm 1$ & +0 & $\mp 1$ & $\mp 1$ & 1 & 1 & 9.91e-05 & 7.72e-05 & 7.9e-05 & 0.000248 & 0.000124 & 0.000333 & 0.000117 & 0.000259 & 0.000277 & 0.000167 & 0.000222 \\
 $\pm 1$ & $\pm 1$ & +0 & $\pm 1$ & 1 & 0 & 0.000114 & 0.000101 & 0.000113 & 0.000335 & 0.000324 & 0.000376 & 0.000147 & 0.00032 & 0.000325 & 0.000316 & 0.000448 \\
 $\pm 1$ & $\pm 1$ & +0 & $\pm 1$ & 1 & 1 & 0.000107 & 9.23e-05 & 0.000102 & 0.000287 & 0.000235 & 0.000339 & 0.000128 & 0.000255 & 0.000262 & 0.000271 & 0.000275 \\
 $\pm 1$ & $\pm 1$ & +0 & $\mp 1$ & 1 & 0 & 0.000124 & 0.000103 & 0.000108 & 0.000324 & 0.000234 & 0.000383 & 0.000143 & 0.000325 & 0.000333 & 0.000198 & 0.000319 \\
 $\pm 1$ & $\pm 1$ & +0 & $\mp 1$ & 1 & 1 & 9.91e-05 & 7.72e-05 & 7.9e-05 & 0.000248 & 0.000124 & 0.000333 & 0.000117 & 0.000259 & 0.000277 & 0.000167 & 0.000222 \\
 $\pm 1$ & $\mp 1$ & +0 & $\pm 1$ & 1 & 0 & 0.00218 & 0.00687 & 0.0136 & 0.0215 & 0.0327 & 0.0346 & 0.0347 & 0.0364 & 0.0331 & 0.0277 & 0.0191 \\
 $\pm 1$ & $\mp 1$ & +0 & $\pm 1$ & 1 & 1 & 0.00221 & 0.00727 & 0.0142 & 0.0219 & 0.0226 & 0.0307 & 0.0352 & 0.0359 & 0.036 & 0.0313 & 0.0202 \\
 $\pm 1$ & $\mp 1$ & +0 & $\mp 1$ & 1 & 0 & 6.75e-05 & 5.77e-05 & 6.21e-05 & 0.000196 & 0.000161 & 0.000211 & 8.24e-05 & 0.000188 & 0.000192 & 0.000136 & 0.00023 \\
 $\pm 1$ & $\mp 1$ & +0 & $\mp 1$ & 1 & 1 & 5.77e-05 & 4.84e-05 & 5.15e-05 & 0.000149 & 0.000125 & 0.000182 & 6.7e-05 & 0.000152 & 0.000155 & 0.000112 & 0.000159 \\
 $\pm 1$ & $\pm 1$ & $\pm 1$ & +0 & 1 & 0 & 6.48e-05 & 5.81e-05 & 6.35e-05 & 0.000202 & 0.000163 & 0.000212 & 8.32e-05 & 0.000188 & 0.000191 & 0.000181 & 0.000282 \\
 $\pm 1$ & $\pm 1$ & $\pm 1$ & +0 & 1 & 1 & 5.91e-05 & 5.27e-05 & 5.54e-05 & 0.000162 & 0.000124 & 0.000186 & 6.88e-05 & 0.000151 & 0.000152 & 0.000148 & 0.00019 \\
 $\pm 1$ & $\pm 1$ & $\mp 1$ & +0 & 1 & 0 & 0.00217 & 0.0072 & 0.0141 & 0.0217 & 0.0223 & 0.0306 & 0.0352 & 0.0358 & 0.0358 & 0.0311 & 0.0197 \\
 $\pm 1$ & $\pm 1$ & $\mp 1$ & +0 & 1 & 1 & 0.00222 & 0.00692 & 0.0136 & 0.0217 & 0.033 & 0.0347 & 0.0348 & 0.0365 & 0.0332 & 0.0276 & 0.0193 \\
 $\pm 1$ & $\mp 1$ & $\pm 1$ & +0 & 1 & 0 & 0.000117 & 0.000104 & 0.000111 & 0.000339 & 0.00028 & 0.000384 & 0.000144 & 0.000322 & 0.000323 & 0.000312 & 0.000447 \\
 $\pm 1$ & $\mp 1$ & $\pm 1$ & +0 & 1 & 1 & 0.000103 & 9.02e-05 & 9.51e-05 & 0.000277 & 0.00017 & 0.000342 & 0.000124 & 0.000255 & 0.000265 & 0.00027 & 0.000316 \\
 $\pm 1$ & $\mp 1$ & $\mp 1$ & +0 & 1 & 0 & 0.000117 & 0.000105 & 0.000112 & 0.000338 & 0.000316 & 0.000384 & 0.000144 & 0.000322 & 0.000328 & 0.000315 & 0.000447 \\
 $\pm 1$ & $\mp 1$ & $\mp 1$ & +0 & 1 & 1 & 0.000103 & 9.01e-05 & 9.51e-05 & 0.000277 & 0.000212 & 0.000342 & 0.000124 & 0.000256 & 0.000267 & 0.000272 & 0.000315 \\
 +0 & +0 & +0 & +0 & 0 & 0 & 0.000213 & 0.000696 & 0.00344 & 0.01 & 0.0168 & 0.0256 & 0.0288 & 0.0324 & 0.0289 & 0.0196 & 0.00804 \\
\hline
\end{tabular} 
\caption{
Approximate models for $|\psi_1\rangle$ (calculated on a reduced set of states),
given as a list of absolute frequencies $\rho_s = P_{\cal M}(s)$ for state $s$,
for a pair of values ($\theta$(rads), $\eta$),
simulating all quantum predictions regarding the observables defined in
(\ref{A1})--(\ref{B2}).
\textit{
The problem is solved in its GEN + NG version, a sufficient condition for either
the GEN or NG cases alone.}
Models obtained for $\mu < 0.00005$.
}\end{table}

\clearpage

\begin{table}[h]
\centering 
\begin{tabular}{ c c c c c c c c c c c c c c c c c}
$A_1$ & $A_2$ & $B_1$ & $B_2$ & $A$ & $B$ & $\theta = 0.1$  & $\theta = 0.2$  & $\theta = 0.3$  & $\theta = 0.4$  & $\theta = 0.5$  & $\theta = 0.6$  & $\theta = 0.7$  & $\theta = 0.8$  & $\theta = 0.9$  & $\theta = 1$  & $\theta = 1.1$  \\
& & & & &  & $\eta = 0.99$  & $\eta = 0.97$  & $\eta = 0.96$  & $\eta = 0.98$  & $\eta = 1$  & $\eta = 0.93$  & $\eta = 0.84$  & $\eta = 0.82$  & $\eta = 0.86$  & $\eta = 0.94$  & $\eta = 0.94$  \\
\hline
 $\pm 1$ & $\pm 1$ & $\pm 1$ & $\pm 1$ & 1 & 1 & 6.77e-05 & 0.000137 & 0.000173 & 7.51e-05 & 0.00522 & 0.0358 & 0.0391 & 0.0486 & 0.0694 & 0.1 & 0.0628 \\
 $\pm 1$ & $\pm 1$ & $\pm 1$ & $\mp 1$ & 1 & 1 & 5.8e-05 & 0.000132 & 0.000143 & 7.43e-05 & 0.00477 & 0.000109 & 0.000156 & 0.00015 & 4.62e-05 & 7.6e-05 & 0.00126 \\
 $\pm 1$ & $\pm 1$ & $\mp 1$ & $\pm 1$ & 1 & 1 & 0.0101 & 0.039 & 0.0847 & 0.152 & 0.215 & 0.165 & 0.085 & 0.0418 & 0.0172 & 0.00108 & 9.96e-05 \\
 $\pm 1$ & $\pm 1$ & $\mp 1$ & $\mp 1$ & 1 & 1 & 0.459 & 0.36 & 0.245 & 0.139 & 0.0325 & 8.82e-05 & 0.000149 & 0.000152 & 4.61e-05 & 7.15e-05 & 9.36e-05 \\
 $\pm 1$ & $\mp 1$ & $\pm 1$ & $\pm 1$ & 1 & 1 & 5.8e-05 & 0.000132 & 0.000143 & 7.43e-05 & 0.00477 & 0.000109 & 0.000156 & 0.00015 & 4.62e-05 & 7.6e-05 & 0.00126 \\
 $\pm 1$ & $\mp 1$ & $\pm 1$ & $\mp 1$ & 1 & 1 & 0.000461 & 0.00262 & 0.00792 & 0.0176 & 0.0154 & 0.000103 & 0.000147 & 0.000152 & 4.65e-05 & 8.23e-05 & 0.000103 \\
 $\pm 1$ & $\mp 1$ & $\mp 1$ & $\pm 1$ & 1 & 1 & 5.94e-05 & 0.000128 & 0.000144 & 5.89e-05 & 0.00725 & 0.000515 & 0.00878 & 0.0559 & 0.145 & 0.282 & 0.32 \\
 $\pm 1$ & $\mp 1$ & $\mp 1$ & $\mp 1$ & 1 & 1 & 0.0101 & 0.039 & 0.0847 & 0.152 & 0.215 & 0.165 & 0.085 & 0.0418 & 0.0172 & 0.00108 & 9.96e-05 \\
 +0 & $\pm 1$ & $\pm 1$ & $\pm 1$ & 0 & 1 & 0.000135 & 0.000304 & 0.000352 & 0.000189 & 2.7e-05 & 0.0149 & 0.0313 & 0.0355 & 0.031 & 0.0127 & 0.000241 \\
 +0 & $\pm 1$ & $\pm 1$ & $\pm 1$ & 1 & 1 & 0.000134 & 0.000272 & 0.000333 & 0.000145 & 4.58e-05 & 0.0165 & 0.0343 & 0.0367 & 0.0288 & 0.0147 & 0.000211 \\
 +0 & $\pm 1$ & $\pm 1$ & $\mp 1$ & 0 & 1 & 8.03e-05 & 0.000161 & 0.000189 & 0.000112 & 2.86e-05 & 0.000234 & 0.000339 & 0.00033 & 9.71e-05 & 0.000166 & 0.0137 \\
 +0 & $\pm 1$ & $\pm 1$ & $\mp 1$ & 1 & 1 & 6.79e-05 & 0.000143 & 0.000153 & 8.34e-05 & 5.44e-05 & 0.00019 & 0.000289 & 0.000267 & 9.54e-05 & 0.000146 & 0.0135 \\
 +0 & $\pm 1$ & $\mp 1$ & $\pm 1$ & 0 & 1 & 0.00222 & 0.00669 & 0.00874 & 0.00425 & 2.69e-05 & 0.00028 & 0.000335 & 0.00034 & 9.59e-05 & 0.000169 & 0.000124 \\
 +0 & $\pm 1$ & $\mp 1$ & $\pm 1$ & 1 & 1 & 0.00208 & 0.00647 & 0.00889 & 0.00469 & 4.54e-05 & 0.000202 & 0.000276 & 0.000277 & 8.87e-05 & 0.000158 & 0.00011 \\
 +0 & $\pm 1$ & $\mp 1$ & $\mp 1$ & 0 & 1 & 0.000131 & 0.000279 & 0.000303 & 0.000189 & 2.85e-05 & 0.000142 & 0.000197 & 0.000195 & 5.28e-05 & 9.32e-05 & 0.000216 \\
 +0 & $\pm 1$ & $\mp 1$ & $\mp 1$ & 1 & 1 & 0.000108 & 0.000243 & 0.000237 & 0.000144 & 5.37e-05 & 0.000113 & 0.000159 & 0.000158 & 4.59e-05 & 7.96e-05 & 0.000191 \\
 $\pm 1$ & +0 & $\pm 1$ & $\pm 1$ & 0 & 1 & 7.55e-05 & 0.00016 & 0.000194 & 0.000114 & 2.85e-05 & 0.000297 & 0.000324 & 0.000337 & 9.5e-05 & 0.000176 & 0.0134 \\
 $\pm 1$ & +0 & $\pm 1$ & $\pm 1$ & 1 & 1 & 6.86e-05 & 0.000145 & 0.000165 & 8.97e-05 & 5.43e-05 & 0.000238 & 0.00029 & 0.000269 & 9.27e-05 & 0.000171 & 0.0138 \\
 $\pm 1$ & +0 & $\pm 1$ & $\mp 1$ & 0 & 1 & 0.000121 & 0.000279 & 0.000321 & 0.000193 & 2.86e-05 & 0.000164 & 0.000196 & 0.000197 & 5.3e-05 & 9.62e-05 & 0.000226 \\
 $\pm 1$ & +0 & $\pm 1$ & $\mp 1$ & 1 & 1 & 0.00011 & 0.000258 & 0.000278 & 0.000157 & 5.44e-05 & 0.000131 & 0.00016 & 0.000158 & 4.62e-05 & 8.55e-05 & 0.000203 \\
 $\pm 1$ & +0 & $\mp 1$ & $\pm 1$ & 0 & 1 & 0.00014 & 0.000301 & 0.000346 & 0.000186 & 2.7e-05 & 0.0163 & 0.0341 & 0.0366 & 0.0287 & 0.0146 & 0.000225 \\
 $\pm 1$ & +0 & $\mp 1$ & $\pm 1$ & 1 & 1 & 0.000126 & 0.000261 & 0.000313 & 0.00013 & 4.64e-05 & 0.015 & 0.0315 & 0.0357 & 0.0311 & 0.0128 & 0.000194 \\
 $\pm 1$ & +0 & $\mp 1$ & $\mp 1$ & 0 & 1 & 0.00205 & 0.00638 & 0.00876 & 0.00457 & 2.67e-05 & 0.000274 & 0.000334 & 0.000336 & 9.58e-05 & 0.00016 & 0.000121 \\
 $\pm 1$ & +0 & $\mp 1$ & $\mp 1$ & 1 & 1 & 0.00226 & 0.00677 & 0.00883 & 0.00437 & 4.42e-05 & 0.000201 & 0.00028 & 0.000275 & 8.89e-05 & 0.000145 & 0.000104 \\
 $\pm 1$ & $\pm 1$ & +0 & $\pm 1$ & 1 & 0 & 0.000135 & 0.000304 & 0.000352 & 0.000189 & 2.7e-05 & 0.0149 & 0.0313 & 0.0355 & 0.031 & 0.0127 & 0.000241 \\
 $\pm 1$ & $\pm 1$ & +0 & $\pm 1$ & 1 & 1 & 0.000134 & 0.000272 & 0.000333 & 0.000145 & 4.58e-05 & 0.0165 & 0.0343 & 0.0367 & 0.0288 & 0.0147 & 0.000211 \\
 $\pm 1$ & $\pm 1$ & +0 & $\mp 1$ & 1 & 0 & 0.000131 & 0.000279 & 0.000303 & 0.000189 & 2.85e-05 & 0.000142 & 0.000197 & 0.000195 & 5.28e-05 & 9.32e-05 & 0.000216 \\
 $\pm 1$ & $\pm 1$ & +0 & $\mp 1$ & 1 & 1 & 0.000108 & 0.000243 & 0.000237 & 0.000144 & 5.37e-05 & 0.000113 & 0.000159 & 0.000158 & 4.59e-05 & 7.96e-05 & 0.000191 \\
 $\pm 1$ & $\mp 1$ & +0 & $\pm 1$ & 1 & 0 & 8.03e-05 & 0.000161 & 0.000189 & 0.000112 & 2.86e-05 & 0.000234 & 0.000339 & 0.00033 & 9.71e-05 & 0.000166 & 0.0137 \\
 $\pm 1$ & $\mp 1$ & +0 & $\pm 1$ & 1 & 1 & 6.79e-05 & 0.000143 & 0.000153 & 8.34e-05 & 5.44e-05 & 0.00019 & 0.000289 & 0.000267 & 9.54e-05 & 0.000146 & 0.0135 \\
 $\pm 1$ & $\mp 1$ & +0 & $\mp 1$ & 1 & 0 & 0.00222 & 0.00669 & 0.00874 & 0.00425 & 2.69e-05 & 0.00028 & 0.000335 & 0.00034 & 9.59e-05 & 0.000169 & 0.000124 \\
 $\pm 1$ & $\mp 1$ & +0 & $\mp 1$ & 1 & 1 & 0.00208 & 0.00647 & 0.00889 & 0.00469 & 4.54e-05 & 0.000202 & 0.000276 & 0.000277 & 8.87e-05 & 0.000158 & 0.00011 \\
 $\pm 1$ & $\pm 1$ & $\pm 1$ & +0 & 1 & 0 & 7.55e-05 & 0.00016 & 0.000194 & 0.000114 & 2.85e-05 & 0.000297 & 0.000324 & 0.000337 & 9.5e-05 & 0.000176 & 0.0134 \\
 $\pm 1$ & $\pm 1$ & $\pm 1$ & +0 & 1 & 1 & 6.86e-05 & 0.000145 & 0.000165 & 8.97e-05 & 5.43e-05 & 0.000238 & 0.00029 & 0.000269 & 9.27e-05 & 0.000171 & 0.0138 \\
 $\pm 1$ & $\pm 1$ & $\mp 1$ & +0 & 1 & 0 & 0.00205 & 0.00638 & 0.00876 & 0.00457 & 2.67e-05 & 0.000274 & 0.000334 & 0.000336 & 9.58e-05 & 0.00016 & 0.000121 \\
 $\pm 1$ & $\pm 1$ & $\mp 1$ & +0 & 1 & 1 & 0.00226 & 0.00677 & 0.00883 & 0.00437 & 4.42e-05 & 0.000201 & 0.00028 & 0.000275 & 8.89e-05 & 0.000145 & 0.000104 \\
 $\pm 1$ & $\mp 1$ & $\pm 1$ & +0 & 1 & 0 & 0.000121 & 0.000279 & 0.000321 & 0.000193 & 2.86e-05 & 0.000164 & 0.000196 & 0.000197 & 5.3e-05 & 9.62e-05 & 0.000226 \\
 $\pm 1$ & $\mp 1$ & $\pm 1$ & +0 & 1 & 1 & 0.00011 & 0.000258 & 0.000278 & 0.000157 & 5.44e-05 & 0.000131 & 0.00016 & 0.000158 & 4.62e-05 & 8.55e-05 & 0.000203 \\
 $\pm 1$ & $\mp 1$ & $\mp 1$ & +0 & 1 & 0 & 0.00014 & 0.000301 & 0.000346 & 0.000186 & 2.7e-05 & 0.0163 & 0.0341 & 0.0366 & 0.0287 & 0.0146 & 0.000225 \\
 $\pm 1$ & $\mp 1$ & $\mp 1$ & +0 & 1 & 1 & 0.000126 & 0.000261 & 0.000313 & 0.00013 & 4.64e-05 & 0.015 & 0.0315 & 0.0357 & 0.0311 & 0.0128 & 0.000194 \\
 +0 & +0 & +0 & +0 & 0 & 0 & 0.000235 & 0.000881 & 0.00159 & 0.000397 & 4.37e-05 & 0.00488 & 0.0256 & 0.0324 & 0.0194 & 0.00346 & 0.00347 \\
\hline
\end{tabular} 
\caption{
Approximate models for $|\psi_2\rangle$ (calculated on a reduced set of states),
given as a list of absolute frequencies $\rho_s = P_{\cal M}(s)$ for state $s$,
for a pair of values ($\theta$(rads), $\eta$),
simulating all quantum predictions regarding the observables defined in
(\ref{A1})--(\ref{B2}).
\textit{
The problem is solved in its GEN + NG version, a sufficient condition for either
the GEN or NG cases alone.}
Models obtained for $\mu < 0.00005$.
}\end{table}

\clearpage

\begin{figure}[ht!] 
\includegraphics[width=2.5 \columnwidth,clip]{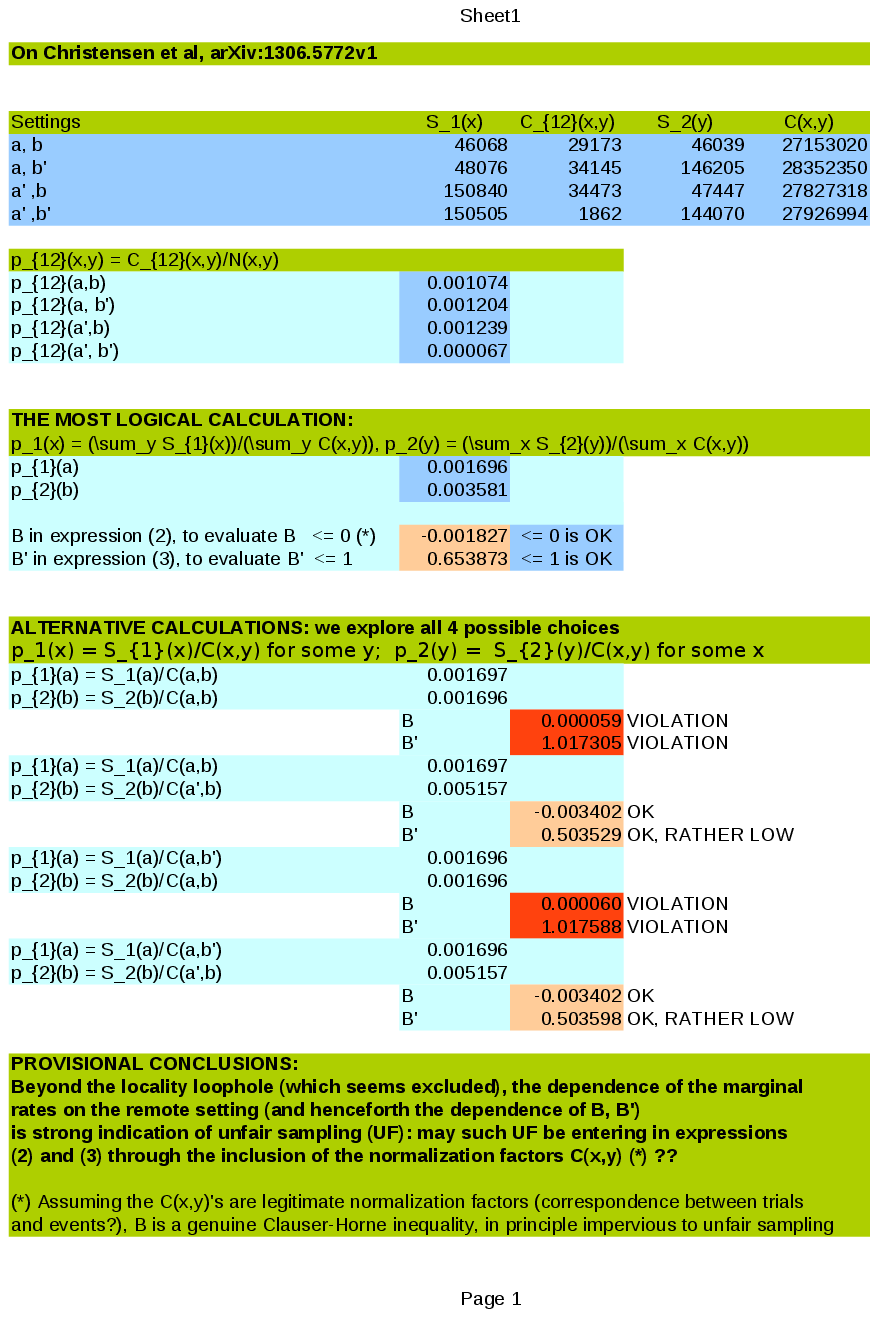} 
\caption{
Preliminary analysis on Christensen \textit{et al}, arXiv.
} \label{Chr_et_al} \end{figure}



\begin{thebibliography}{99}
\bibitem{fs}
From \cite{CHSH69}:
``given a pair of photons emerges from the polarizers, the probability of their joint
detection is independent of the polarizer orientations''.
\bibitem{no_enhancement}
From \cite{CH74}:
``for every emission $\lambda$, the probability of a count with a polarizer in place
is less or equal to the probability with the polarizer removed'', where $\lambda$ is
the (hypothetical) hidden variable expressing the state (at least the initial one)
of the pair of particles.
\bibitem{CH_conv}
In our expression for the CH inequality, 
be aware of:
\begin{table}[ht]
\centering \footnotesize
\begin{tabular}{ c c }
Ours here (and in \cite{CLR09}) & Larsson-Semitecolos in \cite{LS01}   \\ \hline
$A_1$ &  $A_1$ \\
$A_2$ &  $B_1$ \\
$B_1$ &  $A_2$ \\
$B_2$ &  $B_2$ \\  \hline
\end{tabular} \end{table}
\bibitem{lr}
Local realism or ``local causality'': the result of any measurement upon
the system (or its probability) is predetermined by a vector of local
hidden variables.
Local means that no hidden variable can influence the result of any
measurement performed out of its ``light cone''.
\bibitem{note_det}
Indeterminism: it is also possible to give a definition of local realism in terms
of probabilities, such as $P(A=a) = f_A(\lambda,a), P(B=b) = f_B(\lambda,b)$ for
$a,b \in \{\pm 1\}$ (or, equivalently, $P(A=a|B=b,\lambda,a,b)=P(A=a|\lambda,a)$
and the analogous for $B$).\\
That definition is, however, always reducible to one of the kind
$A=A(\lambda^{\prime},a),B=B(\lambda^{\prime},b)$, where the hidden variable has
been now redefined from $\lambda$ to $\lambda^{\prime}=\lambda \oplus \mu$, and
where $\mu$ is an extra set of variables that also describe the state of the setup.
A proof of this is almost trivial: it suffices to define $\mu$ as a label for each
of the possible results for a given $\lambda$.
\bibitem{d_channel}
In a double channel experiment, detections are recorded both for each of the
two orthogonal polarizations, i.e. counts are actually recorded in both the $+1$ and
$-1$ channels;
this type of setup would allow to test the CHSH and CH inequalities simultaneously, 
or even the complete set of quantum predictions for a given state and set of four
observables: see Sec.\ref{Full_M}.
\bibitem{note_LHV_CHSH}
The values of $p,q$ as a function of $\beta,\eta$ are obtained from the following
set of equations:
\begin{eqnarray}
&\beta({\cal M}) = \nonumber\\
&\frac{ \beta({\cal M}_P) P_{{\cal M}_P}(A_i,B_j) \cdot p + \beta({\cal M}_Q) \cdot P_{{\cal M}_Q}(A_i,B_j) \cdot q }
{ P_{{\cal M}_P}(A_i,B_j) \cdot p + P_{{\cal M}_Q}(A_i,B_j) \cdot q},
\end{eqnarray}
which is actually only one restriction due to the symmetry of
$P_{{\cal M}_P}(A_i,B_j),P_{{\cal M}_Q}(A_i,B_j)$ for all $i,j$,
and
\begin{eqnarray}
&\eta({\cal M}) = 
\frac{ P_{{\cal M}_P}(A_i,B_j) \cdot p +P_{{\cal M}_Q}(A_i,B_j) \cdot q }
{ P_{{\cal M}_P}(\Theta) \cdot p + P_{{\cal M}_Q}(\Theta) \cdot q}, \\
&\eta({\cal M}) =  P_{{\cal M}_P}(\Theta) \cdot p +P_{{\cal M}_Q}(\Theta) \cdot q,
\end{eqnarray}
both for $\Theta = A_1,A_2,B_1,B_2$,
and which again turn to just two restrictions, due to the symmetry of
$P_{{\cal M}_P}(\Theta),P_{{\cal M}_Q}(\Theta)$ for all $\Theta$'s.\\
Now, using
$\beta({\cal M}_P) = 2$, $\beta({\cal M}_Q) = 4$, 
$P_{{\cal M}_P}(A_i,B_j) = 1$, 
$P_{{\cal M}_Q}(A_i,B_j) = \tfrac{1}{2}$, 
$P_{{\cal M}_P}(A_i) = P_{{\cal M}_P}(B_j) = 1$ and
$P_{{\cal M}_Q}(A_i) = P_{{\cal M}_Q}(B_j) = \tfrac{1}{2}$,
we arrive to (\ref{eta_beta}) and (\ref{p})--(\ref{q}), of course together
with the complementary condition $r=1-p-q$.
\bibitem{models_lower_heta}
Just for completeness, we provide an immediate way of doing so: it suffices to add to
the model a new subset of states ${\cal M}_{Q^{\prime}}$ with one '0' per state (the
same kind of states as in ${\cal M}_Q$), $\pm 1$ evenly distributed and simulating
$\beta = 2$, for instance:
${\cal M}_{Q^{\prime}} \equiv \{
(\pm 1,\pm 1;\pm 1,0),$ $
(\pm 1,\mp 1;0,\pm 1),$ $
(\pm 1,0;\pm 1,\pm 1),$ $
(0,\pm 1;\pm 1,\mp 1)\}$,
so that now
$p = P({\cal M}_P)$,
$q = P({\cal M}_Q)$,
$q^{\prime}= P({\cal M}_{Q^{\prime}})$, and of course
$r = 1-p-q-q^{\prime} = P({\cal M}_R)$.\\
Adding this new subset ${\cal M}_{Q^{\prime}}$ to the model has the effect of decreasing
$\eta$, while the parameters $p,q,q^{\prime}$ can always be readjusted to leave $\beta$
untouched.
\bibitem{proposal}
I propose my definition is adopted from now on,
and the usual one henceforth referred to as ``average independent errors''.
\bibitem{div}
Looking at (\ref{ng_beta}), 
let us first ignore all probabilities that enter with negative sign, as these cannot
give rise to a positive divergence of $\beta$;
we are left, then, with terms of the kind $P(A_i=a,B_j=b)$ at the numerator, while
the denominator is simply equal to $\eta^2$.
It is now clear that forcing (\ref{cond_joint}) for all $i,j$ sets 
\begin{eqnarray}
\lim_{\eta \rightarrow 0} \beta_{ng}^{CH}(\eta) = constant,
\end{eqnarray}
given that, naturally, $P_{QM}(A_i=a,B_j=b)$ in the right hand side of
(\ref{cond_joint}) are $\eta$-independent.
\bibitem{Glauber}
Glauber establishes a probability of detection proportional to the expectation value
of the field intensity at the entrance of the detector:
\begin{eqnarray}
P(\mathbf{r},t) \propto 
\langle \psi(\mathbf{r},t) | \mathbf{E}^{(-)}\mathbf{E}^{(+)} | \psi(\mathbf{r},t) \rangle,
\end{eqnarray}
where $|\psi\rangle$ represents the state of the fields at all points
of space, and where $\hat{E}^{(-)},\hat{E}^{(+)}$ are local field operators.
\bibitem{CH_fac}
In \cite{CH74}'s original notation, such condition reads:
\begin{eqnarray}
p_{12}(a,\infty|\lambda) &=& p_1(\lambda, a)      \cdot p_2(\lambda, \infty), \label{CH_fac_A}\\
p_{12}(\infty,b|\lambda) &=& p_1(\lambda, \infty) \cdot p_2(\lambda, b),      \label{CH_fac_B}
\end{eqnarray}
where $p_i(\lambda,a)$ is the probability of obtaining a detection at the $i$-th detector
conditioned to the hidden variable $\lambda$ with device orientation $a$, and
$p_i(\lambda,\infty)$ is the analogous in the absence of a polarizer.\\
Conditions (\ref{CH_fac_A})--(\ref{CH_fac_B}) are not in general true for any definition
of the hidden variable $\lambda$:
for instance, suppose that in an LHV model determining results for two measurementes $A,B$,
we have that for $\lambda=\lambda_0$, either $A=B=1$ or $A=B=0$ with equal probability;
it is immediate to see that $P(A=B=1|\lambda_0) \neq P(A=1|\lambda_0) \cdot P(B=1|\lambda_0)$
($\tfrac{1}{2} \neq \tfrac{1}{4}$). 
Nevertheless, one can always build a new hidden variable $\lambda^{\prime} = \lambda \oplus \mu$
(for instance $\mu$ implements a labeling of all the states compatible with $\lambda$)
for which determinism holds, and hence factorization as well.
\bibitem{note_leads}
First, we note that (\ref{n_enh}) can alternatively be written as
\begin{eqnarray}
P_{s}(A_i \cap A)   &=& P_{s}(A_i), \label{n_enh_alt_A} 
\end{eqnarray}
as well as $P_{s}(B_j \cap B) = P_{s}(B_j)$,
both conditions of course $\forall s\in {\cal M}^{\prime}$, which can
therefore be averaged to yield, for instance,
\begin{eqnarray}
P_{{\cal M}^{\prime}}(A_i \cap A) =  P_{{\cal M}^{\prime}}(A_i).
\label{n_enh_alt_A_av}
\end{eqnarray}
Now, we write, dropping model-subscripts for clarity,
\begin{eqnarray}
P(A_1=1|A) 
&=& P(A_1=1|A_1 \cap A) \nonumber\\
&=& P(A_1=1, A_1 \cap A)/P(A_1 \cap A),
\end{eqnarray}
and using the obvious
$P(A_1 \cap A) \leq P(A) = \eta$,
as well as
$P(A_1=1, A_1 \cap A) = P(A_1=1, A)$,
we obtain
\begin{eqnarray}
P(A_1=1|A) 
&\geq& P(A_1=1 \cap A)/P(A) \nonumber\\
&=&    P(A_1=1 \cap A)/\eta, 
\end{eqnarray}
which, using now (\ref{n_enh_alt_A_av}), finally turns into
\begin{eqnarray}
P(A_1=1|A) \geq P(A_1=1)/\eta,
\end{eqnarray}
i.e., we get to (\ref{ne_A});
following the analogous path with $B$'s we can get to (\ref{ne_B}) as well.
\end{thebibliography}
\end{document}